\documentclass[journal,10pt,margin=1in]{IEEEtran}
\usepackage{amsmath,amsfonts,mathtools}
\usepackage{pifont}
\usepackage{algorithmic}
\usepackage{algorithm}
\usepackage{array}
\usepackage{textcomp}
\usepackage{stfloats}
\usepackage{url}
\usepackage{verbatim}
\usepackage{graphicx}
\usepackage{gensymb}
\usepackage{cite}
\usepackage{cases}
\usepackage{glossaries}
\usepackage{multirow}
\usepackage{makecell}
\usepackage{color, colortbl}
\definecolor{mixed}{gray}{0.9}
\definecolor{synth}{gray}{0.8}
\usepackage{multicol}
\usepackage[tableposition=top]{caption}
\usepackage{subcaption}
\usepackage{siunitx}

\usepackage{colortbl}
\usepackage{pgfplots}
\usepackage{pgfplotstable}
\usepgfplotslibrary{fillbetween}
\usepackage{longtable}
\usepackage{xspace}
\usepackage{arydshln}
\usepackage{booktabs}

\usepackage{soul}

\usepackage[font=scriptsize]{subcaption}
\usepackage[font=footnotesize]{caption}

\usepackage[outdir=./]{epstopdf}
\usepackage{tikz}
\pgfplotsset{compat=newest}
\pgfplotsset{plot coordinates/math parser=false}
\newlength\fheight
\newlength\fwidth
\usetikzlibrary{plotmarks,shapes,patterns,decorations.pathreplacing,backgrounds,calc,arrows,arrows.meta,spy,matrix,external}
\usepackage{tikzscale}
\usepackage[utf8]{inputenc}

\usepackage[capitalise]{cleveref}
\crefname{section}{Sec.}{Secs.}
\crefname{figure}{Fig.}{Figs.}

\usepackage{eqparbox}

\newcommand*\diff{\mathop{}\!\mathrm{d}}

\DeclareCaptionLabelFormat{andtable}{#1~#2  \&  \tablename~\thetable}

\newacronym{3gpp}{3GPP}{3rd Generation Partnership Project}
\newacronym{5g}{5G}{5th Generation}
\newacronym{5gc}{5GC}{5G Core}
\newacronym{6g}{6G}{6th Generation}
\newacronym{adc}{ADC}{Analog to Digital Converter}
\newacronym{afbw}{AFBW}{Average Fading Bandwidth}
\newacronym{aimd}{AIMD}{Additive Increase Multiplicative Decrease}
\newacronym{am}{AM}{Acknowledged Mode}
\newacronym{amc}{AMC}{Adaptive Modulation and Coding}
\newacronym{aoa}{AoA}{Angle of Arrival}
\newacronym{aod}{AoD}{Angle of Departure}
\newacronym{ap}{AP}{Access Point}
\newacronym{app}{APP}{Application Layer}
\newacronym{aqm}{AQM}{Active Queue Management}
\newacronym{awgn}{AGWN}{Additive White Gaussian Noise}
\newacronym{balia}{BALIA}{Balanced Link Adaptation}
\newacronym{bdp}{BDP}{Bandwidth-Delay Product}
\newacronym{ber}{BER}{Bit Error Rate}
\newacronym{bler}{BLER}{Block Error Rate}
\newacronym{bf}{BF}{Beamforming}
\newacronym{cad}{CAD}{Computer-Aided Design}
\newacronym{cbr}{CBR}{Constant Bit Rate}
\newacronym{cc}{CC}{Congestion Control}
\newacronym{cdf}{CDF}{Cumulative Distribution Function}
\newacronym{ci}{CI}{Confidence Interval}
\newacronym{cir}{CIR}{Channel Impulse Response}
\newacronym{cn}{CN}{Core Network}
\newacronym{cp}{CP}{Control Plane}
\newacronym{cqi}{CQI}{Channel Quality Information}
\newacronym{crs}{CRS}{Cell Reference Signal}
\newacronym{csirs}{CSI-RS}{Channel State Information - Reference Signal}
\newacronym{d2d}{D2D}{Device-to-Device}
\newacronym{dc}{DC}{Dual Connectivity}
\newacronym{dce}{DCE}{Direct Code Execution}
\newacronym{dci}{DCI}{Downlink Control Information}
\newacronym{dl}{DL}{Downlink}
\newacronym{dmr}{DMR}{Deadline Miss Ratio}
\newacronym{dmrs}{DMRS}{DeModulation Reference Signal}
\newacronym{dray}{D-Ray}{Deterministic Ray}
\newacronym{e2e}{E2E}{End-to-End}
\newacronym{ecn}{ECN}{Explicit Congestion Notification}
\newacronym{ecdf}{ECDF}{Empirical Cumulative Distribution Function}
\newacronym{edf}{EDF}{Earliest Deadline First}
\newacronym{em}{EM}{electromagnetic}
\newacronym{enb}{eNB}{evolved Node Base}
\newacronym{endc}{EN-DC}{E-UTRAN-\gls{nr} \gls{dc}}
\newacronym{epc}{EPC}{Evolved Packet Core}
\newacronym{es}{ES}{Edge Server}
\newacronym{eess}{EESS}{Earth Exploration-Satellite Service}
\newacronym{fdd}{FDD}{Frequency Division Duplexing}
\newacronym{fdma}{FDMA}{Frequency Division Multiple Access}
\newacronym{fray}{F-Ray}{Flashing Ray}
\newacronym{fs}{FS}{Fast Switching}
\newacronym{fss}{FSS}{Fixed Satellite Service}
\newacronym{ftp}{FTP}{File Transfer Protocol}
\newacronym{gmm}{GMM}{Gaussian Mixture Model}
\newacronym{gnb}{gNB}{Next Generation Node Base}
\newacronym{gr}{GR}{Ground Reflection}
\newacronym{harq}{HARQ}{Hybrid Automatic Repeat reQuest}
\newacronym{hetnet}{HetNet}{Heterogeneous Network}
\newacronym{hh}{HH}{Hard Handover}
\newacronym{hol}{HOL}{Head-of-Line}
\newacronym{hpbw}{HPBW}{Half Power Beamwidth}
\newacronym{hqf}{HQF}{Highest-quality-first}
\newacronym{ia}{IA}{Initial Access}
\newacronym{iab}{IAB}{Integrated Access and Backhaul}
\newacronym{ieee}{IEEE}{Institute of Electrical and Electronics Engineers}
\newacronym{imt}{IMT}{International Mobile Telecommunication}
\newacronym{inr}{INR}{Interference to Noise Ratio}
\newacronym{iot}{IoT}{Internet of Things}
\newacronym{ked}{KED}{Knife-Edge Diffraction}
\newacronym{kpi}{KPI}{Key Performance Indicator}
\newacronym{ks}{KS}{Kolmogorov–Smirnov}
\newacronym{lcf}{LCF}{Level Crossing Frequency}
\newacronym{lcr}{LCR}{Level Crossing Rate}
\newacronym{los}{LoS}{Line-of-Sight}
\newacronym{lte}{LTE}{Long Term Evolution}
\newacronym{m2m}{M2M}{Machine to Machine}
\newacronym{mac}{MAC}{Medium Access Control}
\newacronym{mc}{MC}{Multi-Connectivity}
\newacronym{mcl}{MCL}{Minimum Coupling Loss}
\newacronym{mcs}{MCS}{Modulation and Coding Scheme}
\newacronym{mec}{MEC}{Mobile Edge Cloud}
\newacronym{mi}{MI}{Mutual Information}
\newacronym{mib}{MIB}{Master Information Block}
\newacronym{mimo}{MIMO}{Multiple Input, Multiple Output}
\newacronym{mlr}{MLR}{Maximum-local-rate}
\newacronym{mls}{MLS}{Microwave Limb Sounder}
\newacronym[plural=\gls{mme}s,firstplural=Mobility Management Entities (MMEs)]{mme}{MME}{Mobility Management Entity}
\newacronym{mmwave}{mmWave}{millimeter wave}
\newacronym{moi}{MoI}{Method of Images}
\newacronym{mpc}{MPC}{Multi Path Component}
\newacronym{mptcp}{MPTCP}{Multipath TCP}
\newacronym{mr}{MR}{Maximum Rate}
\newacronym{mrdc}{MR-DC}{Multi \gls{rat} \gls{dc}}
\newacronym{mss}{MSS}{Maximum Segment Size}
\newacronym{mtd}{MTD}{Machine-Type Device}
\newacronym{mtu}{MTU}{Maximum Transmission Unit}
\newacronym{nfv}{NFV}{Network Function Virtualization}
\newacronym{nist}{NIST}{National Institute of Standards and Technology}
\newacronym{nlos}{NLoS}{Non-Line-of-Sight}
\newacronym{nr}{NR}{New Radio}
\newacronym{nrmse}{NRMSE}{Normalized Root Mean Square Error}
\newacronym{ns3}{ns-3}{Network Simulator 3}
\newacronym{nsa}{NSA}{Non Stand Alone}
\newacronym{ntn}{NTN}{Non Terrestrial Network}
\newacronym{o2i}{O2I}{Outdoor-to-Indoor}
\newacronym{ofdm}{OFDM}{Orthogonal Frequency Division Multiplexing}
\newacronym{osm}{OSM}{OpenStreetMap}
\newacronym{pa}{PA}{Position-aware}
\newacronym{pbch}{PBCH}{Physical Broadcast Channel}
\newacronym{pdcch}{PDCCH}{Physical Downlonk Control Channel}
\newacronym{pdcp}{PDCP}{Packet Data Convergence Protocol}
\newacronym{pdf}{PDF}{Probability Density Function}
\newacronym{pdsch}{PDSCH}{Physical Downlink Shared Channel}
\newacronym{pdu}{PDU}{Packet Data Unit}
\newacronym{per}{PER}{Packet Error Rate}
\newacronym{pf}{PF}{Proportional Fair}
\newacronym{pgw}{PGW}{Packet Gateway}
\newacronym{phy}{PHY}{Physical}
\newacronym{pl}{PL}{Path Loss}
\newacronym{ppp}{PPP}{Poisson Point Process}
\newacronym{prb}{PRB}{Physical Resource Block}
\newacronym{pss}{PSS}{Primary Synchronization Signal}
\newacronym{pucch}{PUCCH}{Physical Uplink Control Channel}
\newacronym{pusch}{PUSCH}{Physical Uplink Shared Channel}
\newacronym{qd}{QD}{Quasi Deterministic}
\newacronym{rach}{RACH}{Random Access Channel}
\newacronym{ran}{RAN}{Radio Access Network}
\newacronym[firstplural=Radio Access Technologies (RATs)]{rat}{RAT}{Radio Access Technology}
\newacronym{red}{RED}{Random Early Detection}
\newacronym{rf}{RF}{Radio Frequency}
\newacronym{rfi}{RFI}{Radio Frequency Interference}
\newacronym{rlc}{RLC}{Radio Link Control}
\newacronym{rlf}{RLF}{Radio Link Failure}
\newacronym{rr}{RR}{Round Robin}
\newacronym{rray}{R-Ray}{Random Ray}
\newacronym{rrc}{RRC}{Radio Resource Control}
\newacronym{rrm}{RRM}{Radio Resource Management}
\newacronym{rs}{RS}{Remote Sensing}
\newacronym{rsrp}{RSRP}{Reference Signal Received Power}
\newacronym{rsrq}{RSRQ}{Reference Signal Received Quality}
\newacronym{rss}{RSS}{Received Signal Strength}
\newacronym{rssi}{RSSI}{Received Signal Strength Indicator}
\newacronym{rt}{RT}{Ray Tracer}
\newacronym{rtt}{RTT}{Round Trip Time}
\newacronym{rw}{RW}{Receive Window}
\newacronym{rx}{RX}{Receiver}
\newacronym{sa}{SA}{standalone}
\newacronym{sack}{SACK}{Selective Acknowledgment}
\newacronym{sap}{SAP}{Service Access Point}
\newacronym{sch}{SCH}{Secondary Cell Handover}
\newacronym{scm}{SCM}{Spatial Channel Model}
\newacronym{scoot}{SCOOT}{Split Cycle Offset Optimization Technique}
\newacronym{sdma}{SDMA}{Spatial Division Multiple Access}
\newacronym{sf}{SF}{Shadow Fading}
\newacronym{si}{SI}{Study Item}
\newacronym{sib}{SIB}{Secondary Information Block}
\newacronym{sinr}{SINR}{Signal-to-Interference-plus-Noise Ratio}
\newacronym{sir}{SIR}{Signal-to-Interference Ratio}
\newacronym{sm}{SM}{Saturation Mode}
\newacronym{snr}{SNR}{Signal-to-Noise Ratio}
\newacronym{son}{SON}{Self-Organizing Network}
\newacronym{srs}{SRS}{Sounding Reference Signal}
\newacronym{ss}{SS}{Synchronization Signal}
\newacronym{sss}{SSS}{Secondary Synchronization Signal}
\newacronym{sta}{STA}{Station}
\newacronym{subthz}{sub-THz}{sub-TeraHertz}
\newacronym{svd}{SVD}{Singular Value Decomposition}
\newacronym{tb}{TB}{Transport Block}
\newacronym{tcp}{TCP}{Transmission Control Protocol}
\newacronym{udp}{UDP}{User Datagram Protocol}
\newacronym{tdd}{TDD}{Time Division Duplexing}
\newacronym{tdma}{TDMA}{Time Division Multiple Access}
\newacronym{te}{TE}{Transverse Electric}
\newacronym{tfl}{TfL}{Transport for London}
\newacronym{tgad}{TGad}{Task Group ad}
\newacronym{tgay}{TGay}{Task Group ay}
\newacronym{tm}{TM}{Transverse Magnetic}
\newacronym{trp}{TRP}{Transmitter Receiver Pair}
\newacronym{tti}{TTI}{Transmission Time Interval}
\newacronym{ttt}{TTT}{Time-to-Trigger}
\newacronym{tx}{TX}{Transmitter}
\newacronym{ue}{UE}{User Equipment}
\newacronym{ul}{UL}{Uplink}
\newacronym{um}{UM}{Unacknowledged Mode}
\newacronym{uma}{UMa}{Urban Macro}
\newacronym{uml}{UML}{Unified Modeling Language}
\newacronym{utc}{UTC}{Urban Traffic Control}
\newacronym{vm}{VM}{Virtual Machine}
\newacronym{wbf}{WBF}{Wired Bias Function}
\newacronym{wf}{WF}{Wired-first}
\newacronym{wifi}{Wi-Fi}{Wireless Fidelity}
\newacronym{wigig}{WiGig}{Wireless Gigabit}
\newacronym{wlan}{WLAN}{Wireless Local Area Network}
\newacronym{xpr}{XPR}{Cross Polarization Ratio}
\newacronym{sthz}{Sub-THz}{sub-terahertz}
\newacronym{thz}{THz}{terahertz}
\tikzstyle{startstop} = [rectangle, rounded corners, minimum width=2cm, minimum height=0.5cm,text centered, draw=black]
\tikzstyle{io} = [trapezium, trapezium left angle=70, trapezium right angle=110, minimum width=3cm, minimum height=1cm, text centered, draw=black]
\tikzstyle{process} = [rectangle, minimum width=2cm, minimum height=0.5cm, text centered, draw=black, alignb=center]
\tikzstyle{decision} = [ellipse, minimum width=2cm, minimum height=1cm, text centered, draw=black]
\tikzstyle{arrow} = [thick,<->,>=stealth]
\tikzstyle{line} = [thick,>=stealth]
\tikzstyle{darrow} = [thick,<->,>=stealth,dashed]
\tikzstyle{sarrow} = [thick,->,>=stealth]
\tikzstyle{larrow} = [line width=0.1mm,dashdotted,->,>=stealth]

\makeatletter
\def\grd@save@target#1{%
  \def\grd@target{#1}}
\def\grd@save@start#1{%
  \def\grd@start{#1}}
\tikzset{
  grid with coordinates/.style={
    to path={%
      \pgfextra{%
        \edef\grd@@target{(\tikztotarget)}%
        \tikz@scan@one@point\grd@save@target\grd@@target\relax
        \edef\grd@@start{(\tikztostart)}%
        \tikz@scan@one@point\grd@save@start\grd@@start\relax
        \draw[minor help lines] (\tikztostart) grid (\tikztotarget);
        \draw[major help lines] (\tikztostart) grid (\tikztotarget);
        \grd@start
        \pgfmathsetmacro{\grd@xa}{\the\pgf@x/1cm}
        \pgfmathsetmacro{\grd@ya}{\the\pgf@y/1cm}
        \grd@target
        \pgfmathsetmacro{\grd@xb}{\the\pgf@x/1cm}
        \pgfmathsetmacro{\grd@yb}{\the\pgf@y/1cm}
        \pgfmathsetmacro{\grd@xc}{\grd@xa + \pgfkeysvalueof{/tikz/grid with coordinates/major step x}}
        \pgfmathsetmacro{\grd@yc}{\grd@ya + \pgfkeysvalueof{/tikz/grid with coordinates/major step y}}
        \foreach \x in {\grd@xa,\grd@xc,...,\grd@xb}
        \node[anchor=north] at (\x,\grd@ya) {\pgfmathprintnumber{\x}};
        \foreach \y in {\grd@ya,\grd@yc,...,\grd@yb}
        \node[anchor=east] at (\grd@xa,\y) {\pgfmathprintnumber{\y}};
      }
    }
  },
  minor help lines/.style={
    help lines,
    gray,
    line cap =round,
    xstep=\pgfkeysvalueof{/tikz/grid with coordinates/minor step x},
    ystep=\pgfkeysvalueof{/tikz/grid with coordinates/minor step y}
  },
  major help lines/.style={
    help lines,
    line cap =round,
    line width=\pgfkeysvalueof{/tikz/grid with coordinates/major line width},
    xstep=\pgfkeysvalueof{/tikz/grid with coordinates/major step x},
    ystep=\pgfkeysvalueof{/tikz/grid with coordinates/major step y}
  },
  grid with coordinates/.cd,
  minor step x/.initial=.5,
  minor step y/.initial=.2,
  major step x/.initial=1,
  major step y/.initial=1,
  major line width/.initial=1pt,
}
\makeatother

\begin{document}

\title{Modeling Interference for the Coexistence of\\6G Networks and Passive Sensing Systems}

\author{Paolo~Testolina,~\IEEEmembership{Student~Member,~IEEE},  Michele~Polese,~\IEEEmembership{Member,~IEEE},\\Josep M. Jornet,~\IEEEmembership{Senior Member,~IEEE},
    Tommaso~Melodia,~\IEEEmembership{Fellow,~IEEE},
    Michele~Zorzi,~\IEEEmembership{Fellow,~IEEE}
    \thanks{P. Testolina and M. Zorzi are with the Department of Information Engineering, University of Padova, Padova, Italy. Email: \{testolina,zorzi\}@dei.unipd.it. M. Polese, J. M. Jornet, and T. Melodia are with the Institute for the Wireless Internet of Things, Northeastern University, Boston, MA. Email: \{m.polese, j.jornet, melodia\}@northeastern.edu. The work of P.\ Testolina was supported by Fondazione CaRiPaRo under grants ``Dottorati di Ricerca'' 2019. This work also partially funded by NSF with award CNS-2225590.}
}

\markboth{This paper has been submitted to IEEE Transactions on Wireless Communications. Copyright may change without notice.}%
{Testolina \MakeLowercase{\textit{et al.}}: Modeling Interference for the Coexistence of 6G Networks and Passive Sensing Systems}




\makeatletter
\patchcmd{\@maketitle}
{\addvspace{0.5\baselineskip}\egroup}
{\addvspace{-1.5\baselineskip}\egroup}
{}
{}
\makeatother

\maketitle

\begin{abstract}

    Future wireless networks and sensing systems will benefit from access to large chunks of spectrum above 100~GHz, to achieve terabit-per-second data rates in \gls{6g} cellular systems and improve accuracy and reach of Earth exploration and sensing and radio astronomy applications. These are extremely sensitive to interference from artificial signals, thus the spectrum above 100~GHz features several bands which are protected from active transmissions under current spectrum regulations. To provide more agile access to the spectrum for both services, active and passive users will have to coexist without harming passive sensing operations. In this paper, we provide the first, fundamental analysis of \gls{rfi} that large-scale terrestrial deployments introduce in different satellite sensing systems now orbiting the Earth. We develop a geometry-based analysis and extend it into a data-driven model which accounts for realistic propagation, building obstruction, ground reflection, for network topology with up to $10^5$ nodes in more than $85$~km$^2$. We show that the presence of harmful \gls{rfi} depends on several factors, including network load, density and topology, satellite orientation, and building density. The results and methodology provide the foundation for the development of coexistence solutions and spectrum policy towards 6G.

\end{abstract}


\section{Introduction}

Continuously growing user demand is pushing the \gls{6g} of wireless networks into the \gls{sthz} spectrum of 100-300~GHz~\cite{akyildiz2022terahertz}.
%
%
%
%
The \gls{sthz} spectrum offers theoretically orders of magnitude greater bandwidth than typical communication bands 
depending on the composition of the atmosphere and weather conditions~\cite{itu-p-676}. This makes \gls{sthz} attractive for 
wireless networks, despite the significant challenges related to blockage, low transmission power, and small antenna apertures, which require directional antennas to increase the \gls{gnb} coverage~\cite{petrov2020ieee}.

However, this portion of the spectrum is already used by remote (passive) sensing systems in \gls{eess} and radio astronomy, supporting the weather, climate, and astronomy enterprises. Such services can tolerate limited to no interference. For this reason, the spectrum above 100~GHz features a set of channels reserved for passive remote sensing. This results in 12.5~GHz being the largest contiguous bandwidth available for communications under 200~GHz under current spectrum regulations~\cite{polese2021coexistence}.
Without sharing portions of the spectrum between \gls{eess} and the \gls{sthz} terrestrial communication systems, there is little benefit in climbing all the way from 71~GHz into \gls{sthz} or \gls{thz} bands, if the resulting bandwidth will be comparable.
Hence, channels wider than 12.5~GHz are very much desired.
Further, larger chunks of microwave and \gls{sthz} spectrum can also benefit passive sensing systems themselves, e.g., for more precise hyper-spectral remote sensing~\cite{kummerow2022hyperspectral}. The same applies to radio astronomy, to sense molecular shifts in bands outside those traditionally allocated for this use~\cite{polese2021coexistence}.


Therefore, today's {fixed spectrum allocation} is limiting the potential of both communications and remote sensing. While this could be true across the entire radio spectrum, the much more challenging propagation of \gls{sthz} and \gls{thz} signals through the atmosphere and the opportunity to more precisely control the radiation with compact antenna structures opens the door to more {flexible spectrum sharing} strategies~\cite{polese2021coexistence,polese2022dynamic,xing2021terahertz}.
To date, however, the literature presents a gap in the modeling and analysis of the \gls{rfi} caused by terrestrial next-generation wireless networks to \gls{eess} systems, as most analyses focus on single-link evaluations, simplified terrestrial network models, and tractable but simplified channel models~\cite{xing2021terahertz,polese2021coexistence}.
%
Modeling \gls{rfi} is key to the development of tailored coexistence techniques. This is a timely need for the passive sensing and \gls{sthz} networking communities, as \gls{rfi} insights can influence \gls{6g} standardization and next-generations of remote sensing systems, allowing coexistence embedded in the technology rather than layered on top as an afterthought.

In this paper, we fill this gap with the first analysis of 6G terrestrial \gls{rfi} to passive sensing systems, based on a large-scale network model (with up to $10^5$ nodes in more than 85~km$^2$) and actual sensors deployed on multiple \gls{eess} missions. Our contributions are as follows:
\begin{itemize}
    \item We model the channel with aggregated interference, ground reflections, terrestrial and \gls{eess} sensors beam patterns, building obstructions, and \gls{sthz} spreading and absorption losses, extending ITU models with deployment-specific details. Our model is extremely scalable and can also be used to analyze and model~\glspl{ntn}.
    \item We analyze the single-link interference probability, 
          and show how the peculiar geometry of the terrestrial-satellite \gls{rfi} setup can lead to significant interference from the ground reflection due to the combination of the terrestrial and satellite beam gains.
    \item We extend this into a data-driven large-scale simulation with multiple satellites (i.e., TEMPEST-D and the \gls{mls}~\cite{reising2015overview,waters2006earth}) and device deployments and obstructions based on \gls{osm} and 3D models, including urban cellular and backhaul. The area size and the number of buildings (62512 3D polygons) are unprecedented and showcase how we can model large-scale, site-specific deployments.
    \item We show that dense \gls{sthz} networks can affect \gls{eess} satellites operations.
          Specifically, (i) even \glspl{ue} can generate harmful interference, when combined in large numbers; (ii) the secondary reflections (e.g., the ground reflection), although generating lower interference, are significant and not to be neglected; (iii) the attenuation outside the main lobe of directional arrays might not be enough to protect the passive users; whereas (iv) path loss and atmospheric attenuation as well as the building obstruction can shield them more effectively.
\end{itemize}
Therefore, there exist regimes in which interference is significant, and other conditions in which it does not lead to any harm to passive satellites. These insights are a starting point for (i) the design of 6G networks; (ii) passive/active sharing, based on an accurate \gls{rfi} understanding; and (iii) data-driven spectrum policy toward 6G.

\begin{figure*}[t]
    \centering
    \begin{subfigure}[t]{.9\columnwidth}
        \centering
        \includegraphics[width=.8\columnwidth]{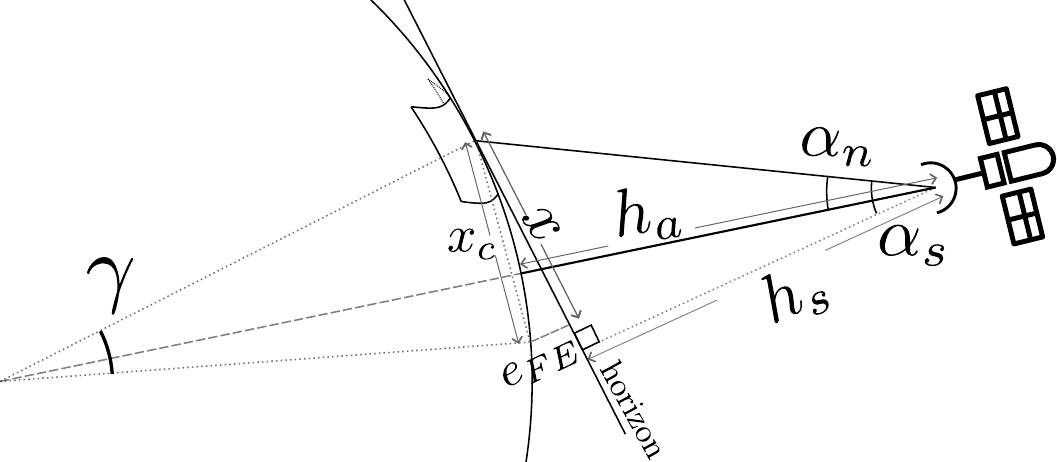}
        \caption{Nadir adjustment.}
        \label{fig:nadir}
    \end{subfigure}
    \hfill
    \begin{subfigure}[t]{.9\columnwidth}
        \setlength\fheight{0.2\columnwidth}
        \setlength\fwidth{0.75\columnwidth}
        \centering
  \tikzsetnextfilename{imgs/geometry/flat_earth_err.tex}%
  \input{imgs/geometry/flat_earth_err.tex}%

        \caption{2D-distance from the ground node to the reflection point (blue) and corresponding flat-Earth approximation error (orange) for different look-angles for a satellite at $400$~km altitude and a ground node at $10$~m from the ground.}
        \label{fig:flat_earth}
    \end{subfigure}
    \setlength\belowcaptionskip{-.5cm}
    \caption{Schematic representation of the geometry of the system (a) and error due to the flat-Earth approximation (b).}
    \label{fig:g2s_geom}
\end{figure*}

\section{Related Works}
\label{sec:related}

\gls{rfi} analyses follow two main approaches.
%
Single-link analysis considers the \gls{rfi} generated by a single interferer and a single victim~\cite{park2019modeling,polese2021coexistence,xing2021terahertz,marcus2014harmful}.
Specifically, \cite{bosso2021ultrabroadband} was among the first studies to numerically characterize \gls{rfi} at sub-THz frequencies.
\cite{polese2021coexistence} analyzes it in several scenarios for a single backhaul, terrestrial link.
Similarly, \cite{xing2021terahertz} considers an urban scenario where a receiver on a rooftop is used as a surrogate for the satellite.
The overall attenuation is measured for different ground transmitter locations.
The authors of both studies conclude that, for a single link in an urban area, harmful interference can be avoided if the satellite or the beam direction remains below certain angles.

The second approach relies on Monte Carlo simulations to aggregate interference from multiple interferers modeled through multiple random variables. \gls{rfi} statistics are derived through multiple iterations.~\cite{guidolin2015study,su2014coexistence} consider a \gls{fss} terrestrial station and compute the aggregated interference produced by the \glspl{gnb} of a nearby network at 3.4-3.6~GHz and 18~GHz.
The authors of~\cite{gasiewski2002impacts} analyze the impact of automotive radars on satellite radiometers in the 22-27~GHz frequency range, and conclude that most realistic vehicle densities would generate harmful \gls{rfi}.
The authors of both studies conclude that coexistence between the two systems at \gls{mmwave} frequencies is possible, provided that the base station deployment and configuration respect some conditions.
\cite{guidolin2015study,hassan2017feasibility,hattab2018interference} consider random deployments of the \glspl{gnb} and \glspl{ue}, realistic antenna and beamforming radiation patterns, and stochastic channel models~\cite{alkhateeb2013hybrid}.
\cite{zhong2020feasibility} provides an analysis of the interference between terrestrial and satellite relays in the 25.25-27.5~GHz bands,
considering \gls{los} propagation and the ITU channel model, and aggregating the interference over extremely wide areas ($\ang{0.5}\times\ang{0.5}$ latitude/longitude).
The authors of~\cite{winter2019statistics} estimate the aggregated \gls{rfi} distribution from a \gls{fs} network to an aircraft at 18~GHz.
In~\cite{cho2018spectral,cho2019modeling,cho2020coexistence}, the coexistence between terrestrial networks and \gls{eess} at \gls{mmwave} frequencies is analyzed.
\textit{In these papers, the building blockage and attenuation, the node locations, and the beamforming orientation are modeled through random variables, that are not representative of a real topology.}

Obstruction by physical obstacles is indeed another key element for \gls{rfi}, particularly at high frequencies.
\cite{ayoubi2023imt} stochastically estimates the aggregated interference from a wide area for the upper 6~GHz band using building statistics from real data for the city of Milan.
%
This paper aims at filling these gaps in the literature, characterizing the \gls{rfi} at sub-THz frequencies with the following original contributions: (i) we propose and validate a ray-based channel model that can scale to footprint-wide areas; (ii) we use it to take a ray-tracing-like approach to an unprecedented scale, in terms of the number of terrestrial nodes, considered area, and building modeling; (iii) thanks to these elements, we analyze in detail the contribution of the multi-path component, in particular of the ground reflection, at the network level; (iv) we perform this analysis using real 3D building models and node distributions.

\section{Propagation Model}
\label{sec:prop_model}

In this section, we extend the channel models from~\cite{jakes1994microwave,han2014multi} to improve their scalability and to embed key elements of propagation at frequencies above 100~GHz.
%
We model the propagation through two paths: the direct or \gls{los} ray and the ray reflected by the ground (\gls{gr}),
to provide key insights on the multipath contribution to the overall \gls{rfi} at the satellite.
Thus, the ray power takes into account (a) the reflection loss (for the reflected ray); (b) the free-space spreading loss; (c) the ``atmospheric'' loss due to the molecular absorption of the atmosphere layers crossed by the ray; and (d) the absorption loss due to obstacles.

\subsection{Ground-to-Satellite Path}
\label{ssec:g2s_geom}
\begin{figure*}[t]
    \centering
    \includegraphics[width=1.7\columnwidth]{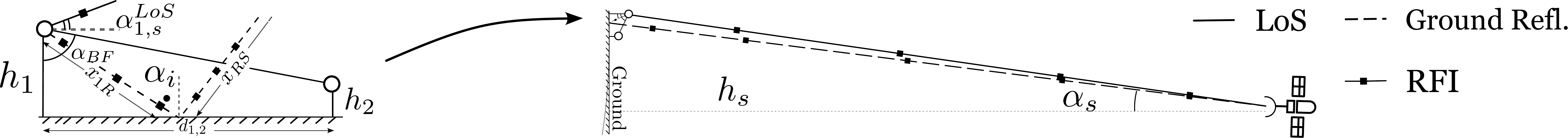}
    \setlength\abovecaptionskip{.1cm}
    \setlength\belowcaptionskip{-.3cm}
    \caption{Geometry of the problem (not to scale).}
    \label{fig:geom_setup}
\end{figure*}
The traditional 2-ray model~\cite{jakes1994microwave} is based on the flat-Earth assumption 
when considering short distances.
The effect of this simplification has been evaluated in several works~\cite{kotz2004experimental,fund2016how,loyka2001using,feuerstein1994path,dottling2001two}.
However, the distance between the satellite and the ground node can be hundreds of kilometers not only on the altitude but also on the longitudinal plane.
Here, we verify whether the flat-Earth assumption holds when considering the path reflected on the ground from the terrestrial node to the satellite.

Consider a satellite at altitude $h_a$, which points the main lobe of its sensor towards the ground with an angle $\alpha_{n}$ with respect to nadir.
By definition, the nadir at a given point is the direction pointing toward the center of the Earth.
As shown in Fig.~\ref{fig:nadir}, we distinguish between the \textit{nadir} angle $\alpha_{n}$, i.e., the angle between the pointing direction and nadir, and the \textit{apparent nadir} angle $\alpha_{s}$ seen from the ground, i.e., the angle between the pointing direction of the satellite and the normal to the horizon.
The relation between the two angles can be expressed as \cite{soler1994determination}
\begin{equation}
    \alpha_{s} = \arcsin{\left(\frac{r}{R}\sin{\alpha_{n}}\right)},
    \label{eq:alpha_sat}
\end{equation}
where $R$ is the Earth radius and $r = R+h_a$ is the distance of the satellite from the center of the Earth, here assumed to be a perfect sphere for simplicity.
Thus, a satellite at the horizon, with $\alpha_{s}=\ang{90}$, has a nadir angle $\alpha_{n}=\arcsin\left(\frac{R}{r}\right)$, that for a satellite at $h_a = 400$~km of altitude corresponds to about \ang{70.21}.
Conversely, the same satellite looking at a \ang{65} angle with respect to nadir has an apparent look-angle $\alpha_{s} = \ang{74.41}$.

From $\alpha_{s}$ it is straightforward to derive the incidence angle $\alpha_i$, the \gls{los} angle $\alpha_{1,s}^{LoS}$ between the ground node and the satellite, the distance $x_{1R}$ ($x_{RS}$) between the ground node (satellite) and the reflection point, and the overall length of the reflected path $x$:
\begin{align}
    \alpha_i           & = \arctan\left(\frac{h_{s}}{h_{s}+h_1}\tan{\alpha_{s}}\right) \simeq \alpha_s\label{eq:alpha_i}          \\
    \alpha_{1,s}^{LoS} & = \arctan\left(\frac{h_{s}-h_1}{\left(h_{s}+h_1\right)\tan{\alpha_{i}}}\right) + \frac{\pi}{2}           \\
                       & \simeq \arctan\left(\frac{1}{\tan{\alpha_{s}}}\right)+ \frac{\pi}{2} = \pi-\alpha_s \label{eq:alpha_los} \\
    x_{1R}             & = h_1\tan{\alpha_i} \;\;\;\;\;\;\;\;\; x_{RS} = h_s\tan{\alpha_i} \nonumber                              \\
    x_{GR}             & = x_{1R} + x_{RS}
\end{align}
We define the flat-Earth error at distance $x$ from node 1 as $e_{FE}(x) = \left|h_{FE}(x)-h_E(x)\right|$, where $h_{FE}(x)$ and $h_E(x)$ are the flat and spherical Earth ground height at distance $x$, respectively.
The error can be computed as
$e_{FE}(x) = x^2 + x_c^2 -2 x x_c \cos{\left(\frac{\gamma}{2}\right)},$
where $x_c= 2R\sin{\left(\frac{\gamma}{2}\right)}$ and $\gamma=\arctan {\left(\frac{x}{R}\right)}$ are defined as in \cref{fig:nadir}.
\cref{fig:flat_earth} shows the distance $x_{1R}$ (blue) of the reflection point from the ground node, and the corresponding flat-Earth error $e_{FE}(x_{1R})$, for a ground node at $10$~m and a satellite at $400$~km altitude at different look-angles $\alpha_{s}$.
As long as the satellite is above the horizon ($\alpha_{s}<\ang{80}$), the ray is reflected on the ground closer than $30$~m from the ground node, a distance small enough so that the flat-Earth approximation holds, with an error in the order of $10^{-5}$~m.
For $\alpha_s > \ang{80}$, the reflection point is so far from the ground node that the curvature of the Earth needs to be accounted for.
In that case, the approximations for the reflected path do not hold, whereas the model for the \gls{los} remains valid.
However, since in this work we are interested in analyzing also the effect of multipath, we limit our analysis to elevation angles $\alpha_s \le \ang{80}$.


\subsection{Ray Power}
As represented in \cref{fig:geom_setup}, in our channel model we consider the direct ray ($j=0$: \gls{los}) and the ray reflected on the ground ($j=1$), representing the two main propagation paths from the source to the receiver. Each ray is associated with a path loss $L_{j}$ that depends on a number of elements, detailed in the following paragraphs.

\paragraph{Reflection Loss} If the ray $j$ is reflected, a reflection loss $L_{R,j}$ is applied, which changes with the incidence angle, with the polarization of the wave, and with the material of the reflecting surface. In this paper, we consider the reflection loss presented in \cite{han2014multi}, derived from \cite{piesiewicz2007scattering}.

The Fresnel reflection coefficients $r_{TE}$ and $r_{TM}$ model the power loss in the specular direction when the wave is reflected on a smooth surface 
for the \gls{te} and \gls{tm} polarized waves, respectively.
We present only the derivation for \gls{te}-polarized waves, without polarization loss, as the formulation for the \gls{tm} mode is analogous.
The reflection coefficient $r=r_{TE}$ can be computed, considering the refraction index $n$ and the absorption coefficient $\alpha$, as
\begin{equation}
    r = \frac{\cos{\alpha_i}-n\sqrt{1-\left(\frac{1}{n}\sin{\alpha_i}\right)^2}}{\cos{\alpha_i}+n\sqrt{1-\left(\frac{1}{n}\sin{\alpha_i}\right)^2}}  = \left|r\right|e^{j\phi_R}
    \label{eq:fresnel}
\end{equation}
where $n=\sqrt{\epsilon}$ is the refractive index and $\phi_R$ is the phase shift that occurs during the reflection.
Relative permittivity coefficients $\epsilon\in \mathbb{C}$ for common building materials in the frequency range of interest are reported in \cite{itu-r-2040}.
For simplicity, throughout this work we assume that $\phi_R = \pi$, i.e., $r = -\left|r\right|$.

\begin{figure*}[t]
    \centering
    \begin{subfigure}[t]{.95\columnwidth}
        \setlength\fheight{0.3\columnwidth}
        \setlength\fwidth{.8\columnwidth}
        \centering
  \tikzsetnextfilename{imgs/geometry/refl_loss}%
%
%
\definecolor{mycolor1}{rgb}{0.00000,0.44700,0.74100}%
\definecolor{mycolor2}{rgb}{0.85000,0.32500,0.09800}%
\definecolor{mycolor3}{rgb}{0.92900,0.69400,0.12500}%
\definecolor{mycolor4}{rgb}{0.49400,0.18400,0.55600}%
\definecolor{mycolor5}{rgb}{0.46600,0.67400,0.18800}%
\definecolor{mycolor6}{rgb}{0.30100,0.74500,0.93300}%
\definecolor{mycolor7}{rgb}{0.63500,0.07800,0.18400}%
\begin{tikzpicture}
\pgfplotsset{every tick label/.append style={font=\scriptsize}}

\pgfplotsset{every tick label/.append style={font=\scriptsize}}

\begin{axis}[%
width=\fwidth,
height=\fheight,
at={(0\fwidth,0\fheight)},
anchor = south west,
scale only axis,
xmin=0,
xmax=90,
xlabel style={font=\footnotesize\color{white!15!black}},
xlabel={$\alpha_i$ [degree]},
ymin=0,
ymax=70,
ylabel style={font=\footnotesize\color{white!15!black}},
ylabel={Reflection Loss $L_R$ [dB]},
axis background/.style={fill=white},
xmajorgrids,
ymajorgrids,
legend style={legend cell align=center, align=center, draw=white!15!black,at={(.6\fwidth,1.2\fheight)},anchor=south, /tikz/every even column/.append style={column sep = 0.2cm},font=\scriptsize},
legend columns = 4
]

\addplot [fill=black, fill opacity=0.1, color=black, forget plot, mark=none, draw=none]
table[row sep=crcr]{%
-10 -10\\
-10 75\\
48.72 75\\
48.72 -10\\
-10 -10\\
};

\addplot [fill=black, fill opacity=0.1, color=black, forget plot, mark=none, draw=none]
table[row sep=crcr]{%
-10 -10\\
-10 75\\
74.41 75\\
74.41 -10\\
-10 -10\\
};

\addplot [color=mycolor1, line width=1.5pt]
  table[row sep=crcr]{%
0	10.5534916904041\\
4	10.5232367720369\\
8	10.4327397558482\\
12	10.2828007549508\\
16	10.0747430213573\\
20	9.81039780066533\\
24	9.49208362926606\\
28	9.12258053249464\\
32	8.70509965381989\\
36	8.24324887474498\\
41	7.60961425057812\\
46	6.92130759706787\\
52	6.03598959871643\\
59	4.94236225753866\\
69	3.3188058736627\\
81	1.37765563965533\\
88	0.29735957403787\\
90	0\\
};
\addlegendentry{0.10 mm}

\addplot [color=mycolor2, line width=1.5pt]
  table[row sep=crcr]{%
0	13.5756182743079\\
3	13.5503166218483\\
6	13.4745871434754\\
9	13.3489547286914\\
12	13.1742890534758\\
15	12.9517969349397\\
18	12.6830117552632\\
21	12.3697800789436\\
24	12.0142456180084\\
28	11.4786196940012\\
32	10.8785694939356\\
36	10.2212564034862\\
40	9.51445164580389\\
45	8.5739784171668\\
51	7.38430134303439\\
61	5.33179099464049\\
69	3.70693029923171\\
75	2.54221249322995\\
80	1.62727968521838\\
84	0.940910816396027\\
88	0.301040448271294\\
90	0\\
};
\addlegendentry{0.15 mm}

\addplot [color=mycolor3, line width=1.5pt]
  table[row sep=crcr]{%
0	17.8065954917733\\
3	17.7697049713679\\
6	17.6593358592197\\
9	17.4763925639075\\
12	17.2223726714109\\
15	16.8993524202515\\
18	16.5099665998673\\
21	16.0573831006124\\
24	15.5452724022476\\
27	14.9777723412769\\
30	14.3594485450233\\
33	13.6952509610516\\
37	12.7474760025139\\
41	11.7408846250736\\
46	10.4212946616531\\
53	8.51014885812488\\
62	6.0586326685217\\
67	4.75337721022539\\
71	3.76038621023014\\
75	2.82563422538354\\
78	2.1691188574069\\
81	1.55515172351127\\
84	0.98713931811713\\
87	0.467972659130623\\
90	0\\
};
\addlegendentry{0.20 mm}

\addplot [color=mycolor4, line width=1.5pt]
  table[row sep=crcr]{%
0	68.5783221013579\\
1	68.5587562801902\\
2	68.5000813969788\\
3	68.402365165602\\
4	68.2657203525699\\
5	68.0903046422086\\
6	67.8763204481509\\
7	67.6240146713635\\
8	67.3336784050085\\
9	67.0056465865005\\
10	66.6402975971871\\
11	66.2380528101429\\
12	65.7993760866322\\
13	65.3247732218571\\
14	64.8147913406703\\
15	64.2700182439931\\
16	63.6910817067386\\
17	63.0786487280992\\
18	62.4334247351159\\
19	61.7561527405033\\
21	60.3086193606375\\
23	58.7427096201604\\
25	57.0656247247314\\
27	55.2850717133388\\
29	53.4092267223023\\
31	51.4466960324074\\
33	49.4064750880163\\
35	47.2979056863856\\
37	45.1306315437709\\
40	41.7913632024894\\
44	37.2275566436536\\
52	28.0296663217597\\
55	24.6634629478114\\
58	21.3958559356096\\
60	19.2884837361452\\
62	17.2489013647957\\
64	15.2861254088068\\
66	13.4087867759805\\
68	11.6250899147449\\
70	9.94277404945572\\
72	8.36907661274468\\
74	6.91069904764177\\
76	5.57377514253149\\
77	4.95261271423249\\
78	4.36384205177021\\
79	3.80805778448618\\
80	3.28581414396751\\
81	2.79762431050284\\
82	2.34395980850228\\
83	1.92524995168317\\
84	1.54188133877049\\
85	1.19419740040708\\
86	0.882497997913617\\
87	0.607039074480994\\
88	0.368032359319528\\
89	0.165645125230029\\
90	0\\
};
\addlegendentry{0.50 mm}

\addplot [color=mycolor5, line width=1.5pt]
  table[row sep=crcr]{%
38	102.578689898386\\
41	94.3662921149838\\
45	83.2205050395917\\
50	69.2703432928106\\
53	61.0489233390607\\
55	55.6840382456597\\
57	50.44155513926\\
59	45.346129416806\\
61	40.4216851517456\\
63	35.6912999247641\\
65	31.1770937387747\\
66	29.0076821450199\\
67	26.9001225496266\\
68	24.856864061614\\
69	22.880276922341\\
70	20.9726496527295\\
71	19.136186298875\\
72	17.3730037793686\\
73	15.6851293375383\\
74	14.074498101697\\
75	12.5429507563635\\
76	11.0922313272947\\
77	9.7239850830391\\
78	8.43975655558778\\
79	7.24098768256366\\
80	6.12901607325175\\
81	5.10507340063002\\
82	4.17028392141673\\
83	3.32566312600231\\
84	2.57211651998385\\
85	1.91043853886728\\
86	1.34131159734729\\
87	0.865305274417381\\
88	0.482875635402209\\
89	0.194364691842708\\
90	0\\
};
\addlegendentry{0.80 mm}


\addplot [color=mycolor7, line width=1.5pt]
  table[row sep=crcr]{%
65	100.699538058554\\
66	93.4031220018236\\
67	86.3272225436864\\
68	79.4803419499707\\
69	72.87070293563\\
70	66.5062386816289\\
71	60.3945832034052\\
72	54.543062082611\\
73	48.9586835733994\\
74	43.648130094079\\
75	38.617750114493\\
76	33.8735504490093\\
77	29.4211889645228\\
78	25.2659677123732\\
79	21.4128264925758\\
80	17.8663368582456\\
81	14.6306965675653\\
82	11.7097244901148\\
83	9.10685597383258\\
84	6.82513867832623\\
85	4.8672288796902\\
86	3.23538825141962\\
87	1.93148112543683\\
88	0.95697223666663\\
89	0.312924954012971\\
90	0\\
};
\addlegendentry{1.50 mm}

\addplot [color=mycolor2, line width=1.5pt]
  table[row sep=crcr]{%
71	105.240666795286\\
72	94.945299368744\\
73	85.1255903515092\\
74	75.7933822597116\\
75	66.9599233298512\\
76	58.6358538421774\\
77	50.8311931835268\\
78	43.5553276654008\\
79	36.8169991121542\\
80	30.624294233239\\
81	24.984634792495\\
82	19.9047685865258\\
83	15.3907612432133\\
84	11.4479888504375\\
85	8.08113142406293\\
86	5.29416722323738\\
87	3.0903679200232\\
88	1.47229462934534\\
89	0.441794804198054\\
90	0\\
};
\addlegendentry{2.00 mm}

\addplot [color=black, dashed, line width=1.5pt]
  table[row sep=crcr]{%
0	12.4787352423136\\
5	12.4498467180747\\
10	12.3632760774275\\
15	12.2193132728571\\
20	12.0184586757356\\
25	11.7614484743485\\
30	11.4492895626866\\
35	11.0833029926287\\
40	10.6651744427466\\
45	10.1970093106859\\
50	9.68138897763117\\
56	9.00442272601678\\
62	8.27009773588297\\
68	7.48612822937424\\
75	6.52075794235073\\
84	5.22441899035854\\
90	4.34294481903252\\
};
\addlegendentry{g=1}
\end{axis}
\end{tikzpicture}%
%

        \caption{Reflection loss $L_R$ at $178$~GHz for different values of the standard deviation $\sigma$ of the surface roughness. $g=1$ (black dashed) marks the separation between rough and smooth surfaces, as confirmed by the much greater losses for $\sigma>0.20$~mm.}
        \label{fig:refl_loss}
    \end{subfigure}
    \hfill
    \begin{subfigure}[t]{.95\columnwidth}
        \setlength\fheight{0.3\columnwidth}
        \setlength\fwidth{.8\columnwidth}
        \centering
  \tikzsetnextfilename{imgs/geometry/losses}%
  \input{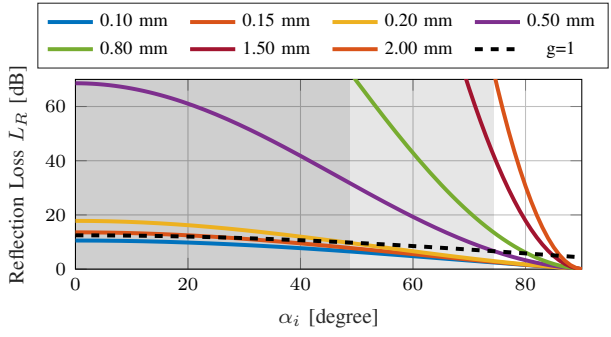}
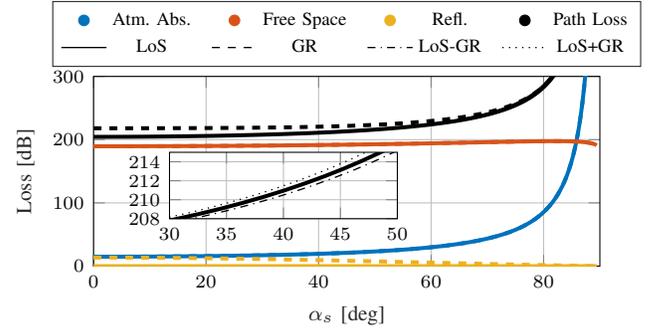%

        \caption{Contribution of the different losses at 178~GHz for a satellite at 400~km of altitude, for different elevation angles $\alpha_{s}$ of the satellite. The free-space loss dominates, with the atmospheric absorption dominating when the satellite is at the horizon and the signal crosses a larger section of the atmosphere.}
        \label{fig:losses}
    \end{subfigure}
    \setlength\belowcaptionskip{-.5cm}
    \caption{Analysis of the reflection loss (a) and of the overall path loss with the individual contributions (b).}
\end{figure*}

When considering outdoor propagation, particularly at high frequencies, reflections occur on rough materials, where scattering becomes relevant.
To include the scattering loss in the specular direction, 
we multiply the Fresnel coefficient by the Rayleigh roughness factor
\begin{align}
    L_{R,j} & = \left(\rho \cdot r\right)^{-1}, \;\;\; \mbox{with} \;\;\;
    \rho = e^{-\frac{g}{2}},
\end{align}
where the roughness $g$ of the material ($g\ll1$: smooth, $g\simeq 1$ moderately rough, $g>1$ very rough) is defined as
$g = \left(\frac{4\pi \sigma \cos{\alpha_i}}{\lambda}\right)^2$,
where $\sigma$ is the standard deviation of the surface roughness~\cite{piesiewicz2007scattering}.
Note that the dependence of $g$ on the incident angle $\alpha_i$ accounts for the effective roughness seen by the incoming wave.
\cref{fig:refl_loss} reports the reflection loss as a function of the incident angles for different values of $\sigma$, computed with the relative permittivity for the concrete $\epsilon=5.24$~\cite{itu-r-2040}, that is used throughout the rest of this work.

\paragraph{Free-Space Loss}
As the $j$-th ray propagates through the space, the signal is attenuated proportionally to the distance $d_j$ and to the center frequency $f_c$.
The free-space loss for ray $j$ is thus computed as
\begin{equation}
    L_{fs,j} = \frac{4\pi f_c d_j}{c}.
\end{equation}
Clearly, the \gls{los} path is the shortest, with length $d_0<d_j, j\in \mathbb{N}^+$. The length difference $\Delta d$ between the \gls{los} and the ground-reflected ray is approximated for the case of ground-to-satellite propagation in \cref{eq:approx}.
An approximation for the phase difference in the two-ray model commonly adopted in the literature~\cite{jakes1994microwave} is
$\Delta d \simeq 2\frac{h_1 h_{s}}{x},$
which holds when $x\gg h_1+h_{s}$.
However, in the considered case, this assumption is not verified, and the approximation is not valid.
Thus, we derived an approximation that can be applied to satellite communication:
\begin{align}
    \Delta d & = \sqrt{x^2+(h_1+h_{s})^2} - \sqrt{x^2+(h_{s}-h_1)^2} =\nonumber                                                                                                             \\
             & = \frac{x^2+(h_1+h_{s})^2 - (x^2+(h_{s}-h_1)^2)}{\sqrt{x^2+(h_1+h_{s})^2} + \sqrt{x^2+(h_{s}-h_1)^2}}\nonumber                                                               \\
             & = \frac{4 h_1}{\sqrt{\left(\frac{x}{h_{s}}\right)^2+\left(\frac{h_1}{h_{s}}+1\right)^2} + \sqrt{\left(\frac{x}{h_{s}}\right)^2+\left(1-\frac{h_1}{h_{s}}\right)^2}}\nonumber \\
             & = \simeq \frac{4 h_1}{\left(\sqrt{\left(\frac{x}{h_{s}}\right)^2 + 1} + \sqrt{\left(\frac{x}{h_{s}}\right)^2 + 1 }\right)}\nonumber                                          \\
             & = \frac{2 h_1}{\sqrt{\left(\frac{x}{h_{s}}\right)^2 + 1}} = \frac{2 h_1}{\sqrt{\left(\tan\left(\alpha_{s}\right)\right)^2 + 1}}
    \label{eq:approx}
\end{align}
where the approximation holds for $h_{s}\gg h_1$, which applies to ground-to-satellite scenarios.

\cref{fig:losses} compares the path loss of the direct and of the reflected ray.
Note from \cref{eq:approx} that $\Delta d$ varies between $0$ (satellite at the horizon) and $2 h_1$ (satellite directly above the ground node), that makes the difference between the free-space loss of the two paths negligible.

\paragraph{Atmospheric Loss}
Electromagnetic waves propagating through the atmosphere interact with the molecules, transferring part of their energy to the medium.
This effect is accounted for through the atmospheric absorption coefficient $L_{A,j}$, which depends on the composition of the atmospheric layers and on the propagation angle of the $j$-th ray~\cite{polese2021coexistence,polese2022dynamic}:
\begin{equation}
    L_{A,j} = \left(\int_{0}^{h_a} \frac{\gamma(h)}{\sqrt{1-\cos^2(\theta(h)})}\diff h\right)^{-1}
\end{equation}
where $\gamma(h) = \gamma_o(h) + \gamma_w(h)$ is the attenuation given by oxygen ($\gamma_o$) and water vapor ($\gamma_w$) at height $h$, $h_a$ is the satellite altitude, and $\theta(h)$ is the local apparent elevation angle at height $h$.

\paragraph{Absorption Loss}
In the sub-THz bands, the transmitted power through materials is negligible~\cite{du2021sub}.
Thus, for large obstacles, e.g., buildings, with multiple, thick, non-reflective layers, a hard, on/off loss can be applied.
Specifically, we model the building blockage as $L_B = +\infty$ if the ray is obstructed, or $L_B = 1$ otherwise.

Factoring in these elements, the overall loss for ray $j$ is
\begin{equation}
    L_{j} = L_{fs,j}L_{R,j}L_{A,j}L_{B,j}.
    \label{eq:tot_pathloss}
\end{equation}
\cref{fig:losses} shows the different contributions to the path loss at $f_c=178$~GHz for a satellite at $400$~km of altitude, for different apparent look-angles $\alpha_{s}$.

\subsection{Ray combining}
\label{ssec:ray_comb}
Considering the propagation of a generic electric signal $E_s\in \mathbb{C}$ from the source to the receiver, the signal $E$ at the receiver antenna is given by the superposition of the electric fields of the \gls{los} and ground-reflected rays
\begin{equation}
    \begin{aligned}
         & E = \sum_{j=0,1}{E_j} =                                                                                                                                            
        \left(\frac{E_s e^{j2\pi \tau_{0} f_c}}{L_{B,0} L_{fs,0} L_{A,0}} + \frac{E_s e^{j2\pi \tau_{1} f_c + \phi_R}}{L_{B,1} L_{fs,1} \left|L_R\right| L_{A,1}}\right)      \\ & = E_s e^{j2\pi \tau_{0} f_c}\left(\frac{1}{L_{B,0} L_{fs,0}L_{A,0}} + \frac{ e^{j2\pi \frac{\Delta d}{\lambda} + \phi_R}}{L_{B,1} L_{fs,1} \left|L_R\right| L_{A,1}}\right)\\
         & \simeq \frac{E_s}{L_{fs,0}L_{A,0}}e^{j2\pi \tau_{0} f_c}\left(\frac{1}{L_{B,0}} + \frac{e^{j2\pi \frac{\Delta d}{\lambda} + \pi}}{L_{B,1}\left|L_R\right|} \right)
        \label{eq:ef_superposition}
    \end{aligned}
\end{equation}
where $\Delta d = d_1-d_0>0$ is the difference between the length of the two paths; and $f_c$ and $\lambda = \frac{c}{f_c}$ are the central frequency and the wavelength of the signal.

The approximations $L_{fs,0}=L_{fs,1}$ and $L_{A,0}=L_{A,1}$ are justified by the fact that for very long propagation distances, e.g., when considering the transmission from the ground to the satellite, the difference in path length $\Delta d$ is small, making the atmospheric and the free-space losses experienced by the two rays almost equal.
On the contrary, this kind of approximation does not hold in general for the phase.
This is particularly true when considering high frequencies/short wavelengths, as the two rays are in phase opposition when
$\Delta d = 2k\frac{\lambda}{2}, k \in \mathbb{N}$,
i.e., a path difference of $\frac{\lambda}{2}$, in the order of millimeters or less for frequencies above 100~GHz, determines whether the rays combine constructively or destructively.

\begin{figure*}[t]
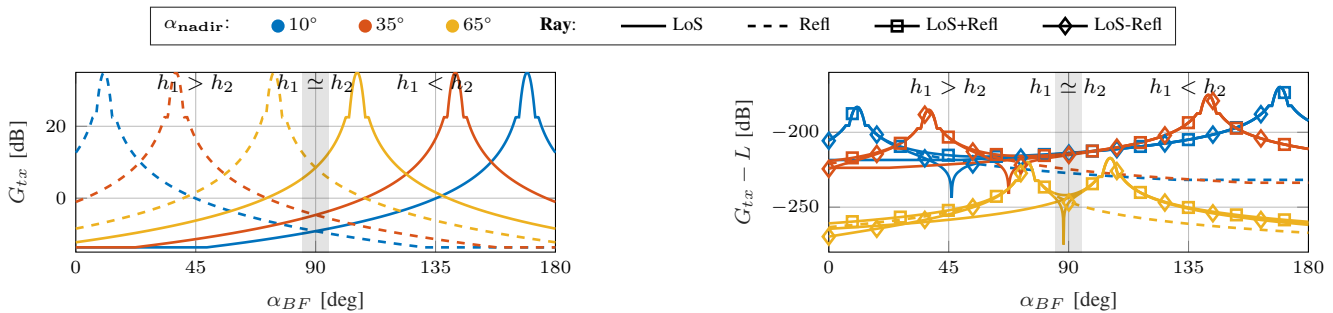

    \centering
    \begin{subfigure}[t]{2\columnwidth}
        \centering
  \tikzsetnextfilename{imgs/geometry/legend_bf.tex}%
  \begin{tikzpicture}

\definecolor{mycolor1}{rgb}{0.00000,0.44700,0.74100}%
\definecolor{mycolor2}{rgb}{0.85000,0.32500,0.09800}%
\definecolor{mycolor3}{rgb}{0.92900,0.69400,0.12500}%

\begin{axis}[%
width=0,
height=0,
at={(0,0)},
xmin=0,
xmax=0,
xtick={},
ymin=0,
ymax=0,
ytick={},
scale only axis,
axis background/.style={fill=white},
legend style={legend cell align=center, align=center, draw=white!15!black,at={(0,0)},anchor=south west, /tikz/every even column/.append style={column sep = 0.5cm},font=\scriptsize},
legend columns = 10
]

\addlegendimage{empty legend};
\addlegendentry{$\mathbf{\alpha_{nadir}}$:}

\addlegendimage{scatter, only marks, color=mycolor1, line width=1.pt};
\addlegendentry{$10\degree$}

\addlegendimage{scatter, only marks, color=mycolor2, line width=1.pt};
\addlegendentry{$35\degree$}

\addlegendimage{scatter, only marks, color=mycolor3, line width=1.pt};
\addlegendentry{$65\degree$}

\addlegendimage{empty legend};
\addlegendentry{\textbf{Ray}:}

\addlegendimage{color=black, line width=1.pt};
\addlegendentry{LoS}

\addlegendimage{color=black, dashed, line width=1.pt};
\addlegendentry{Refl}

\addlegendimage{color=black, line width=1.pt, mark=square, mark options={solid, fill=black, black}};
\addlegendentry{LoS+Refl}

\addlegendimage{color=black, line width=1.pt, mark size=3pt, mark=diamond, mark options={solid, fill=black, black}};
\addlegendentry{LoS-Refl}

\end{axis}
\end{tikzpicture}%
%

    \end{subfigure}\vspace{10pt}\\
    \begin{subfigure}[t]{.9\columnwidth}
        \centering
        \setlength\fheight{0.3\columnwidth}
        \setlength\fwidth{.8\columnwidth}
  \tikzsetnextfilename{imgs/geometry/Gtx.tex}%
  \input{imgs/geometry/Gtx.tex}%

        \caption{Beamforming gain of the terrestrial transmitter in the direction of the satellite as seen by the \gls{los} and by the reflected ray, for different beamforming elevation angles $\alpha_{BF}$.}
        \label{fig:geom_Gtx}
    \end{subfigure}\hspace{0.1\textwidth}%
    \begin{subfigure}[t]{.9\columnwidth}
        \centering
        \setlength\fheight{0.3\columnwidth}
        \setlength\fwidth{.8\columnwidth}
  \tikzsetnextfilename{imgs/geometry/PLplusGtx}%
  \input{imgs/geometry/PLplusGtx}%

        \caption{Path loss to the satellite. LoS+Refl and LoS-Refl represent the bound for constructive and destructive interference, respectively.}
        \label{fig:geom_PL}
    \end{subfigure}
    \setlength\belowcaptionskip{-.5cm}
    \caption{The ground reflection to the satellite is amplified by the main lobe of the \gls{gnb} pointing towards the \glspl{ue} and thus to the ground.}
    \label{fig:geom_metrics}
\end{figure*}
\section{Single Link Analysis}
\label{sec:geometry}
In this section, we present a brief geometry-based analysis of the problem, that serves as the basis for the simulation setup, considering a single terrestrial link and an incumbent \gls{eess}.
We analytically demonstrate that
(i) reflections play a significant role in the \gls{rfi} analysis, and that (ii) narrow beams, while successfully suppressing the direct interference, might amplify the undesired interfering multipath components.


The two factors that come into play when considering the propagation of electromagnetic waves at such high frequencies are (i) their interaction with the environment, described by the propagation model introduced in~\cref{sec:prop_model}, and (ii) the spatial distribution of the power, determined by the radiation and beamforming patterns of the antennas.


Prior work on \gls{rfi} to satellite systems~\cite{xing2021terahertz,polese2021coexistence} shows, often adopting a \gls{los} channel model, that the narrow beamforming used in \gls{mmwave} and \gls{subthz} networks can reduce enough the power that leaks in the direction of the passive user so as not to cause any significant interference.
However, due to the particular geometry of the ground-to-satellite interference, represented in \cref{fig:geom_setup}, reflections can not be neglected.
In this section, we start by analyzing the geometry of a single link in \cref{ssec:bf_ampl} to show that the power received through the ground reflection---accounting for propagation and beamforming---is not negligible, and comparable to the \gls{los} ray in some cases.
We then derive the corresponding probabilities for a single link in~\cref{ssec:single_link}, and provide some considerations on the impact of the frequency band in \cref{ssec:freqs}.

\subsection{Beamforming Amplification}
\label{ssec:bf_ampl}

\begin{figure*}[t]
    \setlength\fheight{0.2\columnwidth}
    \centering
    \begin{subfigure}[t]{\columnwidth}
        \centering
  \tikzsetnextfilename{imgs/geometry/legend_probs}%
  \begin{tikzpicture}

\definecolor{mycolor1}{rgb}{0.00000,0.44700,0.74100}%
\definecolor{mycolor2}{rgb}{0.85000,0.32500,0.09800}%
\definecolor{mycolor3}{rgb}{0.92900,0.69400,0.12500}%

\begin{axis}[%
width=0,
height=0,
at={(0,0)},
xmin=0,
xmax=0,
xtick={},
ymin=0,
ymax=0,
ytick={},
scale only axis,
axis background/.style={fill=white},
legend style={legend cell align=center, align=center, draw=white!15!black,at={(0,0)},anchor=south west, /tikz/every even column/.append style={column sep = 0.5cm}},
legend columns = 8
]

\addlegendimage{empty legend};
\addlegendentry{\footnotesize $h_{tx}$ [m]:}

\addlegendimage{scatter, only marks, color=mycolor1, line width=1.pt};
\addlegendentry{\footnotesize 2}

\addlegendimage{scatter, only marks, color=mycolor2, line width=1.pt};
\addlegendentry{\footnotesize 5}

\addlegendimage{scatter, only marks, color=mycolor3, line width=1.pt};
\addlegendentry{\footnotesize 15}

\addlegendimage{empty legend};
\addlegendentry{\footnotesize $d_{2}$ [m]:}

\addlegendimage{color=black, line width=1.pt, mark=none};
\addlegendentry{\footnotesize 10}

\addlegendimage{color=black, line width=1.pt, mark=diamond};
\addlegendentry{\footnotesize 50}

\addlegendimage{color=black, line width=1.pt, mark=square};
\addlegendentry{\footnotesize 100}
\addlegendentry{}




\end{axis}
\end{tikzpicture}%
%

    \end{subfigure}\\
    \begin{subfigure}[t]{\columnwidth}
        \setlength\fwidth{.8\columnwidth}
        \centering
  \tikzsetnextfilename{imgs/geometry/P_A_LoS}%
%
%
\definecolor{mycolor1}{rgb}{0.00000,0.44700,0.74100}%
\definecolor{mycolor2}{rgb}{0.85000,0.32500,0.09800}%
\definecolor{mycolor3}{rgb}{0.92900,0.69400,0.12500}%
\begin{tikzpicture}
\pgfplotsset{every tick label/.append style={font=\scriptsize}}

\begin{axis}[%
width=\fwidth,
height=\fheight,
at={(0\fwidth,0\fheight)},
scale only axis,
xmin=0,
xmax=90,
xlabel style={font=\footnotesize\color{white!15!black}},
xlabel={$\alpha_{s}$ [deg]},
ymin=0,
ymax=1,
ylabel style={font=\footnotesize\color{white!15!black}},
ylabel={Probability},
axis background/.style={fill=white},
xmajorgrids,
ymajorgrids,
legend style={legend cell align=left, align=left, draw=white!15!black}
]
\addplot [color=mycolor1, line width=1.pt]
  table[row sep=crcr]{%
0	0\\
89.6	0\\
89.7	0.000713574536973738\\
89.8	0.00532699770943168\\
89.9	0.0134406370943765\\
90	0.0243550881332908\\
};

\addplot [color=mycolor1, line width=1.pt, mark repeat=100, mark=diamond, mark phase = 50, mark options={solid,fill=mycolor1,mycolor1}]
  table[row sep=crcr]{%
0	0\\
88.7	0\\
88.8	0.0243152156454585\\
88.9	0.106102031434006\\
89	0.213357182735649\\
89.1	0.331573673802197\\
89.2	0.427070712197292\\
89.3	0.498694509163343\\
89.4	0.554402811859902\\
89.5	0.598970268584679\\
89.6	0.635435291892605\\
89.7	0.665823490155233\\
89.8	0.691537207631356\\
89.9	0.713578118792967\\
90	0.732680784920817\\
};

\addplot [color=mycolor1, line width=1.pt, mark repeat=100, mark=square, mark options={solid,fill=mycolor1,mycolor1}]
  table[row sep=crcr]{%
0	0\\
88.6	0\\
88.7	0.106097269607133\\
88.8	0.331555347677579\\
88.9	0.498670074156848\\
89	0.598939724547421\\
89.1	0.665786836901106\\
89.2	0.713535356098646\\
89.3	0.749347254581679\\
89.4	0.777201405929958\\
89.5	0.79948513429234\\
89.6	0.817717645946303\\
89.7	0.83291174507761\\
89.8	0.845768603815685\\
89.9	0.85678905939649\\
90	0.866340392460415\\
};

\addplot [color=mycolor2, line width=1.pt]
  table[row sep=crcr]{%
0	0\\
90	0\\
};

\addplot [color=mycolor2, line width=1.pt, mark repeat=100, mark=diamond, mark phase = 50, mark options={solid,fill=mycolor2,mycolor2}]
  table[row sep=crcr]{%
0	0\\
90	0\\
};

\addplot [color=mycolor2, line width=1.pt, mark repeat=100, mark=square, mark options={solid,fill=mycolor2,mycolor2}]
  table[row sep=crcr]{%
0	0\\
90	0\\
};

\addplot [color=mycolor3, line width=1.pt]
  table[row sep=crcr]{%
0	0\\
90	0\\
};

\addplot [color=mycolor3, line width=1.pt, mark repeat=100, mark=diamond, mark phase = 50, mark options={solid,fill=mycolor3,mycolor3}]
  table[row sep=crcr]{%
0	0\\
90	0\\
};

\addplot [color=mycolor3, line width=1.pt, mark repeat=100, mark=square, mark options={solid,fill=mycolor3,mycolor3}]
  table[row sep=crcr]{%
0	0\\
90	0\\
};

\end{axis}
\begin{axis}[%
width=0.5\fwidth,
height=0.7\fheight,
at={(.3\fwidth,0.21\fheight)},
scale only axis,
xmin=88,
xmax=90,
ymin=0,
ymax=0.9,
axis background/.style={fill=white},
xtick={88,88.5,89,89.5,90},
xticklabels={88,,89,,90},
xmajorgrids,
ymajorgrids,
legend style={legend cell align=left, align=left, draw=white!15!black}
]
\addplot [color=mycolor1, line width=1.pt]
  table[row sep=crcr]{%
0	0\\
89.6	0\\
89.7	0.000713574536973738\\
89.8	0.00532699770943168\\
89.9	0.0134406370943765\\
90	0.0243550881332908\\
};

\addplot [color=mycolor1, line width=1.pt, mark=diamond, mark phase = 50, mark options={solid,fill=mycolor1,mycolor1}]
  table[row sep=crcr]{%
0	0\\
88.7	0\\
88.8	0.0243152156454585\\
88.9	0.106102031434006\\
89	0.213357182735649\\
89.1	0.331573673802197\\
89.2	0.427070712197292\\
89.3	0.498694509163343\\
89.4	0.554402811859902\\
89.5	0.598970268584679\\
89.6	0.635435291892605\\
89.7	0.665823490155233\\
89.8	0.691537207631356\\
89.9	0.713578118792967\\
90	0.732680784920817\\
};

\addplot [color=mycolor1, line width=1.pt, mark=square, mark options={solid,fill=mycolor1,mycolor1}]
  table[row sep=crcr]{%
0	0\\
88.6	0\\
88.7	0.106097269607133\\
88.8	0.331555347677579\\
88.9	0.498670074156848\\
89	0.598939724547421\\
89.1	0.665786836901106\\
89.2	0.713535356098646\\
89.3	0.749347254581679\\
89.4	0.777201405929958\\
89.5	0.79948513429234\\
89.6	0.817717645946303\\
89.7	0.83291174507761\\
89.8	0.845768603815685\\
89.9	0.85678905939649\\
90	0.866340392460415\\
};

\addplot [color=mycolor2, line width=1.pt]
  table[row sep=crcr]{%
0	0\\
90	0\\
};

\addplot [color=mycolor2, line width=1.pt, mark=diamond, mark phase = 50, mark options={solid,fill=mycolor2,mycolor2}]
  table[row sep=crcr]{%
0	0\\
90	0\\
};

\addplot [color=mycolor2, line width=1.pt, mark=square, mark options={solid,fill=mycolor2,mycolor2}]
  table[row sep=crcr]{%
0	0\\
90	0\\
};

\addplot [color=mycolor3, line width=1.pt]
  table[row sep=crcr]{%
0	0\\
90	0\\
};

\addplot [color=mycolor3, line width=1.pt, mark=diamond, mark phase = 50, mark options={solid,fill=mycolor3,mycolor3}]
  table[row sep=crcr]{%
0	0\\
90	0\\
};

\addplot [color=mycolor3, line width=1.pt, mark=square, mark options={solid,fill=mycolor3,mycolor3}]
  table[row sep=crcr]{%
0	0\\
90	0\\
};

\end{axis}
\end{tikzpicture}%
%

        \setlength\abovecaptionskip{0cm}\caption{$P(A_{LoS}|h_{tx}>h_{rx})$}
        \label{fig:P_ALOS_alpha}
    \end{subfigure}
    \hfill
    \begin{subfigure}[t]{\columnwidth}
        \setlength\fwidth{.8\columnwidth}
        \centering
  \tikzsetnextfilename{imgs/geometry/P_A_GR}%
  \input{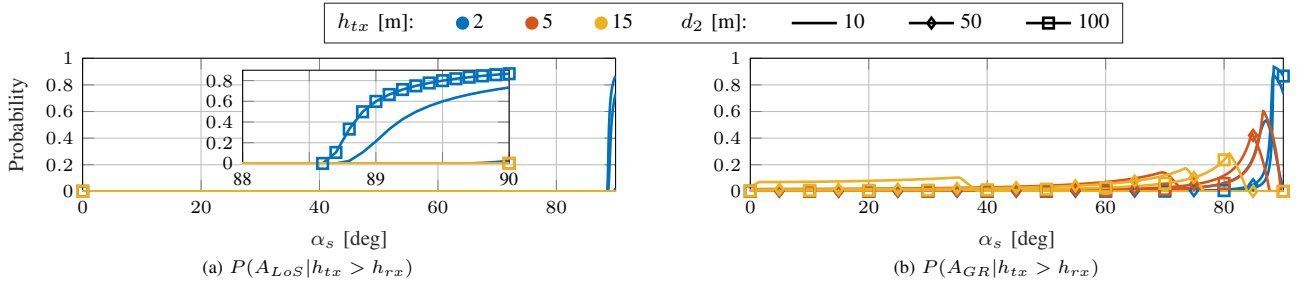}%

        \setlength\abovecaptionskip{0cm}\caption{$P(A_{GR}|h_{tx}>h_{rx})$}
        \label{fig:P_AGR_alpha}
    \end{subfigure}
    \setlength\belowcaptionskip{-.4cm}
    \caption{Probability that the \gls{los} ($P(A_{LoS})$) and the reflected ($P(A_{GR})$) rays are amplified by the main beam of the transmitting ground node.}
    \label{fig:P_A_alpha}
\end{figure*}
Consider the link between a \gls{tx} and a \gls{rx} with heights $h_{tx}$ and $h_{rx}$, respectively, placed in the area illuminated by a passive \gls{eess} satellite with altitude $h_s$.
We define the elevation beamforming angle $\alpha_{BF}$ as the angle between the direction of the beamforming steering angle and the ground.
In the following, we consider geometric beamforming, i.e., the beams of the \gls{tx} and of the \gls{rx} are aligned to the \gls{los} connecting the two, with an inclination $\alpha_{BF}=\alpha_{tx,rx}^{LoS}$, and $\theta_{HB}^{V}$ \gls{hpbw}.

For simplicity, we assume that the nodes and the satellites are aligned and consider the 2D geometry.
Assuming that $h_{s}\gg h_{tx},h_{rx}$, the \gls{los} and the reflected rays emitted by the ground nodes are both amplified with gain $G_{S}$ by the main lobe of the satellite sensor, due to the angular spread~\cite{saggese2022efficient}.
For this reason, in this analysis, we omit it and focus on the beamforming gain of the \gls{tx} node $G_{TX2S}$ toward the satellite.

To evaluate the interplay between the beamforming and the two rays, we consider three representative satellite nadir angles $\alpha_{n}=\left\{\ang{10},\ang{35},\ang{65}\right\}$.
The \gls{tx} is equipped with an antenna array with a \ang{3} \gls{hpbw} and points towards the \gls{los} to the \gls{rx}.
The path losses of the \gls{gr} and the \gls{los} ray are considered separately and combined destructively and constructively.
The full elevation range of the beamforming angle is considered, from \ang{0} (\gls{rx} right below the \gls{tx}) to \ang{180} (\gls{rx} above the \gls{tx}).
\cref{fig:geom_Gtx} shows how the transmitter beamforming gain $G_{TX2S}$ amplifies the \gls{los} (solid) and the reflected ray (dashed) as the beamforming angle changes.
From \cref{eq:alpha_i,eq:alpha_los}, the angular separation between \gls{los} and ground reflection is $\alpha^{LoS}_{tx,s}-\alpha_i\simeq \pi - 2\alpha_s$.
Thus, the only case when both rays are amplified is when $\alpha^{LoS}_{tx,s}-\alpha_i<\theta_{HB}^{V} \implies \frac{\pi-\theta_{HB}^{V}}{2} < \alpha_s$.
Considering the narrow beams that will be adopted at these frequencies, this can happen only when both the satellite and the beamforming direction are at the horizon.
However, as mentioned in~\cref{ssec:g2s_geom}, we only consider satellite angles $\alpha_{s}$ below \ang{80}, thus excluding this case.

We can distinguish three behaviors, according to the geometric characteristics of the link:
\begin{enumerate}
    \item [C1)] $\alpha_{BF}<\ang{90}$ ($h_{tx}>h_{rx}$): In this case, the \gls{tx} focuses its beam towards the ground.
          This is generally the case for the 
          transmission from a \gls{gnb}, placed in high locations for better coverage, to a \gls{ue} at the ground level.
          Indeed, \cref{fig:geom_Gtx} shows that the \gls{los} ray is effectively suppressed, particularly if the satellite is well above the horizon, whereas the ground reflection is greatly amplified.
    \item [C2)] $\alpha_{BF}>\ang{90}$ ($h_{tx}<h_{rx}$): Conversely, in this case, the transmitter points the beam upwards, as can be the case during 
          the communication from a \gls{ue} to a \gls{gnb}.
          Here, the ground reflection is strongly attenuated and is negligible when compared to the \gls{los} ray.
    \item [C3)] $\alpha_{BF}\simeq \ang{90}$ ($h_{tx}\simeq h_{rx}$): Here, the transmitter points at the horizon, e.g., to a node at similar height or very far away.
          We do not consider the latter, as at sub-THz frequencies the link length is short, whereas the former can be the case in \gls{d2d} communications or backhaul links.
          As mentioned above, the angular separation between the two rays makes it so that in this region neither of them is amplified.
\end{enumerate}
In \cref{fig:geom_PL}, the overall amplification of each ray is obtained by subtracting the path loss $L$ from the corresponding transmitter gain $G_{TX2S}$.
The role of beamforming is extremely significant, as it can greatly amplify both the interfering rays, depending on the geometry.
The peak power on the left side of \cref{fig:geom_PL}, i.e., when the ground ray is amplified (case C2), is lower than those on the right side (case C3), and their difference corresponds to the reflection loss.
Finally, note that the superposition of the two rays is relevant only when they have similar amplitude and they have opposite phase (destructive).
Specifically, when the \gls{tx} beam amplifies the reflection enough to compensate for the reflection loss, the \gls{los} and the reflected ray have comparable amplitude and cancel out.
On the other hand, their constructive combination increases the aggregated power by at most 3~dB, when their phase is aligned and they have equal amplitude.

\subsection{Probability of Beamforming Amplification}
\label{ssec:single_link}

Starting from the considerations given in the previous section on the interfering power, we derive the probability of ``beamforming amplification,'' identifying the events when the beam amplifies one of the two interfering rays, and the corresponding probabilities under some simplifying assumptions.

Let us consider only the vertical \gls{hpbw} $\theta_{HB}^{V}$ of the beam generated by \gls{tx}.
Let $\alpha_{tx,s}^{LoS}$ be the angle between the horizontal direction at the \gls{tx} and the \gls{los} connecting the latter to the satellite.
We define the event ``the direct ray is amplified within the 3~dB range of the main lobe'' ($A_{LoS}$, with probability $P(A_{LoS})$) as $\frac{\theta_{HB}^{V}}{2} \ge \lvert\alpha_{BF}-\alpha_{tx,s}^{LoS}\rvert$.
Similarly, for the reflected ray, the event $A_{GR}$ (with probability $P(A_{GR})$) maps to the condition $\frac{\theta_{HB}^{V}}{2} \ge \lvert\alpha_{BF}-\alpha_{i}\rvert$, where $\alpha_i$ is the reflection incident angle (\cref{eq:alpha_i}).
Fixing the elevation angle of the satellite $\alpha_s$, from~\cref{eq:alpha_i} and~\cref{eq:alpha_los} we have:
\begin{align}
     & (A_{LoS}): \nonumber                                                                                                                                                               \\
     & \left|\alpha_{BF}\right| \le \left(\alpha_{tx,s}^{LoS}\pm\frac{\theta_{HB}^{V}}{2}\right) \simeq \left(\pi-\alpha_s\pm\frac{\theta_{HB}^{V}}{2}\right)\coloneqq \theta_{LoS}^{\pm} \\
     & (A_{GR}):\nonumber                                                                                                                                                                 \\
     & \left|\alpha_{BF}\right| \le \left(\alpha_i\pm\frac{\theta_{HB}^{V}}{2}\right) \simeq \left(\alpha_s\pm\frac{\theta_{HB}^{V}}{2}\right)\coloneqq \theta_{GR}^{\pm}
\end{align}
The \gls{pdf} $f^{LoS}_{tx,rx}\left(\alpha_{tx,rx}^{LoS}\right)$ of the \gls{los} angles between the \gls{tx} and the \gls{rx} is fully characterized by the spatial distribution of the ground nodes.
With the assumption of geometric beamforming, we can thus use the \gls{pdf} of $\alpha_{tx,rx}^{LoS}$, easy to derive, in place of $f_{BF}(\alpha_{BF})$.
Thus, 
\begin{align}
    P\left(A_{LoS}\right) & = \int_{\theta_{LoS}^-}^{\theta_{LoS}^+} f_{BF}(\alpha)\diff\alpha = \int_{\theta_{LoS}^-}^{\theta_{LoS}^+} f^{LoS}_{tx,rx}(\alpha)\diff\alpha \nonumber \\
    P\left(A_{GR}\right)  & = \int_{\theta_{GR}^-}^{\theta_{GR}^+} f_{BF}(\alpha)\diff\alpha=\int_{\theta_{GR}^-}^{\theta_{GR}^+} f^{LoS}_{tx,rx}(\alpha)\diff\alpha.
    \label{eq:probs}
\end{align}
We can compute the \gls{los} angle \gls{pdf} starting from its \gls{cdf} $F_{BF}$, that can be derived from the distribution $f_d$ and $f_h$ of the \gls{tx}-\gls{rx} distance and of the node heights, respectively:
\begin{align}
    \alpha_{BF}    & = \arctan\left(\frac{d}{\lvert h_{tx}-h_{rx}\rvert}\right) \coloneqq g(d,h_{tx},h_{rx})\label{eq:alpha_bf}                                              \\
    F_{BF}(\alpha) & = \int\limits_{D_{h_{tx}}}\int\limits_{D_{h_{rx}}}\int\limits_{D_{h_{d}}}f_{h_{tx},h_{rx},d}\left(h_{tx},h_{rx},d\right)\diff h_{tx}\diff h_{rx}\diff d
    \label{eq:f_alpha}
\end{align}
For simplicity, we can fix $h_{tx}$ and limit our analysis to the case C1) identified in~\cref{ssec:bf_ampl} ($h_{tx}>h_{rx}$) and assume that the height of the \gls{rx} $h_{rx}$ and its distance $d$ from the transmitter are statistically independent.
This is representative of a downlink communication from a \gls{gnb} with known height $h_{tx}$ to the served user at unknown distance $d$ and height $h_{rx}$.
Then, from~\cref{eq:f_alpha}
\begin{equation}
    F_{BF}(\alpha) = \int\int\limits_{\alpha_{BF}<\alpha}f_{h_{rx},d}\left(h_{rx},d\right)\diff h_{rx} \diff d,
\end{equation}
which can be computed, if the distributions are known, by subsequently solving the two integrals using~\cref{eq:alpha_bf} to make a change of variable or using the method of transformations.
We compute it for the \gls{tx}-\gls{rx} 2D distance $d$ distributed uniformly in $[d_1,d_2]$, assuming that the links are established between non-overlapping nodes ($d_1>0$) within a maximum radius $d_2$, and $h_{rx}\simeq U[h_1,h_2]$, $0<h_1<h_2<h_{tx}$:
\begin{align}
    F_{BF}(\alpha)= & \int\limits_{h_1}^{h_2}\int\limits_{d_1}^{\min\left(d_2,(h_{tx}-h_{rx})\tan\alpha\right)}f_{h_{rx},d}\left(h_{rx},d\right)\diff h_{rx} \diff d \nonumber \\
    =               & \int\limits_{h_1}^{h_2}(F_d(\min\left(d_2,(h_{tx}-h_{rx})\tan\alpha\right))                                                                              \\
    -               & F_d(d_1))f_{h_{rx}}\left(h_{rx}\right)\diff h_{rx}
\end{align}
\begin{align}
    = & \begin{cases}
        0 \text{\qquad if }  \ang{0}<\alpha<\arctan\left(\frac{d_1}{h_{tx}-h_{rx}}\right)                                               \\
        \frac{\tan\alpha}{2\Delta d}\left(2h_{tx}-(h_2+h_1)\right)                                                                      \\
        \text{\qquad if } \arctan\left(\frac{d_1}{h_{tx}-h_{rx}}\right)<\alpha<\arctan\left(\frac{d_2}{h_{tx}-h_1}\right) \vspace{.5cm} \\
        \frac{h_{tx}-h_1-d_2/\tan\alpha}{\Delta h} + \frac{\tan\alpha((d_2/\tan\alpha)^2-(h_{tx}-h_2)^2)}{2\Delta h \Delta d}           \\\
        \text{\qquad if } \arctan\left(\frac{d_2}{h_{tx}-h_1}\right)<\alpha<\arctan\left(\frac{d_2}{h_{tx}-h_2}\right)\vspace{.5cm}     \\
        1 \text{\qquad if }  \arctan\left(\frac{d_2}{h_{tx}-h_2}\right)<\alpha<\ang{90},
    \end{cases}
\end{align}
where $\Delta h = h_2-h_1$ and $\Delta d = d_2-d_1$.

Then, the probabilities in~\cref{eq:probs} become, for the case under study,
$P\left(A_{LoS}|\text{C1}\right) = F_{BF}(\theta_{LoS}^{+})-F_{BF}(\theta_{LoS}^{-})$ and
$P\left(A_{GR}|\text{C1}\right) = F_{BF}(\theta_{GR}^{+})-F_{BF}(\theta_{GR}^{-})$.
\cref{fig:P_A_alpha} reports their value for different satellite elevation angles $\alpha_{s}$ and $\theta_{HB}^{V}=\ang{3}$.
As expected, for case C1), where $h_{tx}>h_{rx}$, the probability of amplifying the \gls{gr} ray is greater than the corresponding one for \gls{los}.
Specifically, $P\left(A_{LoS}|\text{C1}\right)$ is not negligible only for the edge case when the satellite is at the horizon ($\alpha_{s}\simeq \ang{90}$) and the \gls{tx}'s height is comparable to that of the \gls{rx} ($h_{tx}=2$~m).
On the contrary, $P\left(A_{GR}|\text{C1}\right)$ is significantly larger
\begin{itemize}
    \item when the satellite is above the ground nodes ($\ang{0}<\alpha_{s}\leq \ang{37.5}$) and the \gls{tx} is much higher than the \gls{rx} ($h_{tx}=15$~m), that is within 10~m;
    \item when the satellite is between \ang{37.5} and \ang{75} and the \gls{tx} is much higher ($h_{tx}=15$~m) than an \gls{rx} within 10~m;
    \item when the satellite is at the horizon ($\alpha_{s}>\ang{80}$). In this case, the ground reflection is less representative of practical cases, as explained in \cref{ssec:g2s_geom}.
\end{itemize}

The derivation for cases C2 and C3 is analogous.

\subsection{Frequency}
\label{ssec:freqs}
\begin{figure}[t]
    \centering
    \setlength\fheight{0.3\columnwidth}
    \setlength\fwidth{.7\columnwidth}
  \tikzsetnextfilename{imgs/geometry/freqs.tex}%
  \input{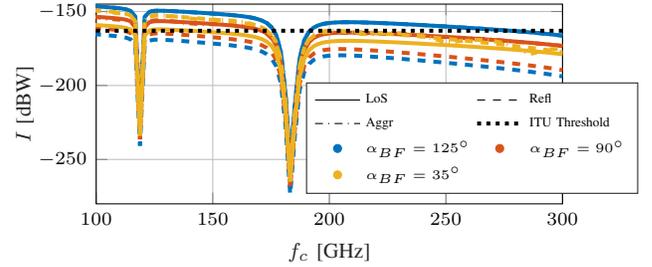}%

    \setlength\abovecaptionskip{0cm}
    \setlength\belowcaptionskip{-.5cm}
    \caption{Interference to the satellite for different \gls{sthz} frequencies. The satellite is at $\alpha_{n}=\ang{35}$ and at 400~km of altitude.}
    \label{fig:freqs}
\end{figure}
In this section, we analyze a basic example to show how the carrier frequency $f_c$ affects the interference to the satellite.
Specifically, we consider a ground node placed at height $h_{tx}=3$~m and a satellite at 400~km of altitude and $\alpha_{n}=\ang{35}$, with antenna gains of $35$~dB and $38.5$~dB and \glspl{hpbw} of $\ang{3}$ and $\ang{2}$, respectively.
In \cref{fig:freqs} we report the interference to the satellite for different frequencies, considering the \gls{los} and the \gls{gr} separately, as well as their superposition.
For the ground node, we considered three beamforming angles, $\ang{90}$ (horizon, case C3), $\ang{125}$ (upwards, case C2), and $\ang{35}$ (downwards, case C1).
The power of each ray is computed by combining the path loss $L$ derived in \cref{eq:tot_pathloss} with the transmitter and satellite gain $G_{TX2S}(\theta_{AoD},\phi_{AoD})$ and $G_{S}(\theta_{AoA},\phi_{AoA})=G_S$ of the \gls{tx} and \gls{rx} at the \gls{aod} and \gls{aoa}, respectively:
\begin{align}
    I_{LoS} = P_{tx}  & + G_{TX2S}(\theta_{AoD}^{LoS},\phi_{AoD}^{LoS}) + L_{LoS} +  G_{S}    \\
    I_{refl} = P_{tx} & + G_{TX2S}(\theta_{AoD}^{refl},\phi_{AoD}^{refl}) + L_{LoS} +  G_{S},
\end{align}
where $P_{tx}=0$~dBW.
The power of the two rays is then combined according to~\cref{eq:ef_superposition}.

First, we observe the presence of the two absorption peaks at 118 and 183~GHz, due to the presence of the oxygen molecules and of the water vapor, respectively.
Secondly, we observe how the reflected ray can be amplified by the beam of the ground node, as explained in \cref{ssec:bf_ampl}.
Specifically, when the beamforming angle is $\ang{125}$, the \gls{los} ray is amplified, whereas the reflected ray is attenuated, and viceversa when the beamforming angle is $\ang{35}$.
Furthermore, we observe that, in the considered frequency band, the reflection loss remains almost constant, whereas the \gls{los} path loss increases with the frequency and dictates the overall trend.

Finally, the ITU threshold~\cite{itu-r-2017} is exceeded particularly for frequencies below the second absorption peak.
The only exception is when the ground node steers its beam toward the satellite, thus amplifying the \gls{los} ray.

According to this observation, we selected three frequencies for the remainder of this work, 164, 178, and 240~GHz, that are representative of the three regions of the spectrum, i.e., below, close to and above the oxygen absorption peak, respectively.
Accordingly, for our simulations we selected two satellites that operate in those bands, i.e., TEMPEST-D~\cite{reising2015overview} for the former two and the \gls{mls} instrument on board of the Aura mission~\cite{waters2006earth} for the latter.

\section{Simulation Setup}
\label{sec:simulation}
In this and the following section, we report the setup and results of an extensive analysis of the \gls{rfi} in different scenarios.
The mathematical analysis in \cref{ssec:single_link} allows us (i) to gain a deep understanding of the simulation results, identifying the most significant elements that contribute to the overall power that reaches an incumbent satellite.
Further, we (ii) accurately quantify the \gls{rfi} in a realistic and complex scenario.
Finally, we (iii) provide useful insights and guidelines for the study and simulation of the \gls{rfi} at \gls{sthz} frequencies, that will hopefully help the study and design of coexistence solutions.

\subsection{Scenarios}
For this analysis, we consider two representative outdoor scenarios, the \textit{Urban Cellular} and the \textit{Backhaul} scenarios, both set in the city of Boston, MA, USA.
The two present inherently different characteristics, e.g., different types of nodes and beamforming angle distributions, thus producing different interference patterns.

\begin{figure}[t]
    \centering
    \includegraphics[width=.5\columnwidth]{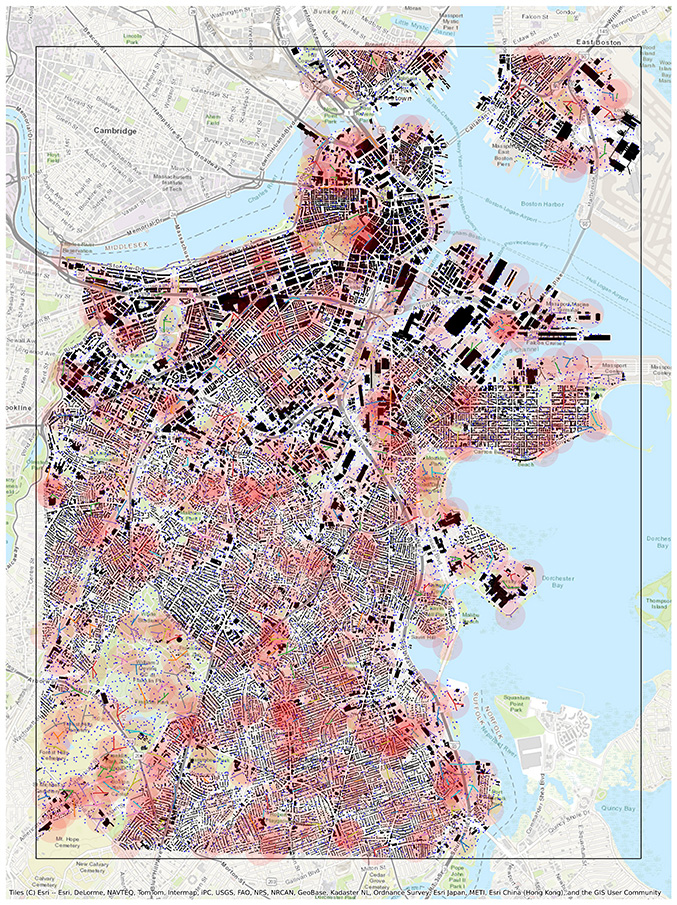}
    \captionof{figure}{Map of Boston with interfering nodes ($\lambda_g=10$). The red areas represent the coverage of each \gls{gnb} ($d_{MAX}=200$~m in the Urban scenario).}
    \label{fig:urban_cell}
\end{figure}
\begin{table}
    \centering
    \footnotesize
    \renewcommand*{\arraystretch}{0.9}
    \begin{tabular}[t]{lll}
        \toprule
                            & \textbf{\gls{gnb}} (Urban/Backhaul) & \textbf{\gls{ue}} (Urban) \\ \midrule
        $P_{TX}$ [dBm]      & 10/30                               & 10                        \\ 
        $G_{MAX}$ [dB]      & 35                                  & 24.5                      \\ 
        $\theta_{HB}$ [deg] & 3                                   & 10                        \\ 
        $N^{SEC}$           & 3                                   & 1                         \\ 
        Height [m]          & $\{3, 5, 8, 10, 15\}$               & $\mathcal{U}([1.6,1.8])$  \\ \bottomrule
    \end{tabular}
    \captionof{table}{Network parameters for the simulation campaign.}
    \label{tab:node_char}
    \centering
    \footnotesize
    \renewcommand*{\arraystretch}{0.9}
    \begin{tabular}[t]{lll}\toprule
                                         & TEMPEST \cite{reising2015overview} & Aura \gls{mls}\cite{waters2006earth} \\ \midrule
        $\theta_{HB}$ [deg]              & 1.68/1.72                          & 0.066 \cite{cofield2006design}       \\ 
        $f_c$ [GHz]                      & 164/178                            & 24.5                                 \\ 
        Altitude [km]                    & 400                                & 705                                  \\ 
        Scan Mode                        & Conical                            & Limb                                 \\ 
        $I_{th}$ [dBW] \cite{itu-r-2017} & $-$163                             & $-$194                               \\ \bottomrule
    \end{tabular}
    \setlength\abovecaptionskip{-.5cm}
    \captionof{table}{Specifications of the considered satellites.}
    \label{tab:sat_char}
\end{table}

\subsubsection{Map}
We use data from the city of Boston to (i) select realistic locations for the terrestrial nodes and to (ii) characterize the path obstruction from the terrestrial network to the satellite.
To do that, we extract terrain data and the road network from \gls{osm} using OSMnx~\cite{boeing2017osmnx} to reject the nodes spawned in invalid locations (e.g., on water surfaces) and to move indoor nodes outside the building footprint, respectively, as explained in the following paragraphs.
To analyze the obstruction of the paths from the terrestrial network to the satellite, we used the 3D model of the buildings published by the Boston Planning \& Development Agency\footnote{\url{www.bostonplans.org/3d-data-maps/3d-smart-model/3d-data-download}}.


\subsubsection{Node Characteristics}
We define two types of nodes: the \glspl{gnb} and the \glspl{ue} (see \cref{tab:node_char}).
The former represents a generic fixed node that can transmit 
using a large antenna array and thus a narrow beam ($\theta_{HB}=\ang{3}$).
Each \gls{gnb} has $N_{gNB}^{SEC}=3$ \ang{120}-wide angular sectors, according to the current \gls{3gpp} guidelines.
The \glspl{ue} represent mobile nodes with more limited beamforming capabilities ($\theta_{HB}=\ang{10}$).
A single sector is available to the \gls{ue}.
For both \glspl{gnb} and \gls{ue}, in each sector a single link at a time can be active.

\subsubsection{Node Placement}
First, $N_{gNB}$ tentative \gls{gnb} locations are generated on the map according to a \gls{ppp} with a given density $\lambda_g$.
Then, the nodes are placed according to an iterative procedure loosely based on a Rejection Sampling process: the \glspl{gnb} that overlap with the building footprint (i.e., indoor) or that are in invalid areas (e.g., water bodies) are projected to the nearest street using OSMnx~\cite{boeing2017osmnx}.
The points for which this is not possible (e.g., the nearest street is an underground road, or two points overlap) are rejected, an equal number of new points are generated, and the procedure is iterated until all the nodes are successfully moved outside.
Finally, for the Urban Cellular scenario, $N_{UE}$ \glspl{ue} are placed on the map with the same iterative procedure.
According to~\cite{3GPP38913}, we deterministically set $N_{UE}=N_{gNB}\times N_{gNB}^{SEC} \times 10$.
Thus, the number of nodes for each scenario is fully characterized by $\lambda_g$.

Three \gls{gnb} densities were considered, $\lambda_g\in \{10, 45, 100\}$~\glspl{gnb}/km\textsuperscript{2}.
In addition, we consider the number and location of \glspl{gnb} that were approved by the City of Boston\footnote{The data is available in the Boston Open Data archive at \url{https://tinyurl.com/2xjn9m43}} as a fourth scenario named \textit{Real}, with $\lambda_g\simeq 18$~\glspl{gnb}/km\textsuperscript{2}.

\subsubsection{Attachment}
In the Urban Cellular scenario, the \glspl{ue} are assigned to \glspl{gnb} according to the following algorithm: first, each \gls{ue} is tentatively assigned to the closest \gls{gnb}. If the distance $d$ between the \gls{ue} and the nearest \gls{gnb} is larger than $d_{max}=200$ m, the \gls{ue} can not be served by any \gls{gnb} and remains unattached, and the process is terminated. Otherwise, if the \gls{los} between the \gls{ue} and the \gls{gnb} is unobstructed by the buildings, the assignment is confirmed and the process is terminated. If, on the contrary, the \gls{los} is obstructed, the \gls{ue} is assigned to the second-closest \gls{gnb}, and the procedure is repeated from the distance-check step.
Each \gls{ue} in coverage is thus assigned a single \gls{gnb}, and each sector of the \gls{gnb} is assigned a list of \glspl{ue} that it can serve.

In the Backhaul scenario, links are established between \glspl{gnb}.
First, a list with all the \gls{gnb} pairs that are in \gls{los} is computed.
Then, for each link, a random direction for the data flow is picked.
For each sector of the \glspl{gnb}, a single link can be active at any given time.

\subsubsection{Beamforming}
The beam of each node is simulated using the ITU antenna pattern shown in~\cref{fig:geom_Gtx}.
The choice is justified both by the ITU recommendations and by computational efficiency, as computing the beamforming vectors and gains for the large areas and the corresponding number of nodes considered in this analysis is extremely demanding from a computational point of view.
In both scenarios, geometric beamforming is considered, i.e., the beams of the \gls{tx} and of the \gls{rx} are aligned to the \gls{los} direction connecting the two.

\subsection{Satellite Model}
For the analysis, we consider a conical-scan and a limb radiometer.
The conical scan mode sounds the Earth’s surface, projecting its footprint on the ground.
The limb scan mode views the edge of the atmosphere and terminates in space rather than at the surface.
We model them using the specifications of the TEMPEST-D satellites~\cite{reising2015overview} and of the \gls{mls} on the Aura mission~\cite{waters2006earth}.
The most relevant characteristics for this study are reported in \cref{tab:sat_char}.

Given the importance of the geometry and of the obstruction by the buildings, we consider three representative nadir angles $\alpha_{n}$, as in~\cref{sec:geometry}, and sample the azimuth space $\alpha_{az}$ with a sampling step of \ang{10} for both satellites.
Specifically, for a given nadir angle $\alpha_{n}$, the location of the satellite is determined by considering a conical scan and assuming that its main lobe is centered in the center of the considered area.
That is, we fix $\alpha_{n}$ and thus determine the horizontal $x$ and the vertical distance $h_{sat}$ from the center of the ground network, according to the geometry and notation shown in \cref{fig:g2s_geom}.
The same location is also used for the limb satellite.
However, for the \gls{rfi} computation, the \gls{mls} main beam points 10~km above the network, while the TEMPEST one illuminates it, according to the respective scan mode.

\begin{figure*}[t]
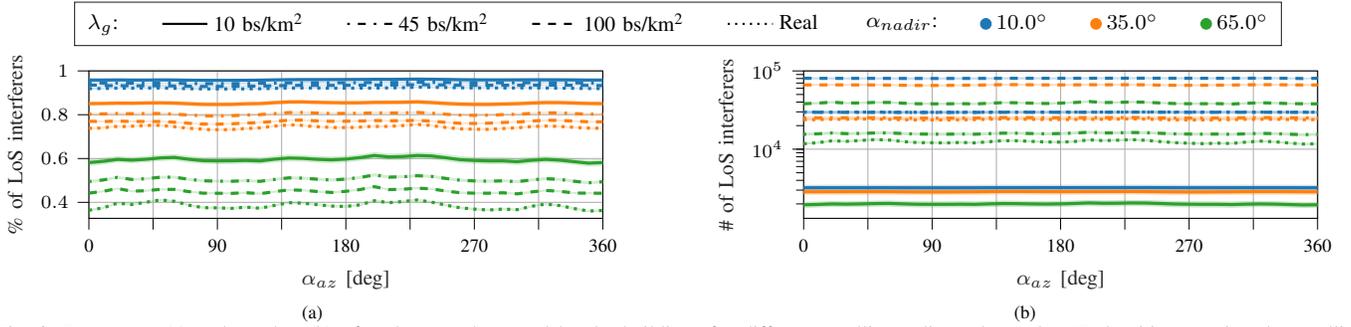

    \setlength\fheight{0.4\columnwidth}
    \centering
    \begin{subfigure}[t]{2\columnwidth}
        \centering
  \tikzsetnextfilename{imgs/results/n_interf/legend_n_interferers}%
  \begin{tikzpicture}

\definecolor{crimson2143940}{RGB}{214,39,40}
\definecolor{darkgray176}{RGB}{176,176,176}
\definecolor{darkorange25512714}{RGB}{255,127,14}
\definecolor{darkturquoise23190207}{RGB}{23,190,207}
\definecolor{forestgreen4416044}{RGB}{44,160,44}
\definecolor{goldenrod18818934}{RGB}{188,189,34}
\definecolor{gray127}{RGB}{127,127,127}
\definecolor{lightgray204}{RGB}{204,204,204}
\definecolor{mediumpurple148103189}{RGB}{148,103,189}
\definecolor{orchid227119194}{RGB}{227,119,194}
\definecolor{sienna1408675}{RGB}{140,86,75}
\definecolor{steelblue31119180}{RGB}{31,119,180}

\begin{axis}[%
width=0,
height=0,
at={(0,0)},
xmin=0,
xmax=0,
xtick={},
ymin=0,
ymax=0,
ytick={},
scale only axis,
axis background/.style={fill=white},
legend style={legend cell align=center, align=center, draw=white!15!black,at={(0,0)},anchor=south west, /tikz/every even column/.append style={column sep = 0.5cm},font=\footnotesize},
legend columns = 9
]   
    \addlegendimage{empty legend}
    \addlegendentry{$\lambda_g$:}
    \addlegendimage{thick,black}
    \addlegendentry{10 bs/km\textsuperscript{2}}
    \addlegendimage{thick,black, dash pattern=on 1pt off 3pt on 3pt off 3pt}
    \addlegendentry{45 bs/km\textsuperscript{2}}
    \addlegendimage{thick,black, dashed}
    \addlegendentry{100 bs/km\textsuperscript{2}}
    \addlegendimage{thick,black, dotted}
    \addlegendentry{Real}
    \addlegendimage{empty legend}
    \addlegendentry{$\alpha_{nadir}$:}
    \addlegendimage{scatter, only marks, steelblue31119180}
    \addlegendentry{\ang{10.0}}
    \addlegendimage{scatter, only marks, darkorange25512714}
    \addlegendentry{\ang{35.0}}
    \addlegendimage{scatter, only marks, forestgreen4416044}
    \addlegendentry{\ang{65.0}}

\end{axis}
\end{tikzpicture}%
%

    \end{subfigure}\\
    \begin{subfigure}[t]{.95\columnwidth}
        \setlength\fwidth{\columnwidth}
        \centering
  \tikzsetnextfilename{imgs/results/n_interf/perc_interferers_los}%
  \input{imgs/results/n_interf/perc_interferers_los}%

        \setlength\abovecaptionskip{-10pt}
        \caption{}
        \label{fig:perc_interf}
    \end{subfigure}
    \hfill
    \begin{subfigure}[t]{.95\columnwidth}
        \setlength\fwidth{\columnwidth}
        \centering
  \tikzsetnextfilename{imgs/results/n_interf/n_interferers_los}%
  \input{imgs/results/n_interf/n_interferers_los}%

        \setlength\abovecaptionskip{-10pt}
        \caption{}
        \label{fig:n_interf}
    \end{subfigure}
    \setlength\abovecaptionskip{0cm}
    \setlength\belowcaptionskip{-.5cm}
    \caption{Percentage (a) and number (b) of nodes not obstructed by the buildings for different satellite nadir angles and \gls{gnb} densities, varying the satellite azimuth location.}
    \label{fig:n_perc_interf}
\end{figure*}
\begin{figure*}[t]
    \def\tw{.8}
    \def\th{.15}
    \centering
    \setlength\fheight{\th\columnwidth}
    \setlength\fwidth{\tw\columnwidth}
    \begin{tabular}{ccc}
                                                                    & \footnotesize{\gls{gnb}}                          & \footnotesize{\gls{ue}}                          \\
        \rotatebox{90}{\qquad \footnotesize{$\alpha_{n}=\ang{10}$}} & %
  \tikzsetnextfilename{imgs/results/gtx/hist_gtx_nadir10_gnb}%
%
%
\definecolor{mycolor1}{rgb}{0.00000,0.44700,0.74100}%
\definecolor{mycolor2}{rgb}{0.85000,0.32500,0.09800}%
\begin{tikzpicture}
\pgfplotsset{every tick label/.append style={font=\scriptsize}}

    \begin{axis}[%
            width=0.951\fwidth,
            height=\fheight,
            at={(0\fwidth,0\fheight)},
            scale only axis,
            xmin=-14,
            xmax=35,
            ymin=0,
            ymax=0.45,
            axis background/.style={fill=white},
            xmajorgrids,
            ymajorgrids,
            every axis plot/.append style={thick},
            tick label style={font=\footnotesize},
            xlabel={$G_{TX2S}$ [dBi]},
            ylabel={Estimated PDF},
            xlabel style={font=\footnotesize\color{white!15!black}},
ylabel style={font=\footnotesize\color{white!15!black}}
        ]
        \addplot[ybar interval, fill=mycolor1, fill opacity=0.3, draw=black, area legend] table[row sep=crcr] {%
                x	y\\
                -14	0.15513812869374\\
                -13.5	0.056468180090976\\
                -13	0.0779903261773113\\
                -12.5	0.111178049933153\\
                -12	0.163867838618706\\
                -11.5	0.244900884108898\\
                -11	0.325735955060309\\
                -10.5	0.377993501296139\\
                -10	0.370350284694972\\
                -9.5	0.116376851325796\\
                -9	0.116376851325796\\
            };
        \addplot[ybar interval, fill=mycolor1, fill opacity=0.3, dashed, draw=black, area legend] table[row sep=crcr] {%
                x	y\\
                -11.5	0.000475135953149358\\
                -11	0.0225576758130178\\
                -10.5	0.0785725635924118\\
                -10	0.158010920350821\\
                -9.5	0.333432193205981\\
                -9	0.398456171967964\\
                -8.5	0.256620453498106\\
                -8	0.171109990673088\\
                -7.5	0.119039379558262\\
                -7	0.0856886957684927\\
                -6.5	0.0633970096554482\\
                -6	0.048500131194146\\
                -5.5	0.0379240308721508\\
                -5	0.0303331155339014\\
                -4.5	0.0246428175063704\\
                -4	0.0203017685852861\\
                -3.5	0.0171043063284087\\
                -3	0.0144040000617384\\
                -2.5	0.0123213426048763\\
                -2	0.0107407067138329\\
                -1.5	0.00930481802232726\\
                -1	0.00804395775621996\\
                -0.5	0.00714873594256563\\
                0	0.00631316520393156\\
                0.5	0.00558831203480283\\
                1	0.0050299027105372\\
                1.5	0.00450536132043784\\
                2	0.00411463060946847\\
                2.5	0.00364530100787973\\
                3	0.00335862893665123\\
                3.5	0.00301758295549174\\
                4	0.00267199479045318\\
                4.5	0.00245173562137884\\
                5	0.00223638612678867\\
                5.5	0.00202208030551699\\
                6	0.00187570144769249\\
                6.5	0.00174119988622491\\
                7	0.00160024518973172\\
                7.5	0.00148016395960837\\
                8	0.00137792807326881\\
                8.5	0.00126887156129854\\
                9	0.00118849401615048\\
                9.5	0.0010768944691929\\
                10	0.00101290700503242\\
                10.5	0.000959032882324363\\
                11	0.000907569497985784\\
                11.5	0.000878670036800509\\
                12	0.000787047279136309\\
                12.5	0.000728851466912248\\
                13	0.000690691242478386\\
                13.5	0.000650002682681417\\
                14	0.000623440461744596\\
                14.5	0.000601332226941618\\
                15	0.000562657515660612\\
                15.5	0.000538902922946775\\
                16	0.000493378186786655\\
                16.5	0.000457187711996143\\
                17	0.000449558607034211\\
                17.5	0.000421056035702447\\
                18	0.000366388133287373\\
                18.5	0.000349189572968567\\
                19	0.000327022539668773\\
                19.5	0.000298402371343377\\
                20	0.000269135419552999\\
                20.5	0.00025274533856542\\
                21	0.000228873148857949\\
                21.5	0.000205956434723746\\
                22	0.000190948118411353\\
                22.5	0.00104680433844709\\
                23	4.0438666185499e-05\\
                23.5	4.19086286059097e-05\\
                24	4.86851553640031e-05\\
                24.5	3.84542169179445e-05\\
                25	4.4010674867097e-05\\
                25.5	3.94978902364361e-05\\
                26	3.63227714083489e-05\\
                26.5	4.51425459308133e-05\\
                27	4.15558376250111e-05\\
                27.5	4.43928650964038e-05\\
                28	4.19380278543179e-05\\
                28.5	3.83513195485157e-05\\
                29	3.96742857268854e-05\\
                29.5	3.78662319497802e-05\\
                30	4.47897549499147e-05\\
                30.5	4.49808500645681e-05\\
                31	4.06738601727647e-05\\
                31.5	4.24231154530534e-05\\
                32	3.66608627650434e-05\\
                32.5	3.76457375867186e-05\\
                33	4.80971703958388e-05\\
                33.5	4.85087598735538e-05\\
                34	4.11442481472961e-05\\
                34.5	4.45839602110572e-05\\
                35	4.45839602110572e-05\\
            };
    \end{axis}
    \begin{axis}[%
        width=0.35\fwidth,
        height=0.5\fheight,
        at={(0.55\fwidth,0.35\fheight)},
        scale only axis,
        xmin=20,
        xmax=35,
        ymin=0,
        ymax=0.0012,
        axis background/.style={fill=white},
        xmajorgrids,
        ymajorgrids,
        every axis plot/.append style={thick},
        tick label style={font=\footnotesize},
    ]
    \addplot[ybar interval, fill=mycolor1, fill opacity=0.3, dashed, draw=black, area legend] table[row sep=crcr] {%
            x	y\\
            0	0.00631316520393156\\
            0.5	0.00558831203480283\\
            1	0.0050299027105372\\
            1.5	0.00450536132043784\\
            2	0.00411463060946847\\
            2.5	0.00364530100787973\\
            3	0.00335862893665123\\
            3.5	0.00301758295549174\\
            4	0.00267199479045318\\
            4.5	0.00245173562137884\\
            5	0.00223638612678867\\
            5.5	0.00202208030551699\\
            6	0.00187570144769249\\
            6.5	0.00174119988622491\\
            7	0.00160024518973172\\
            7.5	0.00148016395960837\\
            8	0.00137792807326881\\
            8.5	0.00126887156129854\\
            9	0.00118849401615048\\
            9.5	0.0010768944691929\\
            10	0.00101290700503242\\
            10.5	0.000959032882324363\\
            11	0.000907569497985784\\
            11.5	0.000878670036800509\\
            12	0.000787047279136309\\
            12.5	0.000728851466912248\\
            13	0.000690691242478386\\
            13.5	0.000650002682681417\\
            14	0.000623440461744596\\
            14.5	0.000601332226941618\\
            15	0.000562657515660612\\
            15.5	0.000538902922946775\\
            16	0.000493378186786655\\
            16.5	0.000457187711996143\\
            17	0.000449558607034211\\
            17.5	0.000421056035702447\\
            18	0.000366388133287373\\
            18.5	0.000349189572968567\\
            19	0.000327022539668773\\
            19.5	0.000298402371343377\\
            20	0.000269135419552999\\
            20.5	0.00025274533856542\\
            21	0.000228873148857949\\
            21.5	0.000205956434723746\\
            22	0.000190948118411353\\
            22.5	0.00104680433844709\\
            23	4.0438666185499e-05\\
            23.5	4.19086286059097e-05\\
            24	4.86851553640031e-05\\
            24.5	3.84542169179445e-05\\
            25	4.4010674867097e-05\\
            25.5	3.94978902364361e-05\\
            26	3.63227714083489e-05\\
            26.5	4.51425459308133e-05\\
            27	4.15558376250111e-05\\
            27.5	4.43928650964038e-05\\
            28	4.19380278543179e-05\\
            28.5	3.83513195485157e-05\\
            29	3.96742857268854e-05\\
            29.5	3.78662319497802e-05\\
            30	4.47897549499147e-05\\
            30.5	4.49808500645681e-05\\
            31	4.06738601727647e-05\\
            31.5	4.24231154530534e-05\\
            32	3.66608627650434e-05\\
            32.5	3.76457375867186e-05\\
            33	4.80971703958388e-05\\
            33.5	4.85087598735538e-05\\
            34	4.11442481472961e-05\\
            34.5	4.45839602110572e-05\\
            35	4.45839602110572e-05\\
        };
    \end{axis}
\end{tikzpicture}%
%
 & %
  \tikzsetnextfilename{imgs/results/gtx/hist_gtx_nadir10_ue}%
%
%
\definecolor{mycolor1}{rgb}{0.00000,0.44700,0.74100}%
\definecolor{mycolor2}{rgb}{0.85000,0.32500,0.09800}%
\begin{tikzpicture}
\pgfplotsset{every tick label/.append style={font=\scriptsize}}

    \begin{axis}[%
            width=0.951\fwidth,
            height=\fheight,
            at={(0\fwidth,0\fheight)},
            scale only axis,
            xmin=-8.5,
            xmax=25,
            ymin=0,
            ymax=0.4,
            axis background/.style={fill=white},
            xmajorgrids,
            ymajorgrids,
            every axis plot/.append style={thick},
            tick label style={font=\footnotesize},
            xlabel={$G_{TX2S}$ [dBi]},
            xlabel style={font=\footnotesize\color{white!15!black}},
            legend style={font=\tiny},
        ]

        \addplot[ybar interval, fill=mycolor2, fill opacity=0.3, draw=black, area legend] table[row sep=crcr] {%
                x	y\\
                -7	0.000305682429540772\\
                -6.5	0.133513100446373\\
                -6	0.206243914621862\\
                -5.5	0.210442920364219\\
                -5	0.20978033717773\\
                -4.5	0.222965811219858\\
                -4	0.266145954521415\\
                -3.5	0.187510272128959\\
                -3	0.125677975830719\\
                -2.5	0.0879013660706881\\
                -2	0.0638115944415085\\
                -1.5	0.0477710270947035\\
                -1	0.0368348700216431\\
                -0.5	0.0290257026268568\\
                0	0.0233575426944626\\
                0.5	0.0191327856242077\\
                1	0.0158853730695196\\
                1.5	0.013418177218212\\
                2	0.0114055097071537\\
                2.5	0.00978150439478177\\
                3	0.00845313197412648\\
                3.5	0.00733735875169308\\
                4	0.00641381112787333\\
                4.5	0.00565513965314501\\
                5	0.0049925466622285\\
                5.5	0.00442441932685838\\
                6	0.00396921938334953\\
                6.5	0.00352905452886822\\
                7	0.00316827121877003\\
                7.5	0.0028732560059635\\
                8	0.00258236355477164\\
                8.5	0.00233024761294002\\
                9	0.00207342554607859\\
                9.5	0.00188344516128125\\
                10	0.00172128974072268\\
                10.5	0.00156471794142346\\
                11	0.0014140631139065\\
                11.5	0.0012951746303408\\
                12	0.00116903087068747\\
                12.5	0.00108642857198707\\
                13	0.000995639576620227\\
                13.5	0.000916130574684218\\
                14	0.000825733756403189\\
                14.5	0.00521724942576144\\
                15	0.000170023473342783\\
                15.5	0.000160170024061608\\
                16	0.000169077346123248\\
                16.5	0.000163513333718206\\
                17	0.000168371427368776\\
                17.5	0.000168748897813875\\
                18	0.000163920217444742\\
                18.5	0.000167572366556422\\
                19	0.000168165534398721\\
                19.5	0.000172273589372664\\
                20	0.000171778465801819\\
                20.5	0.000175082557749835\\
                21	0.000176646363879533\\
                21.5	0.000164037870570488\\
                22	0.000166837034520513\\
                22.5	0.000168410645077357\\
                23	0.000166915469937677\\
                23.5	0.000165493828001587\\
                24	0.000170391139360738\\
                24.5	0.000170391139360738\\
            };
        \addplot[ybar interval, fill=mycolor2, fill opacity=0.3, dashed, draw=black, area legend] table[row sep=crcr] {%
                x	y\\
                -8.5	0.31109857647782\\
                -8	0.138089655069248\\
                -7.5	0.216602845851161\\
                -7	0.333557754219586\\
                -6.5	0.267898187034208\\
                -6	0.227909904403134\\
                -5.5	0.20863896459602\\
                -5	0.184256962211343\\
                -4.5	0.11073861232761\\
                -4	0.00120853780987005\\
                -3.5	0.00120853780987005\\
            };
    \end{axis}
    \begin{axis}[%
        width=0.6\fwidth,
        height=0.6\fheight,
        at={(0.32\fwidth,0.3\fheight)},
        scale only axis,
        xmin=10,
        xmax=25,
        ymin=0,
        ymax=0.006,
        axis background/.style={fill=white},
        xmajorgrids,
        ymajorgrids,
        every axis plot/.append style={thick},
        tick label style={font=\footnotesize},
        legend style={font=\tiny},
    ]
    \addlegendimage{ybar,ybar legend,draw=black,fill=black,fill opacity=0.5}
    \addlegendentry{LoS}
    
    \addlegendimage{ybar,ybar legend,draw=black,fill=black,fill opacity=0.5,dashed}
    \addlegendentry{Refl}
    \addplot[ybar interval, fill=mycolor2, fill opacity=0.3, draw=black, area legend] table[row sep=crcr] {%
            x	y\\
            0	0.0233575426944626\\
            0.5	0.0191327856242077\\
            1	0.0158853730695196\\
            1.5	0.013418177218212\\
            2	0.0114055097071537\\
            2.5	0.00978150439478177\\
            3	0.00845313197412648\\
            3.5	0.00733735875169308\\
            4	0.00641381112787333\\
            4.5	0.00565513965314501\\
            5	0.0049925466622285\\
            5.5	0.00442441932685838\\
            6	0.00396921938334953\\
            6.5	0.00352905452886822\\
            7	0.00316827121877003\\
            7.5	0.0028732560059635\\
            8	0.00258236355477164\\
            8.5	0.00233024761294002\\
            9	0.00207342554607859\\
            9.5	0.00188344516128125\\
            10	0.00172128974072268\\
            10.5	0.00156471794142346\\
            11	0.0014140631139065\\
            11.5	0.0012951746303408\\
            12	0.00116903087068747\\
            12.5	0.00108642857198707\\
            13	0.000995639576620227\\
            13.5	0.000916130574684218\\
            14	0.000825733756403189\\
            14.5	0.00521724942576144\\
            15	0.000170023473342783\\
            15.5	0.000160170024061608\\
            16	0.000169077346123248\\
            16.5	0.000163513333718206\\
            17	0.000168371427368776\\
            17.5	0.000168748897813875\\
            18	0.000163920217444742\\
            18.5	0.000167572366556422\\
            19	0.000168165534398721\\
            19.5	0.000172273589372664\\
            20	0.000171778465801819\\
            20.5	0.000175082557749835\\
            21	0.000176646363879533\\
            21.5	0.000164037870570488\\
            22	0.000166837034520513\\
            22.5	0.000168410645077357\\
            23	0.000166915469937677\\
            23.5	0.000165493828001587\\
            24	0.000170391139360738\\
            24.5	0.000170391139360738\\
        };
    \end{axis}
\end{tikzpicture}%
%
 \\
        \rotatebox{90}{\qquad \footnotesize{$\alpha_{n}=\ang{35}$}} & %
  \tikzsetnextfilename{imgs/results/gtx/hist_gtx_nadir35_gnb}%
%
%
\definecolor{mycolor1}{rgb}{0.00000,0.44700,0.74100}%
\definecolor{mycolor2}{rgb}{0.85000,0.32500,0.09800}%
\begin{tikzpicture}
\pgfplotsset{every tick label/.append style={font=\scriptsize}}

    \begin{axis}[%
            width=0.951\fwidth,
            height=\fheight,
            at={(0\fwidth,0\fheight)},
            scale only axis,
            xmin=-14,
            xmax=35,
            ymin=0,
            ymax=0.2,
            axis background/.style={fill=white},
            xmajorgrids,
            ymajorgrids,
            every axis plot/.append style={thick},
            tick label style={font=\footnotesize},
            xlabel={$G_{TX2S}$ [dBi]},
            ylabel={Estimated PDF},
            scaled y ticks=base 10:2,
            xlabel style={font=\footnotesize\color{white!15!black}},
ylabel style={font=\footnotesize\color{white!15!black}}
           ]
        \addplot[ybar interval, fill=mycolor1, fill opacity=0.3, draw=black, area legend] table[row sep=crcr] {%
                x	y\\
                -14	0.0537503519327622\\
                -13.5	0.0283830919371821\\
                -13	0.0414771274609522\\
                -12.5	0.0576654195511394\\
                -12	0.0727254560474277\\
                -11.5	0.0857305116561917\\
                -11	0.0966692597658375\\
                -10.5	0.106489376427018\\
                -10	0.116679366021893\\
                -9.5	0.125364431122218\\
                -9	0.134098617582886\\
                -8.5	0.143601992152243\\
                -8	0.15107917488428\\
                -7.5	0.15813343027801\\
                -7	0.162851881253944\\
                -6.5	0.162663261901119\\
                -6	0.153881357746087\\
                -5.5	0.121558522854472\\
                -5	0.0271973694243373\\
                -4.5	0.0271973694243373\\
            };
        \addplot[ybar interval, fill=mycolor1, fill opacity=0.3, dashed, draw=black, area legend] table[row sep=crcr] {%
                x	y\\
                -12.5	0.00265017000583863\\
                -12	0.00975418605332634\\
                -11.5	0.0197978399957123\\
                -11	0.0304302345315297\\
                -10.5	0.0398062576832906\\
                -10	0.0475018911445778\\
                -9.5	0.0554618446376496\\
                -9	0.0618689544254048\\
                -8.5	0.0677874355787674\\
                -8	0.0747762676300707\\
                -7.5	0.0808181949521239\\
                -7	0.0882461103981484\\
                -6.5	0.0961907833888306\\
                -6	0.106678822691978\\
                -5.5	0.123011643735872\\
                -5	0.164696513761371\\
                -4.5	0.186957836188025\\
                -4	0.138945441456111\\
                -3.5	0.103982971002325\\
                -3	0.0794556800146337\\
                -2.5	0.0621258878387082\\
                -2	0.0496368740439306\\
                -1.5	0.0403218895156304\\
                -1	0.0333537372165377\\
                -0.5	0.0281195154314347\\
                0	0.023841704390711\\
                0.5	0.0205093318736875\\
                1	0.0175920970436842\\
                1.5	0.0152588011576575\\
                2	0.0134878541651644\\
                2.5	0.011781802403147\\
                3	0.0103610909183569\\
                3.5	0.00920789593242685\\
                4	0.00825479283658665\\
                4.5	0.00730663731587212\\
                5	0.00653446120535118\\
                5.5	0.00589606419078472\\
                6	0.00528074427873565\\
                6.5	0.00480547826648725\\
                7	0.0043178989071564\\
                7.5	0.00391456035295\\
                8	0.00354816183999096\\
                8.5	0.00317202105129287\\
                9	0.00286638320813571\\
                9.5	0.00264547953784729\\
                10	0.00239258340143719\\
                10.5	0.00218878777884459\\
                11	0.00198353289954357\\
                11.5	0.00182546065381469\\
                12	0.00165485269807636\\
                12.5	0.00147737928696701\\
                13	0.00136213275120943\\
                13.5	0.00122579648159449\\
                14	0.00110932000089696\\
                14.5	0.00100800589228402\\
                15	0.000939240157110248\\
                15.5	0.000860412499489868\\
                16	0.00078183500016236\\
                16.5	0.000719559482920307\\
                17	0.000648313011342226\\
                17.5	0.000587927578979646\\
                18	0.000550209267487807\\
                18.5	0.000502297005561457\\
                19	0.000462125753031184\\
                19.5	0.000416437120374802\\
                20	0.00038306183480087\\
                20.5	0.000349137590750914\\
                21	0.000314810313895777\\
                21.5	0.000287001050338235\\
                22	0.000266050293310251\\
                22.5	0.00141430465295606\\
                23	5.5354471139276e-05\\
                23.5	5.52710850416522e-05\\
                24	5.30405069302151e-05\\
                24.5	5.5931224981174e-05\\
                25	5.73487886407789e-05\\
                25.5	5.48333080291271e-05\\
                26	5.31794837595881e-05\\
                26.5	5.6111894859359e-05\\
                27	5.06153612576558e-05\\
                27.5	5.82660357146408e-05\\
                28	5.36936980282683e-05\\
                28.5	5.73557374822475e-05\\
                29	5.445807058982e-05\\
                29.5	5.90304082761925e-05\\
                30	5.64871322986661e-05\\
                30.5	5.30752511375584e-05\\
                31	6.07954140092299e-05\\
                31.5	5.40341912602322e-05\\
                32	6.08579535824478e-05\\
                32.5	5.40828331505128e-05\\
                33	5.63134112619498e-05\\
                33.5	5.34365908939282e-05\\
                34	5.29640696740599e-05\\
                34.5	5.2644422966502e-05\\
                35	5.2644422966502e-05\\
            };
    \end{axis}
    \begin{axis}[%
            width=0.35\fwidth,
            height=0.5\fheight,
            at={(0.55\fwidth,0.35\fheight)},
            scale only axis,
            xmin=20,
            xmax=35,
            ymin=0,
            ymax=0.0015,
            axis background/.style={fill=white},
            xmajorgrids,
            ymajorgrids,
            every axis plot/.append style={thick},
            tick label style={font=\footnotesize},
        ]
        \addplot[ybar interval, fill=mycolor1, fill opacity=0.3, dashed, draw=black, area legend] table[row sep=crcr] {%
                x	y\\
                0	0.023841704390711\\
                0.5	0.0205093318736875\\
                1	0.0175920970436842\\
                1.5	0.0152588011576575\\
                2	0.0134878541651644\\
                2.5	0.011781802403147\\
                3	0.0103610909183569\\
                3.5	0.00920789593242685\\
                4	0.00825479283658665\\
                4.5	0.00730663731587212\\
                5	0.00653446120535118\\
                5.5	0.00589606419078472\\
                6	0.00528074427873565\\
                6.5	0.00480547826648725\\
                7	0.0043178989071564\\
                7.5	0.00391456035295\\
                8	0.00354816183999096\\
                8.5	0.00317202105129287\\
                9	0.00286638320813571\\
                9.5	0.00264547953784729\\
                10	0.00239258340143719\\
                10.5	0.00218878777884459\\
                11	0.00198353289954357\\
                11.5	0.00182546065381469\\
                12	0.00165485269807636\\
                12.5	0.00147737928696701\\
                13	0.00136213275120943\\
                13.5	0.00122579648159449\\
                14	0.00110932000089696\\
                14.5	0.00100800589228402\\
                15	0.000939240157110248\\
                15.5	0.000860412499489868\\
                16	0.00078183500016236\\
                16.5	0.000719559482920307\\
                17	0.000648313011342226\\
                17.5	0.000587927578979646\\
                18	0.000550209267487807\\
                18.5	0.000502297005561457\\
                19	0.000462125753031184\\
                19.5	0.000416437120374802\\
                20	0.00038306183480087\\
                20.5	0.000349137590750914\\
                21	0.000314810313895777\\
                21.5	0.000287001050338235\\
                22	0.000266050293310251\\
                22.5	0.00141430465295606\\
                23	5.5354471139276e-05\\
                23.5	5.52710850416522e-05\\
                24	5.30405069302151e-05\\
                24.5	5.5931224981174e-05\\
                25	5.73487886407789e-05\\
                25.5	5.48333080291271e-05\\
                26	5.31794837595881e-05\\
                26.5	5.6111894859359e-05\\
                27	5.06153612576558e-05\\
                27.5	5.82660357146408e-05\\
                28	5.36936980282683e-05\\
                28.5	5.73557374822475e-05\\
                29	5.445807058982e-05\\
                29.5	5.90304082761925e-05\\
                30	5.64871322986661e-05\\
                30.5	5.30752511375584e-05\\
                31	6.07954140092299e-05\\
                31.5	5.40341912602322e-05\\
                32	6.08579535824478e-05\\
                32.5	5.40828331505128e-05\\
                33	5.63134112619498e-05\\
                33.5	5.34365908939282e-05\\
                34	5.29640696740599e-05\\
                34.5	5.2644422966502e-05\\
                35	5.2644422966502e-05\\
            };
    \end{axis}
\end{tikzpicture}%
%
 & %
  \tikzsetnextfilename{imgs/results/gtx/hist_gtx_nadir35_ue}%
%
%
\definecolor{mycolor1}{rgb}{0.00000,0.44700,0.74100}%
\definecolor{mycolor2}{rgb}{0.85000,0.32500,0.09800}%
\begin{tikzpicture}
\pgfplotsset{every tick label/.append style={font=\scriptsize}}

    \begin{axis}[%
            width=0.951\fwidth,
            height=\fheight,
            at={(0\fwidth,0\fheight)},
            scale only axis,
            xmin=-8.5,
            xmax=25,
            ymin=0,
            ymax=0.8,
            axis background/.style={fill=white},
            xmajorgrids,
            ymajorgrids,
            every axis plot/.append style={thick},
            tick label style={font=\footnotesize},
            xlabel={$G_{TX2S}$ [dBi]},
            xlabel style={font=\footnotesize\color{white!15!black}}
        ]
        \addplot[ybar interval, fill=mycolor2, fill opacity=0.3, draw=black, area legend] table[row sep=crcr] {%
                x	y\\
                -8.5	0.243240024279126\\
                -8	0.12131424937605\\
                -7.5	0.1103151987566\\
                -7	0.100585875470368\\
                -6.5	0.0925747068544472\\
                -6	0.0860267876866914\\
                -5.5	0.0805630377151415\\
                -5	0.0759827633055882\\
                -4.5	0.072250048933454\\
                -4	0.0689775120677756\\
                -3.5	0.0662803187345241\\
                -3	0.0641614464465438\\
                -2.5	0.0623553659378057\\
                -2	0.0612283159223849\\
                -1.5	0.0607653045651835\\
                -1	0.0610764048390747\\
                -0.5	0.0631433316928628\\
                0	0.0689325574166575\\
                0.5	0.0849533653422915\\
                1	0.0690175775732898\\
                1.5	0.0516862260586027\\
                2	0.039455627467125\\
                2.5	0.0307928608672722\\
                3	0.0244883748079729\\
                3.5	0.0196753677991567\\
                4	0.0160969488377386\\
                4.5	0.0133753097764682\\
                5	0.0112503642891383\\
                5.5	0.00955751500230563\\
                6	0.00812233295731517\\
                6.5	0.00697329337233041\\
                7	0.00607329898615861\\
                7.5	0.00526091530865\\
                8	0.00458549490483542\\
                8.5	0.00403030577452295\\
                9	0.00357472313808258\\
                9.5	0.00315091531380561\\
                10	0.00277340839353572\\
                10.5	0.00247773557529855\\
                11	0.00222256851739307\\
                11.5	0.00199583496929996\\
                12	0.00177657185031899\\
                12.5	0.00160042821500671\\
                13	0.0014376196664005\\
                13.5	0.00128133152335309\\
                14	0.00116126940038609\\
                14.5	0.00732236446565731\\
                15	0.000217080381938896\\
                15.5	0.000213257116419345\\
                16	0.000212515634621613\\
                16.5	0.000211966474665169\\
                17	0.000216005233332186\\
                17.5	0.000221313779577818\\
                18	0.0002141051862255\\
                18.5	0.000213210773806986\\
                19	0.000217024770804066\\
                19.5	0.000214554709565374\\
                20	0.000213711274020455\\
                20.5	0.000213305776162321\\
                21	0.00020955897595316\\
                21.5	0.000209748980663829\\
                22	0.000211827446828094\\
                22.5	0.000210620221776163\\
                23	0.000209744346402593\\
                23.5	0.000212096233979772\\
                24	0.000215152529264795\\
                24.5	0.000215152529264795\\
            };
        \addplot[ybar interval, fill=mycolor2, fill opacity=0.3, dashed, draw=black, area legend] table[row sep=crcr] {%
                x	y\\
                -8.5	0.720364268689624\\
                -8	0.104300886964581\\
                -7.5	0.096718871793273\\
                -7	0.0910878018198037\\
                -6.5	0.0868523720573303\\
                -6	0.0834562695671905\\
                -5.5	0.0807943823531642\\
                -5	0.0787432212561004\\
                -4.5	0.0769608426764529\\
                -4	0.0757346380076305\\
                -3.5	0.074648920968172\\
                -3	0.0736077275935366\\
                -2.5	0.0727327165096872\\
                -2	0.0710956359225663\\
                -1.5	0.0684629074802358\\
                -1	0.0636040236023157\\
                -0.5	0.0533362501373552\\
                0	0.0273852793120513\\
                0.5	0.00011298328892929\\
                1	0.00011298328892929\\
            };
    \end{axis}
    \begin{axis}[%
            width=0.68\fwidth,
            height=0.6\fheight,
            at={(0.23\fwidth,0.3\fheight)},
            scale only axis,
            xmin=10,
            xmax=25,
            ymin=0,
            ymax=0.008,
            axis background/.style={fill=white},
            xmajorgrids,
            ymajorgrids,
            every axis plot/.append style={thick},
            tick label style={font=\footnotesize},
        ]
        \addplot[ybar interval, fill=mycolor2, fill opacity=0.3, draw=black, area legend] table[row sep=crcr] {%
                x	y\\
                0	0.0689325574166575\\
                0.5	0.0849533653422915\\
                1	0.0690175775732898\\
                1.5	0.0516862260586027\\
                2	0.039455627467125\\
                2.5	0.0307928608672722\\
                3	0.0244883748079729\\
                3.5	0.0196753677991567\\
                4	0.0160969488377386\\
                4.5	0.0133753097764682\\
                5	0.0112503642891383\\
                5.5	0.00955751500230563\\
                6	0.00812233295731517\\
                6.5	0.00697329337233041\\
                7	0.00607329898615861\\
                7.5	0.00526091530865\\
                8	0.00458549490483542\\
                8.5	0.00403030577452295\\
                9	0.00357472313808258\\
                9.5	0.00315091531380561\\
                10	0.00277340839353572\\
                10.5	0.00247773557529855\\
                11	0.00222256851739307\\
                11.5	0.00199583496929996\\
                12	0.00177657185031899\\
                12.5	0.00160042821500671\\
                13	0.0014376196664005\\
                13.5	0.00128133152335309\\
                14	0.00116126940038609\\
                14.5	0.00732236446565731\\
                15	0.000217080381938896\\
                15.5	0.000213257116419345\\
                16	0.000212515634621613\\
                16.5	0.000211966474665169\\
                17	0.000216005233332186\\
                17.5	0.000221313779577818\\
                18	0.0002141051862255\\
                18.5	0.000213210773806986\\
                19	0.000217024770804066\\
                19.5	0.000214554709565374\\
                20	0.000213711274020455\\
                20.5	0.000213305776162321\\
                21	0.00020955897595316\\
                21.5	0.000209748980663829\\
                22	0.000211827446828094\\
                22.5	0.000210620221776163\\
                23	0.000209744346402593\\
                23.5	0.000212096233979772\\
                24	0.000215152529264795\\
                24.5	0.000215152529264795\\
            };
        \addplot[ybar interval, fill=mycolor2, fill opacity=0.3, dashed, draw=black, area legend] table[row sep=crcr] {%
                x	y\\
                0	0.0273852793120513\\
                0.5	0.00011298328892929\\
                1	0.00011298328892929\\
            };
    \end{axis}
\end{tikzpicture}%
%
 \\
        \rotatebox{90}{\qquad \footnotesize{$\alpha_{n}=\ang{65}$}} & %
  \tikzsetnextfilename{imgs/results/gtx/hist_gtx_nadir65_gnb}%
  \input{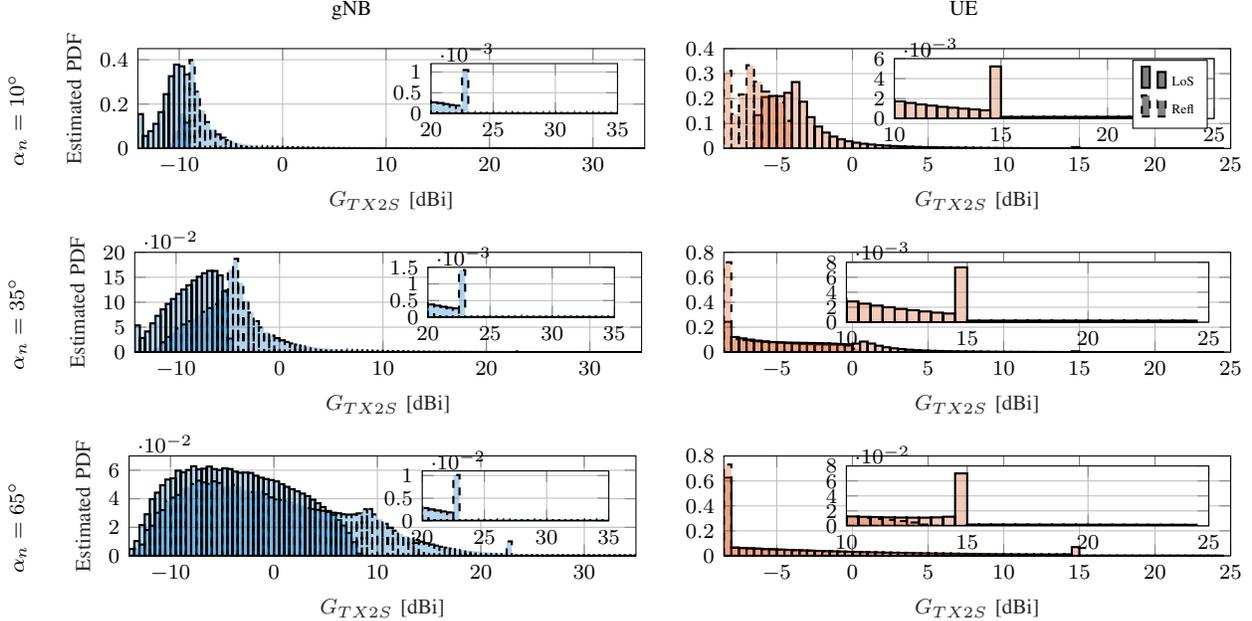}%
 & %
  \tikzsetnextfilename{imgs/results/gtx/hist_gtx_nadir65_ue}%
%
%
\definecolor{mycolor1}{rgb}{0.00000,0.44700,0.74100}%
\definecolor{mycolor2}{rgb}{0.85000,0.32500,0.09800}%
\begin{tikzpicture}
\pgfplotsset{every tick label/.append style={font=\scriptsize}}

    \begin{axis}[%
            width=0.951\fwidth,
            height=\fheight,
            at={(0\fwidth,0\fheight)},
            scale only axis,
            xmin=-8.5,
            xmax=25,
            ymin=0,
            ymax=0.8,
            axis background/.style={fill=white},
            xmajorgrids,
            ymajorgrids,
            every axis plot/.append style={thick},
            tick label style={font=\footnotesize},
            xlabel={$G_{TX2S}$ [dBi]},
            xlabel style={font=\footnotesize\color{white!15!black}},
ylabel style={font=\footnotesize\color{white!15!black}}
        ]
        \addplot[ybar interval, fill=mycolor2, fill opacity=0.3, draw=black, area legend] table[row sep=crcr] {%
                x	y\\
                -8.5	0.626490298934994\\
                -8	0.066481630171105\\
                -7.5	0.0636307144567578\\
                -7	0.0600543398991821\\
                -6.5	0.0576741512764437\\
                -6	0.0545146634577837\\
                -5.5	0.052465878657232\\
                -5	0.0497436095097041\\
                -4.5	0.0475352336703902\\
                -4	0.0455644349961102\\
                -3.5	0.0428489645469465\\
                -3	0.0412905612102469\\
                -2.5	0.0393171472571323\\
                -2	0.0371536315343557\\
                -1.5	0.035499327796344\\
                -1	0.0340414676866798\\
                -0.5	0.0321685687289349\\
                0	0.0305597487706175\\
                0.5	0.0293779122214177\\
                1	0.0281023183109715\\
                1.5	0.0265666927953298\\
                2	0.0252059573146559\\
                2.5	0.0242186600396374\\
                3	0.0232908366178438\\
                3.5	0.022172369891713\\
                4	0.0209512399971167\\
                4.5	0.0199293743309063\\
                5	0.0191358951393403\\
                5.5	0.0183635845638965\\
                6	0.0176156250832353\\
                6.5	0.0167550340896591\\
                7	0.015977813770927\\
                7.5	0.0152291921293833\\
                8	0.0146038966689627\\
                8.5	0.0140972845639421\\
                9	0.013545168147107\\
                9.5	0.0130082968296615\\
                10	0.0125651418092216\\
                10.5	0.0120839228040934\\
                11	0.0116810697694643\\
                11.5	0.0113420818953085\\
                12	0.0111024412521754\\
                12.5	0.0109570610072382\\
                13	0.0109583365652174\\
                13.5	0.0113673928667194\\
                14	0.0122410191902283\\
                14.5	0.0700519734183162\\
                15	0.00177349526339125\\
                15.5	0.00175185440850008\\
                16	0.00172310430692383\\
                16.5	0.0017220135380281\\
                17	0.00169563232984186\\
                17.5	0.00168132657496005\\
                18	0.00165469898132913\\
                18.5	0.00163709268809461\\
                19	0.00161247467574687\\
                19.5	0.00159723214287227\\
                20	0.00157333788776828\\
                20.5	0.0015732916904974\\
                21	0.00154816550817001\\
                21.5	0.00152344483522031\\
                22	0.0015079174191753\\
                22.5	0.0014893384167373\\
                23	0.00148552970840493\\
                23.5	0.00146537486572699\\
                24	0.00145270911396134\\
                24.5	0.00145270911396134\\
            };
        \addplot[ybar interval, fill=mycolor2, fill opacity=0.3, dashed, draw=black, area legend] table[row sep=crcr] {%
                x	y\\
                -8.5	0.731478440018565\\
                -8	0.0648671972443759\\
                -7.5	0.0620355252598644\\
                -7	0.0596939934867905\\
                -6.5	0.0570995188230857\\
                -6	0.0546130431126329\\
                -5.5	0.0523691390054989\\
                -5	0.0498559689720239\\
                -4.5	0.0480867803208845\\
                -4	0.0456777158373325\\
                -3.5	0.0438828825678258\\
                -3	0.0422731745953015\\
                -2.5	0.0401121227269717\\
                -2	0.0386813496171322\\
                -1.5	0.0372269081055129\\
                -1	0.0354160777477073\\
                -0.5	0.0340977770268497\\
                0	0.0329021839569885\\
                0.5	0.0315061922590898\\
                1	0.0300758015599925\\
                1.5	0.0289724464744689\\
                2	0.0279642090031675\\
                2.5	0.0268400779783241\\
                3	0.0256383021737104\\
                3.5	0.0245499124374361\\
                4	0.0237016201839737\\
                4.5	0.0227764582382857\\
                5	0.0218009900311268\\
                5.5	0.0207469864634759\\
                6	0.0198243782002614\\
                6.5	0.0189448334930501\\
                7	0.0179999478792124\\
                7.5	0.0170974148967093\\
                8	0.0161263200311936\\
                8.5	0.0150815918489289\\
                9	0.0140707313992476\\
                9.5	0.0130308154326993\\
                10	0.0119252916419595\\
                10.5	0.0107977189204357\\
                11	0.00943618782009672\\
                11.5	0.00807730023025798\\
                12	0.00643888065388315\\
                12.5	0.00450596117520786\\
                13	0.00169937687529801\\
                13.5	4.54273163631149e-07\\
                14	4.54273163631149e-07\\
            };
    \end{axis}
    \begin{axis}[%
            width=0.68\fwidth,
            height=0.6\fheight,
            at={(0.23\fwidth,0.3\fheight)},
            scale only axis,
            xmin=10,
            xmax=25,
            ymin=0,
            ymax=0.08,
            axis background/.style={fill=white},
            xmajorgrids,
            ymajorgrids,
            every axis plot/.append style={thick},
            tick label style={font=\footnotesize},
        ]
        \addplot[ybar interval, fill=mycolor2, fill opacity=0.3, draw=black, area legend] table[row sep=crcr] {%
                x	y\\
                0	0.0305597487706175\\
                0.5	0.0293779122214177\\
                1	0.0281023183109715\\
                1.5	0.0265666927953298\\
                2	0.0252059573146559\\
                2.5	0.0242186600396374\\
                3	0.0232908366178438\\
                3.5	0.022172369891713\\
                4	0.0209512399971167\\
                4.5	0.0199293743309063\\
                5	0.0191358951393403\\
                5.5	0.0183635845638965\\
                6	0.0176156250832353\\
                6.5	0.0167550340896591\\
                7	0.015977813770927\\
                7.5	0.0152291921293833\\
                8	0.0146038966689627\\
                8.5	0.0140972845639421\\
                9	0.013545168147107\\
                9.5	0.0130082968296615\\
                10	0.0125651418092216\\
                10.5	0.0120839228040934\\
                11	0.0116810697694643\\
                11.5	0.0113420818953085\\
                12	0.0111024412521754\\
                12.5	0.0109570610072382\\
                13	0.0109583365652174\\
                13.5	0.0113673928667194\\
                14	0.0122410191902283\\
                14.5	0.0700519734183162\\
                15	0.00177349526339125\\
                15.5	0.00175185440850008\\
                16	0.00172310430692383\\
                16.5	0.0017220135380281\\
                17	0.00169563232984186\\
                17.5	0.00168132657496005\\
                18	0.00165469898132913\\
                18.5	0.00163709268809461\\
                19	0.00161247467574687\\
                19.5	0.00159723214287227\\
                20	0.00157333788776828\\
                20.5	0.0015732916904974\\
                21	0.00154816550817001\\
                21.5	0.00152344483522031\\
                22	0.0015079174191753\\
                22.5	0.0014893384167373\\
                23	0.00148552970840493\\
                23.5	0.00146537486572699\\
                24	0.00145270911396134\\
                24.5	0.00145270911396134\\
            };
        \addplot[ybar interval, fill=mycolor2, fill opacity=0.3, dashed, draw=black, area legend] table[row sep=crcr] {%
                x	y\\
                0	0.0329021839569885\\
                0.5	0.0315061922590898\\
                1	0.0300758015599925\\
                1.5	0.0289724464744689\\
                2	0.0279642090031675\\
                2.5	0.0268400779783241\\
                3	0.0256383021737104\\
                3.5	0.0245499124374361\\
                4	0.0237016201839737\\
                4.5	0.0227764582382857\\
                5	0.0218009900311268\\
                5.5	0.0207469864634759\\
                6	0.0198243782002614\\
                6.5	0.0189448334930501\\
                7	0.0179999478792124\\
                7.5	0.0170974148967093\\
                8	0.0161263200311936\\
                8.5	0.0150815918489289\\
                9	0.0140707313992476\\
                9.5	0.0130308154326993\\
                10	0.0119252916419595\\
                10.5	0.0107977189204357\\
                11	0.00943618782009672\\
                11.5	0.00807730023025798\\
                12	0.00643888065388315\\
                12.5	0.00450596117520786\\
                13	0.00169937687529801\\
                13.5	4.54273163631149e-07\\
                14	4.54273163631149e-07\\
            };
    \end{axis}
\end{tikzpicture}%
%

    \end{tabular}
    \setlength\abovecaptionskip{0cm}
    \setlength\belowcaptionskip{-.5cm}
    \caption{Estimated \gls{pdf} of the $G_{TX2S}$ of the interfering rays for different satellite nadir angles in the Urban scenario.}
    \label{fig:uc_gtx}
\end{figure*}

\section{Numerical Results}
In this section, we present the results obtained through numerical simulations employing the setup presented in \cref{sec:simulation}.
Specifically, in \cref{ssec:res_urban} we report the results for the Urban Cellular scenario, evaluating the effect of the obstruction on the number of interfering nodes, the distribution of their gain, the aggregated power, the impact of different carrier frequencies and of different network load factors.
Similarly, in \cref{ssec:res_backahul} we report the beamforming gain of the ground nodes toward the satellite and the aggregated interference of the backhaul network.

\subsection{Urban Cellular}
\label{ssec:res_urban}
\paragraph{Number of Interfering Nodes}
\label{par:n_interf}
\cref{fig:perc_interf} and \cref{fig:n_interf} show the percentage and the number of ground nodes in \gls{los} with the satellite, respectively, thus potentially interfering.

The $x$ axis reports the azimuth of the satellite.
Although the obstruction by the buildings plays a fundamental role, we can observe that the azimuth position of the satellite $\alpha_{az}$ does not significantly affect the considered metrics, as the fluctuations are limited.
This suggests that the building distribution is homogeneous in the azimuth domain, with some variations visible when considering large angles ($\alpha_{n}\simeq \ang{65}$).
Considering the percentage of (un)obstructed nodes allows us to draw conclusions that are independent of the absolute number of nodes.
~\cref{fig:perc_interf} shows that the percentage of ground nodes that can interfere with the satellite is strongly correlated to the nadir angle: when the satellite is above the city ($\alpha_{n}\simeq \ang{10}$), the interfering signal propagates in the vertical direction, limiting the number of buildings that block it.
Thus, almost 100\% ground nodes are in \gls{los} with the satellite.
As the satellite moves towards the horizon, the percentage decreases, with almost 80\% of the ground nodes able to reach it when it is at $\alpha_{n}=\ang{35}$, and as little as 50\% when it is at $\alpha_{n}=\ang{65}$.

On the contrary, the absolute number of potential interferers reported in~\cref{fig:n_interf} is grouped by \gls{gnb} density, as expected.
However, the effect of the $\alpha_{n}$ angle is still clearly visible, as for each density the number of potential interferers decreases as the satellite approaches the horizon.
\begin{figure*}[t]
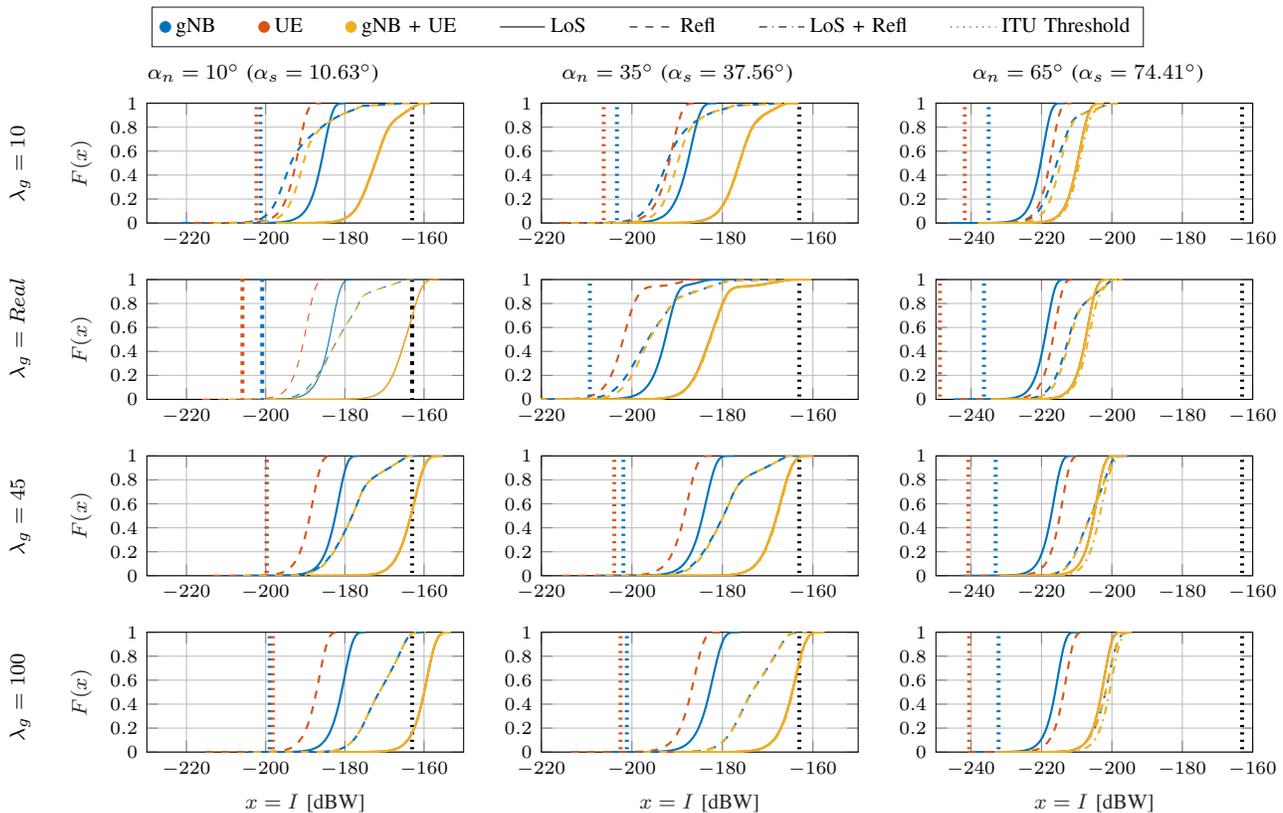

    \setlength\fheight{0.18\columnwidth}
    \setlength\fwidth{0.5\columnwidth}
    \centering
    \begin{tabular}{cccc}
        \multicolumn{4}{c}{%
  \tikzsetnextfilename{imgs/results/I_dBW_178/legend_IdbW}%
  \begin{tikzpicture}

    \definecolor{mycolor1}{rgb}{0.00000,0.44700,0.74100}%
    \definecolor{mycolor2}{rgb}{0.85000,0.32500,0.09800}%
    \definecolor{mycolor3}{rgb}{0.92900,0.69400,0.12500}%

    \begin{axis}[%
            width=0,
            height=0,
            at={(0,0)},
            xmin=0,
            xmax=0,
            xtick={},
            ymin=0,
            ymax=0,
            ytick={},
            scale only axis,
            axis background/.style={fill=white},
            legend style={legend cell align=center, align=center, draw=white!15!black,at={(0,0)},anchor=south west, /tikz/every even column/.append style={column sep = 0.5cm},font=\footnotesize},
            legend columns = 10
        ]


        \addlegendimage{scatter,only marks,color=mycolor1}
        \addlegendentry{gNB}

        \addlegendimage{scatter,only marks,color=mycolor2}
        \addlegendentry{UE}

        \addlegendimage{scatter,only marks,color=mycolor3}
        \addlegendentry{gNB + UE}

        \addlegendimage{draw=black,fill=black}
        \addlegendentry{LoS}

        \addlegendimage{draw=black,fill=black,dashed}
        \addlegendentry{Refl}

        \addlegendimage{draw=black,fill=black,dashdotted}
        \addlegendentry{LoS + Refl}
        
        \addlegendimage{draw=black,fill=black,dotted}
        \addlegendentry{ITU Threshold}
    \end{axis}
\end{tikzpicture}%
%
}                                                                                                                                                                                                       \\
                                                                    & \footnotesize{$\alpha_{n}=\ang{10}$ ($\alpha_{s} = \ang{10.63}$)} & \footnotesize{$\alpha_{n}=\ang{35}$ ($\alpha_{s} =  \ang{37.56}$)} & \footnotesize{$\alpha_{n}=\ang{65}$ ($\alpha_{s} = \ang{74.41}$)} \\
        \rotatebox{90}{\qquad \footnotesize{$\lambda_g=10$}}        & %
  \tikzsetnextfilename{imgs/results/I_dBW_178/I_dbWnadir10_dens10}%
  \input{imgs/results/I_dBW_178/I_dbWnadir10_dens10}%
            & %
  \tikzsetnextfilename{imgs/results/I_dBW_178/I_dbWnadir35_dens10}%
  \input{imgs/results/I_dBW_178/I_dbWnadir35_dens10}%
             & %
  \tikzsetnextfilename{imgs/results/I_dBW_178/I_dbWnadir65_dens10}%
  \input{imgs/results/I_dBW_178/I_dbWnadir65_dens10}%
            \\
        \rotatebox{90}{\qquad\footnotesize{$\lambda_g=Real$}}       & %
  \tikzsetnextfilename{imgs/results/I_dBW_178/I_dbWnadir10_densreal_data}%
  \input{imgs/results/I_dBW_178/I_dbWnadir10_densreal_data}%
     & %
  \tikzsetnextfilename{imgs/results/I_dBW_178/I_dbWnadir35_densreal_data}%
  \input{imgs/results/I_dBW_178/I_dbWnadir35_densreal_data}%
      & %
  \tikzsetnextfilename{imgs/results/I_dBW_178/I_dbWnadir65_densreal_data}%
  \input{imgs/results/I_dBW_178/I_dbWnadir65_densreal_data}%
     \\
        \rotatebox{90}{\qquad\footnotesize{$\lambda_g=45$}}         & %
  \tikzsetnextfilename{imgs/results/I_dBW_178/I_dbWnadir10_dens45}%
  \input{imgs/results/I_dBW_178/I_dbWnadir10_dens45}%

                                                                    & %
  \tikzsetnextfilename{imgs/results/I_dBW_178/I_dbWnadir35_dens45}%
  \input{imgs/results/I_dBW_178/I_dbWnadir35_dens45}%
            & %
  \tikzsetnextfilename{imgs/results/I_dBW_178/I_dbWnadir65_dens45}%
  \input{imgs/results/I_dBW_178/I_dbWnadir65_dens45}%
                                                                                 \\
        \rotatebox{90}{\qquad \quad \footnotesize{$\lambda_g=100$}} & %
  \tikzsetnextfilename{imgs/results/I_dBW_178/I_dbWnadir10_dens100}%
  \input{imgs/results/I_dBW_178/I_dbWnadir10_dens100}%

                                                                    & %
  \tikzsetnextfilename{imgs/results/I_dBW_178/I_dbWnadir35_dens100}%
  \input{imgs/results/I_dBW_178/I_dbWnadir35_dens100}%
           & %
  \tikzsetnextfilename{imgs/results/I_dBW_178/I_dbWnadir65_dens100}%
  \input{imgs/results/I_dBW_178/I_dbWnadir65_dens100}%
                                                                                \\
    \end{tabular}
    \setlength\abovecaptionskip{0cm}
    \setlength\belowcaptionskip{-.5cm}
    \caption{\gls{ecdf} of the \gls{rfi} power at the incumbent satellite, for different nadir angles and different \gls{gnb} densities. For these results, we consider the TEMPEST satellite at 178~GHz.}
    \label{fig:Idbw}
\end{figure*}
\paragraph{Interfering Nodes Gain}
\label{par:inter_gtx}

As shown in~\cref{ssec:single_link}, although the \gls{tx} focuses the emitted power toward the \gls{rx} through narrow beams, there is a non-zero probability that also the interfering rays to the satellite are amplified by the beamforming configuration of the \gls{tx}.
\cref{fig:uc_gtx} reports the estimated \gls{pdf} of the transmitter gain experienced by the interfering \gls{los} and reflected rays that reach the satellite.
For all the nadir angles, the vast majority of the rays are successfully suppressed ($G_{TX2S}<0$).

In particular, the \glspl{ue} suppress the \gls{rfi} with the minimum gain with high probability, corresponding to the peak probability in $-8.5$~dBi, that is clearly visible for $\alpha_{n}=\ang{35}$ and $\alpha_{n}=\ang{65}$.
On the contrary, the same behavior is not present when considering the \glspl{gnb}, where a broader interval of $G_{TX2S}$ has a high probability.
This is due to the fact that the interference, and hence the gain reported in~\cref{fig:uc_gtx}, is computed only for the rays emitted by the angular sector containing the satellite.
Given that the \glspl{ue} have a single sector, all the rays are included for the PDF estimation, even when the antenna is pointing in the opposite direction.
On the contrary, only the transmitter gain of the rays in the \gls{gnb} sector containing the satellite is considered, based on the realistic assumption that the power leakage by a sector to the adjacent ones can be effectively suppressed.
Thus, the \gls{gnb} gains reported in \cref{fig:uc_gtx} are computed only in the \ang{120} angular sector containing the satellite instead of the whole \ang{360} azimuth domain, thus increasing the probability of the antenna pointing toward the satellite in the azimuth plane.

Furthermore, \cref{fig:uc_gtx} shows a distinct distribution for the \gls{los} and for the ground-reflected rays when considering the \glspl{gnb} and the \glspl{ue}.
Specifically, the \gls{los} rays of the \glspl{ue} have a greater probability of being amplified than the \gls{gr} rays, and vice-versa for the \glspl{gnb}.
The zoom on the tails of the \gls{pdf} highlights this behavior, with the highest gain being experienced by the \gls{gr} (\gls{los}) rays of the \glspl{gnb} (\glspl{ue}).
This trend can be traced back to the analysis presented in \cref{ssec:single_link}: the \glspl{ue} point upwards to the \glspl{gnb} (case C2), amplifying the \gls{gr} more than the \gls{los}.
Conversely, the \glspl{gnb} point to the \glspl{ue} and thus toward the ground, increasing the probability of amplifying the \gls{gr}.

Moreover, all the estimated \glspl{pdf} have multiple peaks, typical of the multimodal distributions.
Again, referring to the analysis of \cref{ssec:single_link}, we can trace this behavior back to the discrete distribution of the \gls{gnb} heights in relation to the beam characteristics and the satellite positions. This favors some beamforming angles, and thus some $G_{TX2S}$.

Finally, also in this analysis, the elevation of the satellite plays a fundamental role.
Firstly, we can observe the dramatic change in the shape of the distributions in the six plots of \cref{fig:uc_gtx}.
For $\alpha_{n}=\ang{10}$, both the \gls{los} and the \gls{gr} rays are strongly suppressed.
Looking at \cref{fig:geom_Gtx}, this can be explained by the fact that the maximum amplification with this satellite elevation is obtained with extreme beamforming angles ($\alpha_{BF}\simeq \ang{10}$ or $\alpha_{BF}\simeq \ang{172}$), which are not common.
On the contrary, when the satellite is at a lower elevation angle, the interfering \gls{los} and \gls{gr} rays are more often aligned with the beams, resulting in a more uniform distribution of the transmitter gain.
However, a satellite at lower elevation angles also implies that terrestrial signals need to traverse a longer portion of the atmosphere, as we discuss next, and have a higher blockage probability, as discussed above.

\begin{table*}[t]
    \centering
    \begin{subtable}[t]{0.7\columnwidth}
        \centering
        \scriptsize
        \[
            \begin{array}{cc|cc}
                                                                                                  & \multicolumn{1}{c}{} & \multicolumn{2}{c}{\text{$\alpha_{n}$ [deg]}}                     \\
                                                                                                  &                      & 10                                            & 35                \\
                \cline{2-4}                                                                       & 10                   & 0/0.0371/0.0368                               & 0/0/0             \\
                \smash{\rotatebox[origin=c]{90}{$\lambda_g$ [\glspl{gnb}/km\textsuperscript{2}]}} & 45                   & 0/0.4537/0.4537                               & 0/0.012/0.012     \\
                                                                                                  & Real                 & 0/0.3102/0.3102                               & 0/1.85e-4/1.71e-4 \\
                                                                                                  & 100                  & 0/0.8385/0.8385                               & 0/0.2473/0.2469
            \end{array}
        \]
        \caption{\gls{los} (\gls{gnb}/\gls{ue}/combined)}
    \end{subtable}\hfill
    \begin{subtable}[t]{0.7\columnwidth}
        \centering
        \scriptsize
        \[
            \begin{array}{c|cc}
                \multicolumn{1}{c}{} & \multicolumn{2}{c}{\text{$\alpha_{n}$ [deg]}}                   \\
                                     & 10                                            & 35              \\\hline
                10                   & 0/0/0                                         & 0/0/0           \\
                45                   & 8e-4/0/7e-4                                   & 1.5e-4/0/1e-5   \\
                Real                 & 1.3e-4/0/1.15e-4                              & 0/0/0           \\
                100                  & 0.0124/0/0.0124                               & 0.0011/0/7.5e-4
            \end{array}
        \]
        \caption{Ground Reflection (\gls{gnb}/\gls{ue}/combined)}
    \end{subtable}\hfill
    \begin{subtable}[t]{0.40\columnwidth}
        \centering
        \scriptsize
        \[
            \begin{array}{c|cc}
                \multicolumn{1}{c}{} & \multicolumn{2}{c}{\text{$\alpha_{n}$ [deg]}}          \\
                                     & 10                                            & 35     \\\hline
                10                   & 0.0368                                        & 0      \\
                45                   & 0.4548                                        & 0.0148 \\
                Real                 & 0.3120                                        & 3.3e-4 \\
                100                  & 0.8415                                        & 0.27
            \end{array}
        \]
        \caption{Aggregated}
    \end{subtable}
    \setlength\abovecaptionskip{-0.5cm}
    \setlength\belowcaptionskip{-0.1cm}
    \caption{Probability that the \gls{rfi} is greater than the ITU threshold for the cellular scenario, for the TEMPEST satellite at 178~GHz. For $\alpha_{n} = 65$ degrees, the probability is always 0.}
    \label{tab:P_th}
\end{table*}
\paragraph{Aggregated Power}
\cref{fig:Idbw} reports the \glspl{ecdf} of the aggregated \gls{rfi} power at the satellite for different nadir angles and for different \gls{gnb} densities.
For each Monte Carlo iteration, the total \gls{rfi} is obtained by combining the electric fields at the receiver according to \cref{eq:ef_superposition}, thus accounting also for the phase difference.
In particular, to analyze the different contributions to the overall power, we sum all the interfering power generated by the \gls{los} (solid) and by the reflected (dashed) rays, aggregating the \glspl{gnb} (blue) and of the \glspl{ue} (orange).
The \gls{los} and reflected rays of the individual ground nodes are then combined according to \cref{eq:ef_superposition} (dash-dotted).
Similarly, the overall \gls{rfi} power (yellow) is obtained by combining all the aforementioned contributions.
In each Monte Carlo iteration, we assume that all the \gls{gnb} sectors either transmitting or receiving (with probability $P_{TX}=P_{RX}=0.5$) to a randomly selected \gls{ue} among the one assigned to it.
The sectors without assigned users remain silent.

First, we can observe that the power delivered by the \gls{los} rays of the \glspl{ue} to the satellite dominates the other contributions in almost all of the considered cases.
This is due (i) to the fact that we consider a single \gls{gnb} sector, as previously explained, as we assume that the \glspl{gnb} can effectively suppress the inter-sector leakage.
The same assumption does not hold for the \glspl{ue}, due to the constantly-changing orientation of the hand-held devices.
Furthermore, (ii) the \glspl{ue}' beam amplifies their \gls{los} ray to the satellite.
On the contrary, the \gls{gr} ray of the \glspl{ue} is effectively suppressed.
Conversely, the reflection from the \glspl{gnb} to the satellite is stronger than the corresponding direct ray in almost all scenarios.

Secondly, as the satellite approaches the horizon, the overall \gls{rfi} decreases.
This is due to (i) the lower number of interfering nodes due to the obstruction by the buildings, as illustrated at the beginning of this section, and to (ii) the greater path loss due to the larger distance of the satellite from the ground network.
Furthermore, we observe that for higher nadir angles, the impact of the reflections is also increased, taking over that of the direct rays.
Again, this is due to the considerations on the transmitter gain given in the previous paragraph.

Thirdly, we notice that, as expected, the overall \gls{rfi} increases with the number of nodes.
For reference, \cref{fig:n_interf} reports the potential number of interferers for each density and nadir angle, although not all the nodes are simultaneously active.
In \cref{fig:Idbw}, we report also the average interference generated by a single ground node (dotted).
Note that the gap between the \gls{ecdf} and the corresponding average interference is due to the aggregation of the \gls{rfi} over the network.
Considering the number in \cref{fig:n_interf}, one might expect a larger difference, as we aggregate the power of several thousands of ground nodes, that would correspond to tens of dBs if we just summed the electrical field intensities at the satellite.
However, due to the phase difference among the received signals, the overall interference is reduced, as a large percentage of signals superimpose destructively, canceling out.

Finally, \cref{tab:P_th} reports the probability that the total aggregated \gls{rfi} at the satellite is greater than the ITU threshold for the acceptable \gls{rfi} at the considered frequency ($I_{th} = -163$~dBW~\cite{itu-r-2017}).
The greatest interference is observed when the satellite is orbiting over the terrestrial network ($\alpha_{n}=\ang{10}$) and with $\lambda_g=100$ \glspl{gnb}/km\textsuperscript{2}.
For the same satellite position, even as little as $\lambda_g=10$ \glspl{gnb}/km\textsuperscript{2} is enough to cause significant interference to the satellite.
On the contrary, when the satellite is at the horizon, the \gls{rfi} is well below the safety levels.

\paragraph{Frequency}
\begin{figure}[t]
    \centering
    \setlength\fheight{0.25\columnwidth}
    \setlength\fwidth{0.7\columnwidth}
  \tikzsetnextfilename{imgs/results/freqs/aggrIdbWECDF_freqs.tex}%
  \input{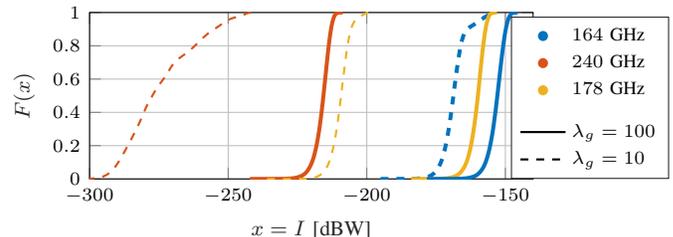}%

    \setlength\abovecaptionskip{-.5cm}
    \setlength\belowcaptionskip{-.5cm}
    \caption{Aggregated \gls{rfi} observed by TEMPEST at 164~GHz and 178~GHz and by Aura at 240~GHz.}
    \label{fig:freqs_cellular}
\end{figure}
\begin{figure}[t]
    \setlength\fheight{0.15\columnwidth}
    \centering
    \begin{subfigure}[t]{\columnwidth}
        \centering
  \tikzsetnextfilename{imgs/results/duty_cycle/legend_duty_cycle}%
  \begin{tikzpicture}
\definecolor{mycolor1}{rgb}{0.00000,0.44700,0.74100}%
\definecolor{mycolor2}{rgb}{0.85000,0.32500,0.09800}%
\definecolor{mycolor3}{rgb}{0.92900,0.69400,0.12500}%
\begin{axis}[%
width=0,
height=0,
at={(0,0)},
xmin=0,
xmax=0,
xtick={},
ymin=0,
ymax=0,
ytick={},
scale only axis,
axis background/.style={fill=white},
legend style={font=\footnotesize,legend cell align=center, align=center, draw=white!15!black,at={(0,0)},anchor=south west, /tikz/every even column/.append style={column sep = 0.5cm}},
legend columns = 5
]
\addlegendimage{empty legend};
\addlegendentry{$\mathbf{\alpha_{nadir}}$:}

\addlegendimage{scatter,only marks, mark=*, draw=mycolor1, fill=mycolor1};
\addlegendentry{\ang{10}}

\addlegendimage{scatter,only marks, mark=*,draw=mycolor2, fill=mycolor2};
\addlegendentry{\ang{35}}

\addlegendimage{scatter,only marks, mark=*, draw=mycolor3, fill=mycolor3};
\addlegendentry{\ang{65}}

\addlegendimage{color=black, dotted, line width=1.5pt};
\addlegendentry{ITU Threshold}

\addlegendimage{empty legend};
\addlegendentry{$\mathbf{\lambda_g}$:}

\addlegendimage{scatter,only marks, mark=*,draw=black, fill=black};
\addlegendentry{10}

\addlegendimage{scatter,only marks, mark=square*,draw=black, fill=black};
\addlegendentry{45}

\addlegendimage{scatter, only marks, mark=diamond*,draw=black, fill=black};
\addlegendentry{100}

\addlegendimage{scatter, only marks, mark=x,draw=black, fill=black};
\addlegendentry{Real}
\end{axis}
\end{tikzpicture}%
%

    \end{subfigure}\\
    \begin{subfigure}[t]{.48\columnwidth}
        \setlength\fwidth{.85\columnwidth}
        \centering
  \tikzsetnextfilename{imgs/results/duty_cycle/mean_idbw}%
%
%
\definecolor{mycolor1}{rgb}{0.00000,0.44700,0.74100}%
\definecolor{mycolor2}{rgb}{0.85000,0.32500,0.09800}%
\definecolor{mycolor3}{rgb}{0.92900,0.69400,0.12500}%
\begin{tikzpicture}
\pgfplotsset{every tick label/.append style={font=\scriptsize}}

\begin{axis}[%
width=0.951\fwidth,
height=\fheight,
at={(0\fwidth,0\fheight)},
scale only axis,
xmin=-220,
xmax=-150,
xlabel style={font=\footnotesize\color{white!15!black}},
xlabel={$E[I]$ [dBW]},
ymin=0,
ymax=1,
ylabel style={font=\footnotesize\color{white!15!black}},
ylabel={$\rho$},
axis background/.style={fill=white},
xmajorgrids,
ymajorgrids
]
\addplot[only marks, mark=*, mark options={}, mark size=1.5000pt, draw=mycolor1, fill=mycolor1, forget plot] table[row sep=crcr]{%
x	y\\
-174.501695896096	0.2\\
};
\addplot[only marks, mark=square*, mark options={}, mark size=1.0607pt, draw=mycolor1, fill=mycolor1, forget plot] table[row sep=crcr]{%
x	y\\
-166.952100335653	0.2\\
};
\addplot[only marks, mark=diamond*, mark options={}, mark size=2.5883pt, draw=mycolor1, fill=mycolor1, forget plot] table[row sep=crcr]{%
x	y\\
-163.179730936968	0.2\\
};
\addplot[only marks, mark=x, mark options={}, mark size=1.5000pt, draw=mycolor1, fill=mycolor1, forget plot] table[row sep=crcr]{%
x	y\\
-168.391238688468	0.2\\
};
\addplot[only marks, mark=*, mark options={}, mark size=1.5000pt, draw=mycolor2, fill=mycolor2, forget plot] table[row sep=crcr]{%
x	y\\
-178.565384782111	0.2\\
};
\addplot[only marks, mark=square*, mark options={}, mark size=1.0607pt, draw=mycolor2, fill=mycolor2, forget plot] table[row sep=crcr]{%
x	y\\
-171.314260730388	0.2\\
};
\addplot[only marks, mark=diamond*, mark options={}, mark size=2.5883pt, draw=mycolor2, fill=mycolor2, forget plot] table[row sep=crcr]{%
x	y\\
-167.856387820001	0.2\\
};
\addplot[only marks, mark=x, mark options={}, mark size=1.5000pt, draw=mycolor2, fill=mycolor2, forget plot] table[row sep=crcr]{%
x	y\\
-173.008064328891	0.2\\
};
\addplot[only marks, mark=*, mark options={}, mark size=1.5000pt, draw=mycolor3, fill=mycolor3, forget plot] table[row sep=crcr]{%
x	y\\
-212.119773162465	0.2\\
};
\addplot[only marks, mark=square*, mark options={}, mark size=1.0607pt, draw=mycolor3, fill=mycolor3, forget plot] table[row sep=crcr]{%
x	y\\
-206.622703608214	0.2\\
};
\addplot[only marks, mark=diamond*, mark options={}, mark size=2.5883pt, draw=mycolor3, fill=mycolor3, forget plot] table[row sep=crcr]{%
x	y\\
-204.093356357013	0.2\\
};
\addplot[only marks, mark=x, mark options={}, mark size=1.5000pt, draw=mycolor3, fill=mycolor3, forget plot] table[row sep=crcr]{%
x	y\\
-209.886030861019	0.2\\
};
\addplot[only marks, mark=*, mark options={}, mark size=1.5000pt, draw=mycolor1, fill=mycolor1, forget plot] table[row sep=crcr]{%
x	y\\
-171.360515799275	0.5\\
};
\addplot[only marks, mark=square*, mark options={}, mark size=1.0607pt, draw=mycolor1, fill=mycolor1, forget plot] table[row sep=crcr]{%
x	y\\
-164.374649246782	0.5\\
};
\addplot[only marks, mark=diamond*, mark options={}, mark size=2.5883pt, draw=mycolor1, fill=mycolor1, forget plot] table[row sep=crcr]{%
x	y\\
-161.030059451828	0.5\\
};
\addplot[only marks, mark=x, mark options={}, mark size=1.5000pt, draw=mycolor1, fill=mycolor1, forget plot] table[row sep=crcr]{%
x	y\\
-165.695442110052	0.5\\
};
\addplot[only marks, mark=*, mark options={}, mark size=1.5000pt, draw=mycolor2, fill=mycolor2, forget plot] table[row sep=crcr]{%
x	y\\
-175.526372413047	0.5\\
};
\addplot[only marks, mark=square*, mark options={}, mark size=1.0607pt, draw=mycolor2, fill=mycolor2, forget plot] table[row sep=crcr]{%
x	y\\
-168.926801520562	0.5\\
};
\addplot[only marks, mark=diamond*, mark options={}, mark size=2.5883pt, draw=mycolor2, fill=mycolor2, forget plot] table[row sep=crcr]{%
x	y\\
-165.669206557833	0.5\\
};
\addplot[only marks, mark=x, mark options={}, mark size=1.5000pt, draw=mycolor2, fill=mycolor2, forget plot] table[row sep=crcr]{%
x	y\\
-170.457190836857	0.5\\
};
\addplot[only marks, mark=*, mark options={}, mark size=1.5000pt, draw=mycolor3, fill=mycolor3, forget plot] table[row sep=crcr]{%
x	y\\
-209.278670463768	0.5\\
};
\addplot[only marks, mark=square*, mark options={}, mark size=1.0607pt, draw=mycolor3, fill=mycolor3, forget plot] table[row sep=crcr]{%
x	y\\
-204.368579532443	0.5\\
};
\addplot[only marks, mark=diamond*, mark options={}, mark size=2.5883pt, draw=mycolor3, fill=mycolor3, forget plot] table[row sep=crcr]{%
x	y\\
-201.994161171594	0.5\\
};
\addplot[only marks, mark=x, mark options={}, mark size=1.5000pt, draw=mycolor3, fill=mycolor3, forget plot] table[row sep=crcr]{%
x	y\\
-207.519239725579	0.5\\
};
\addplot[only marks, mark=*, mark options={}, mark size=1.5000pt, draw=mycolor1, fill=mycolor1, forget plot] table[row sep=crcr]{%
x	y\\
-169.644677508421	1\\
};
\addplot[only marks, mark=square*, mark options={}, mark size=1.0607pt, draw=mycolor1, fill=mycolor1, forget plot] table[row sep=crcr]{%
x	y\\
-162.729518051451	1\\
};
\addplot[only marks, mark=diamond*, mark options={}, mark size=2.5883pt, draw=mycolor1, fill=mycolor1, forget plot] table[row sep=crcr]{%
x	y\\
-159.572464045474	1\\
};
\addplot[only marks, mark=x, mark options={}, mark size=1.5000pt, draw=mycolor1, fill=mycolor1, forget plot] table[row sep=crcr]{%
x	y\\
-163.904693611562	1\\
};
\addplot[only marks, mark=*, mark options={}, mark size=1.5000pt, draw=mycolor2, fill=mycolor2, forget plot] table[row sep=crcr]{%
x	y\\
-174.098719348618	1\\
};
\addplot[only marks, mark=square*, mark options={}, mark size=1.0607pt, draw=mycolor2, fill=mycolor2, forget plot] table[row sep=crcr]{%
x	y\\
-167.29859245456	1\\
};
\addplot[only marks, mark=diamond*, mark options={}, mark size=2.5883pt, draw=mycolor2, fill=mycolor2, forget plot] table[row sep=crcr]{%
x	y\\
-164.095790532433	1\\
};
\addplot[only marks, mark=x, mark options={}, mark size=1.5000pt, draw=mycolor2, fill=mycolor2, forget plot] table[row sep=crcr]{%
x	y\\
-178.501611595872	1\\
};
\addplot[only marks, mark=*, mark options={}, mark size=1.5000pt, draw=mycolor3, fill=mycolor3, forget plot] table[row sep=crcr]{%
x	y\\
-208.196856031388	1\\
};
\addplot[only marks, mark=square*, mark options={}, mark size=1.0607pt, draw=mycolor3, fill=mycolor3, forget plot] table[row sep=crcr]{%
x	y\\
-202.85700071039	1\\
};
\addplot[only marks, mark=diamond*, mark options={}, mark size=2.5883pt, draw=mycolor3, fill=mycolor3, forget plot] table[row sep=crcr]{%
x	y\\
-200.304890626005	1\\
};
\addplot[only marks, mark=x, mark options={}, mark size=1.5000pt, draw=mycolor3, fill=mycolor3, forget plot] table[row sep=crcr]{%
x	y\\
-205.789849763953	1\\
};
\addplot [color=black, dotted, line width=1.5pt, forget plot]
  table[row sep=crcr]{%
-163	-0.01\\
-163	1.09999999999999\\
};
\end{axis}
\end{tikzpicture}%
%

        \caption{Mean}
        \label{fig:mean_idbw}
    \end{subfigure}
    \hfill
    \begin{subfigure}[t]{.48\columnwidth}
        \setlength\fwidth{.85\columnwidth}
        \centering
  \tikzsetnextfilename{imgs/results/duty_cycle/p95_idbw}%
%
%
\definecolor{mycolor1}{rgb}{0.00000,0.44700,0.74100}%
\definecolor{mycolor2}{rgb}{0.85000,0.32500,0.09800}%
\definecolor{mycolor3}{rgb}{0.92900,0.69400,0.12500}%
\begin{tikzpicture}
\pgfplotsset{every tick label/.append style={font=\scriptsize}}

\begin{axis}[%
width=0.951\fwidth,
height=\fheight,
at={(0\fwidth,0\fheight)},
scale only axis,
xmin=-220,
xmax=-150,xlabel style={font=\footnotesize\color{white!15!black}},
xlabel={$I_{95}$ [dBW]},
ymin=0,
ymax=1,
ylabel style={font=\color{white!15!black}},
axis background/.style={fill=white},
xmajorgrids,
ymajorgrids
]
\addplot[only marks, mark=*, mark options={}, mark size=1.5000pt, draw=mycolor1, fill=mycolor1, forget plot] table[row sep=crcr]{%
x	y\\
-169.643611793722	0.2\\
};
\addplot[only marks, mark=square*, mark options={}, mark size=1.0607pt, draw=mycolor1, fill=mycolor1, forget plot] table[row sep=crcr]{%
x	y\\
-161.77766072629	0.2\\
};
\addplot[only marks, mark=diamond*, mark options={}, mark size=2.5883pt, draw=mycolor1, fill=mycolor1, forget plot] table[row sep=crcr]{%
x	y\\
-159.667583189346	0.2\\
};
\addplot[only marks, mark=x, mark options={}, mark size=1.5000pt, draw=mycolor1, fill=mycolor1, forget plot] table[row sep=crcr]{%
x	y\\
-162.703185073697	0.2\\
};
\addplot[only marks, mark=*, mark options={}, mark size=1.5000pt, draw=mycolor2, fill=mycolor2, forget plot] table[row sep=crcr]{%
x	y\\
-174.594133320529	0.2\\
};
\addplot[only marks, mark=square*, mark options={}, mark size=1.0607pt, draw=mycolor2, fill=mycolor2, forget plot] table[row sep=crcr]{%
x	y\\
-166.656820603866	0.2\\
};
\addplot[only marks, mark=diamond*, mark options={}, mark size=2.5883pt, draw=mycolor2, fill=mycolor2, forget plot] table[row sep=crcr]{%
x	y\\
-164.392469484169	0.2\\
};
\addplot[only marks, mark=x, mark options={}, mark size=1.5000pt, draw=mycolor2, fill=mycolor2, forget plot] table[row sep=crcr]{%
x	y\\
-167.476895264141	0.2\\
};
\addplot[only marks, mark=*, mark options={}, mark size=1.5000pt, draw=mycolor3, fill=mycolor3, forget plot] table[row sep=crcr]{%
x	y\\
-208.028498417037	0.2\\
};
\addplot[only marks, mark=square*, mark options={}, mark size=1.0607pt, draw=mycolor3, fill=mycolor3, forget plot] table[row sep=crcr]{%
x	y\\
-202.00879672406	0.2\\
};
\addplot[only marks, mark=diamond*, mark options={}, mark size=2.5883pt, draw=mycolor3, fill=mycolor3, forget plot] table[row sep=crcr]{%
x	y\\
-200.209061488038	0.2\\
};
\addplot[only marks, mark=x, mark options={}, mark size=1.5000pt, draw=mycolor3, fill=mycolor3, forget plot] table[row sep=crcr]{%
x	y\\
-206.11711702998	0.2\\
};
\addplot[only marks, mark=*, mark options={}, mark size=1.5000pt, draw=mycolor1, fill=mycolor1, forget plot] table[row sep=crcr]{%
x	y\\
-164.215398862424	0.5\\
};
\addplot[only marks, mark=square*, mark options={}, mark size=1.0607pt, draw=mycolor1, fill=mycolor1, forget plot] table[row sep=crcr]{%
x	y\\
-160.582948414373	0.5\\
};
\addplot[only marks, mark=diamond*, mark options={}, mark size=2.5883pt, draw=mycolor1, fill=mycolor1, forget plot] table[row sep=crcr]{%
x	y\\
-157.878444099194	0.5\\
};
\addplot[only marks, mark=x, mark options={}, mark size=1.5000pt, draw=mycolor1, fill=mycolor1, forget plot] table[row sep=crcr]{%
x	y\\
-161.348911375738	0.5\\
};
\addplot[only marks, mark=*, mark options={}, mark size=1.5000pt, draw=mycolor2, fill=mycolor2, forget plot] table[row sep=crcr]{%
x	y\\
-168.321792303188	0.5\\
};
\addplot[only marks, mark=square*, mark options={}, mark size=1.0607pt, draw=mycolor2, fill=mycolor2, forget plot] table[row sep=crcr]{%
x	y\\
-165.232908809099	0.5\\
};
\addplot[only marks, mark=diamond*, mark options={}, mark size=2.5883pt, draw=mycolor2, fill=mycolor2, forget plot] table[row sep=crcr]{%
x	y\\
-162.577619422821	0.5\\
};
\addplot[only marks, mark=x, mark options={}, mark size=1.5000pt, draw=mycolor2, fill=mycolor2, forget plot] table[row sep=crcr]{%
x	y\\
-166.216220557084	0.5\\
};
\addplot[only marks, mark=*, mark options={}, mark size=1.5000pt, draw=mycolor3, fill=mycolor3, forget plot] table[row sep=crcr]{%
x	y\\
-205.151421292712	0.5\\
};
\addplot[only marks, mark=square*, mark options={}, mark size=1.0607pt, draw=mycolor3, fill=mycolor3, forget plot] table[row sep=crcr]{%
x	y\\
-200.552101078501	0.5\\
};
\addplot[only marks, mark=diamond*, mark options={}, mark size=2.5883pt, draw=mycolor3, fill=mycolor3, forget plot] table[row sep=crcr]{%
x	y\\
-198.570132243466	0.5\\
};
\addplot[only marks, mark=x, mark options={}, mark size=1.5000pt, draw=mycolor3, fill=mycolor3, forget plot] table[row sep=crcr]{%
x	y\\
-203.400755205132	0.5\\
};
\addplot[only marks, mark=*, mark options={}, mark size=1.5000pt, draw=mycolor1, fill=mycolor1, forget plot] table[row sep=crcr]{%
x	y\\
-163.572178959773	1\\
};
\addplot[only marks, mark=square*, mark options={}, mark size=1.0607pt, draw=mycolor1, fill=mycolor1, forget plot] table[row sep=crcr]{%
x	y\\
-159.306746762706	1\\
};
\addplot[only marks, mark=diamond*, mark options={}, mark size=2.5883pt, draw=mycolor1, fill=mycolor1, forget plot] table[row sep=crcr]{%
x	y\\
-156.511650198553	1\\
};
\addplot[only marks, mark=x, mark options={}, mark size=1.5000pt, draw=mycolor1, fill=mycolor1, forget plot] table[row sep=crcr]{%
x	y\\
-160.231452999169	1\\
};
\addplot[only marks, mark=*, mark options={}, mark size=1.5000pt, draw=mycolor2, fill=mycolor2, forget plot] table[row sep=crcr]{%
x	y\\
-168.433633895574	1\\
};
\addplot[only marks, mark=square*, mark options={}, mark size=1.0607pt, draw=mycolor2, fill=mycolor2, forget plot] table[row sep=crcr]{%
x	y\\
-163.959073060817	1\\
};
\addplot[only marks, mark=diamond*, mark options={}, mark size=2.5883pt, draw=mycolor2, fill=mycolor2, forget plot] table[row sep=crcr]{%
x	y\\
-161.074240567779	1\\
};
\addplot[only marks, mark=x, mark options={}, mark size=1.5000pt, draw=mycolor2, fill=mycolor2, forget plot] table[row sep=crcr]{%
x	y\\
-172.675976245644	1\\
};
\addplot[only marks, mark=*, mark options={}, mark size=1.5000pt, draw=mycolor3, fill=mycolor3, forget plot] table[row sep=crcr]{%
x	y\\
-204.236116539197	1\\
};
\addplot[only marks, mark=square*, mark options={}, mark size=1.0607pt, draw=mycolor3, fill=mycolor3, forget plot] table[row sep=crcr]{%
x	y\\
-199.356247592859	1\\
};
\addplot[only marks, mark=diamond*, mark options={}, mark size=2.5883pt, draw=mycolor3, fill=mycolor3, forget plot] table[row sep=crcr]{%
x	y\\
-197.164294292541	1\\
};
\addplot[only marks, mark=x, mark options={}, mark size=1.5000pt, draw=mycolor3, fill=mycolor3, forget plot] table[row sep=crcr]{%
x	y\\
-201.617330516522	1\\
};
\addplot [color=black, dotted, line width=1.5pt, forget plot]
  table[row sep=crcr]{%
-163	-0.01\\
-163	1.09999999999999\\
};
\end{axis}
\end{tikzpicture}%
%

        \caption{$95^{\text{th}}$ percentile}
        \label{fig:p95_idbw}
    \end{subfigure}
    \setlength\abovecaptionskip{0cm}    \setlength\belowcaptionskip{-.6cm}
    \caption{Aggregated interference with different network load factors.}
    \label{fig:duty_cycle}
\end{figure}
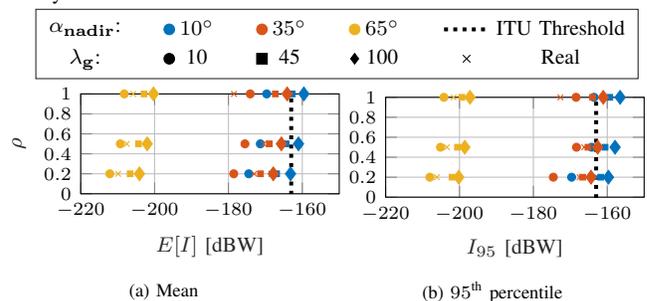

The results presented so far in this section were obtained considering $f_c=178$~GHz, where the propagation is characterized by a strong absorption by the atmospheric gases.
Here, we compare the \gls{rfi} observed by TEMPEST at $164$~GHz, which is characterized by a lower absorption, at 178~GHz, and at $240$~GHz, that is used by Aura.
The results for the latter are analyzed in depth in the next paragraph.
\cref{fig:freqs_cellular} reports the aggregated \gls{rfi} for the three frequencies.
Specifically, at 240~GHz it is well below the ITU threshold ($-194$~dBW for limb scanners), even when considering a high \gls{gnb} density ($\lambda_g=100$~\glspl{gnb}/km\textsuperscript{2}) and  $\alpha_{n}=\ang{10}$.
Conversely, for 164~GHz, even when the \gls{rfi} is the weakest ($\alpha_{n}=\ang{65}$ and $\lambda_g=10$~\glspl{gnb}/km\textsuperscript{2}), it is above the threshold with a probability of 0.09.
We can thus conclude that transmitting at frequencies close to the absorption peaks can indeed help mitigate the \gls{rfi}, whereas transmitting in lower-absorption bands can almost certainly cause significant interference to conical-scan satellites.

\paragraph{MLS}
As reported in \cref{fig:freqs}, the \gls{rfi} at the satellite antenna at $240$~GHz can be higher than at $178$~GHz.
The large difference in interfering power observed in~\cref{fig:freqs_cellular} is due to the different altitudes (705~km Aura vs 400~km TEMPEST) and scan modes of the satellites.
While the TEMPEST satellite has a conical scanning sensor, the Aura satellite observes the limb layers of the atmosphere.
Thus, the beam of TEMPEST amplifies the interference coming from the ground with maximum gain, whereas the \gls{mls} strongly attenuates it.

\setcounter{figure}{13}
\begin{figure}[b]
    \centering
    \begin{minipage}[t]{.95\columnwidth}
        \setlength\fheight{0.2\columnwidth}
        \centering
        \begin{minipage}[t]{\columnwidth}
            \centering
            \setlength\fwidth{\columnwidth}
  \tikzsetnextfilename{imgs/results/I_dBW_178/backhaul_legend.tex}%
  \begin{tikzpicture}

    \definecolor{mycolor1}{rgb}{0.00000,0.44700,0.74100}%
    \definecolor{mycolor2}{rgb}{0.85000,0.32500,0.09800}%
    \definecolor{mycolor3}{rgb}{0.92900,0.69400,0.12500}%
    \definecolor{mycolor4}{rgb}{0.49400,0.18400,0.55600}%
    \begin{axis}[%
            width=0,
            height=0,
            at={(0,0)},
            xmin=0,
            xmax=0,
            xtick={},
            ymin=0,
            ymax=0,
            ytick={},
            scale only axis,
            axis background/.style={fill=white},
            legend style={legend cell align=center, align=center, draw=white!15!black,at={(0,0)},anchor=south west, /tikz/every even column/.append style={column sep = 0.4cm},font=\scriptsize},
            legend columns = 5
        ]
        \addlegendimage{empty legend}
        \addlegendentry{$\lambda_g$:}
        
        \addlegendimage{scatter,only marks,color=mycolor1}
        \addlegendentry{10}

        \addlegendimage{scatter,only marks,color=mycolor2}
        \addlegendentry{45}

        \addlegendimage{scatter,only marks,color=mycolor3}
        \addlegendentry{100}

        \addlegendimage{scatter,only marks,color=mycolor4}
        \addlegendentry{Real}

        \addlegendimage{empty legend}
        \addlegendentry{$\alpha_{nadir}$:}
        
        \addlegendimage{draw=black,fill=black}
        \addlegendentry{\ang{10}}

        \addlegendimage{draw=black,fill=black,dashed}
        \addlegendentry{\ang{35}}

        \addlegendimage{draw=black,fill=black,dashdotted}
        \addlegendentry{\ang{65}}
        
        \addlegendimage{draw=black,fill=black,dotted}
        \addlegendentry{ITU Thr.}
    \end{axis}
\end{tikzpicture}%
%

        \end{minipage}
        \begin{minipage}[t]{\columnwidth}
            \setlength\fwidth{.8\columnwidth}
            \centering
  \tikzsetnextfilename{imgs/results/I_dBW_178/aggrIdbWECDF_backhaul}%
  \input{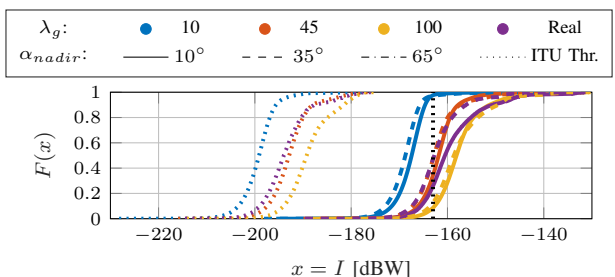}%

        \end{minipage}
        \setlength\belowcaptionskip{-.3cm}
        \setlength\abovecaptionskip{-.4cm}
        \captionof{figure}{Distribution of the aggregated \gls{rfi} in the Backhaul scenarios.}
        \label{fig:I_dbW_backhaul}
    \end{minipage}
\end{figure}

\setcounter{figure}{12}
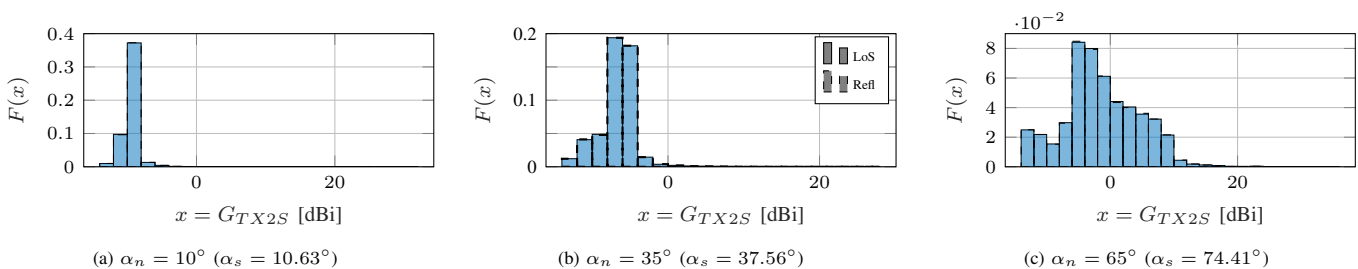
\begin{figure*}[b]
    \def\tw{.85}
    \def\cw{.65}
    \setlength\fheight{0.20\columnwidth}
    \begin{subfigure}[t]{\cw\columnwidth}
        \setlength\fwidth{\tw\columnwidth}
        \centering
  \tikzsetnextfilename{imgs/results/gtx_backhaul/hist_gtx_nadir10}%
%
%
\definecolor{mycolor1}{rgb}{0.00000,0.44700,0.74100}%
\begin{tikzpicture}
\pgfplotsset{every tick label/.append style={font=\scriptsize}}

\begin{axis}[%
width=0.951\fwidth,
height=\fheight,
at={(0\fwidth,0\fheight)},
scale only axis,
xmin=-16.3,
xmax=34.3,
ymin=0,
ymax=0.4,
      xlabel style={font=\footnotesize\color{white!15!black}},
      xlabel={$x=G_{TX2S}$ [dBi]},
      ylabel style={font=\footnotesize\color{white!15!black}},
      ylabel={$F(x)$},
axis background/.style={fill=white},
xmajorgrids,
ymajorgrids,
]
\addplot[ybar interval, fill=mycolor1, fill opacity=0.3, draw=black, area legend] table[row sep=crcr] {%
x	y\\
-14	0.00978173748729918\\
-12	0.0974145118058227\\
-10	0.372059463584857\\
-8	0.0130933370318321\\
-6	0.00416732437622315\\
-4	0.00121395304643404\\
-2	0.000777374382583136\\
0	0.000358611346524305\\
2	0.000392326943206077\\
4	0.000145206944799906\\
6	0.000123943130983561\\
8	0.000119345549617864\\
10	0.000139651533983023\\
12	9.31010226553484e-05\\
14	1.55168371092247e-05\\
16	9.96142629234181e-06\\
18	4.78914725593356e-05\\
20	2.33710386089558e-05\\
22	1.49421394385127e-05\\
24	0\\
26	4.9807131461709e-06\\
28	1.14939534142405e-06\\
30	2.29879068284811e-06\\
32	2.29879068284811e-06\\
};
\addplot[ybar interval, fill=mycolor1, fill opacity=0.3, dashed, draw=black, area legend] table[row sep=crcr] {%
x	y\\
-14	0.00995893593576872\\
-12	0.0962944260456049\\
-10	0.372759445347784\\
-8	0.0132477391393634\\
-6	0.00411713411298096\\
-4	0.00129651794512633\\
-2	0.000744999747133025\\
0	0.000398457051693672\\
2	0.000370296865828783\\
4	0.000136394913848988\\
6	0.000158041859445808\\
8	0.000113598572910744\\
10	0.00014559007658038\\
12	0.000119920247288576\\
14	2.54782634015665e-05\\
16	1.30264805361393e-05\\
18	4.38685888643514e-05\\
20	3.92710074986552e-05\\
22	9.19516273139244e-06\\
24	0\\
26	4.59758136569622e-06\\
28	1.91565890237342e-06\\
30	1.14939534142405e-06\\
32	1.14939534142405e-06\\
};
\end{axis}
\end{tikzpicture}%
%

        \caption{$\alpha_{n}=\ang{10}$ ($\alpha_{s} = \ang{10.63}$)}
    \end{subfigure}
    \hfill
    \begin{subfigure}[t]{\cw\columnwidth}
        \setlength\fwidth{\tw\columnwidth}
        \centering
  \tikzsetnextfilename{imgs/results/gtx_backhaul/hist_gtx_nadir35}%
%
%
\definecolor{mycolor1}{rgb}{0.00000,0.44700,0.74100}%
\begin{tikzpicture}
\pgfplotsset{every tick label/.append style={font=\scriptsize}}

\begin{axis}[%
width=0.951\fwidth,
height=\fheight,
at={(0\fwidth,0\fheight)},
scale only axis,
xmin=-16.1,
xmax=30.1,
ymin=0,
ymax=0.2,
      xlabel style={font=\footnotesize\color{white!15!black}},
      xlabel={$x=G_{TX2S}$ [dBi]},
      ylabel style={font=\footnotesize\color{white!15!black}},
      ylabel={$F(x)$},
axis background/.style={fill=white},
xmajorgrids,
ymajorgrids,
y tick label style={
    /pgf/number format/.cd,
        fixed,
        fixed zerofill,
        precision=1,
    /tikz/.cd,
    legend style={font=\tiny},
},
yticklabels = {,0,,0.1,,0.2}
]
\addplot[ybar interval, fill=mycolor1, fill opacity=0.3, draw=black, area legend, forget plot] table[row sep=crcr] {%
x	y\\
-14	0.0120958534413663\\
-12	0.0407090926366669\\
-10	0.0487858937008538\\
-8	0.193817555711192\\
-6	0.181673810797267\\
-4	0.0144216549147379\\
-2	0.00370967346444614\\
0	0.00209802962987937\\
2	0.000976602908429972\\
4	0.00058101934508986\\
6	0.00041014257099815\\
8	0.000219917641992469\\
10	0.000168961115189336\\
12	8.10323715703959e-05\\
14	0.000105361239630538\\
16	2.75854881941773e-05\\
18	2.35626044991931e-05\\
20	4.40601547545888e-06\\
22	5.13396585836078e-05\\
24	3.25662013403482e-06\\
26	3.5248123803671e-05\\
28	3.5248123803671e-05\\
};
\addplot[ybar interval, fill=mycolor1, fill opacity=0.3, dashed, draw=black, area legend, thick] table[row sep=crcr, forget plot] {%
x	y\\
-14	0.0120590727904407\\
-12	0.0411661688507732\\
-10	0.0478815111330433\\
-8	0.193792652145461\\
-6	0.181844687571358\\
-4	0.0144647572400413\\
-2	0.00392537665685338\\
0	0.00203519601788153\\
2	0.000934649978467994\\
4	0.000728525080572613\\
6	0.000480255686825018\\
8	0.000175665921347643\\
10	0.000177006882579304\\
12	7.73926196558864e-05\\
14	0.000112066045788845\\
16	2.98842788770254e-05\\
18	2.95011470965507e-05\\
20	2.68192246332279e-06\\
22	6.16842166564243e-05\\
24	5.74697670712027e-06\\
26	1.55168371092247e-05\\
28	1.55168371092247e-05\\
};
\addlegendimage{ybar,ybar legend,draw=black,fill=black,fill opacity=0.5}
\addlegendentry{LoS}

\addlegendimage{ybar,ybar legend,draw=black,fill=black,fill opacity=0.5,dashed}
\addlegendentry{Refl}
\end{axis}
\end{tikzpicture}%
%

        \caption{$\alpha_{n}=\ang{35}$ ($\alpha_{s} = \ang{37.56}$)}
    \end{subfigure}
    \hfill
    \begin{subfigure}[t]{\cw\columnwidth}
        \setlength\fwidth{\tw\columnwidth}
        \centering
  \tikzsetnextfilename{imgs/results/gtx_backhaul/hist_gtx_nadir65}%
%
%
\definecolor{mycolor1}{rgb}{0.00000,0.44700,0.74100}%
\begin{tikzpicture}
\pgfplotsset{every tick label/.append style={font=\scriptsize}}

\begin{axis}[%
width=0.951\fwidth,
height=\fheight,
at={(0\fwidth,0\fheight)},
scale only axis,
xmin=-16.5,
xmax=38.5,
ymin=0,
ymax=0.09,
      xlabel style={font=\footnotesize\color{white!15!black}},
      xlabel={$x=G_{TX2S}$ [dBi]},
      ylabel style={font=\footnotesize\color{white!15!black}},
      ylabel={$F(x)$},
axis background/.style={fill=white},
xmajorgrids,
ymajorgrids,
]
\addplot[ybar interval, fill=mycolor1, fill opacity=0.3, draw=black, area legend] table[row sep=crcr] {%
x	y\\
-14	0.0249309596531585\\
-12	0.0218356379987035\\
-10	0.0154003650479604\\
-8	0.0297860055753337\\
-6	0.0840916788374862\\
-4	0.0799042400427882\\
-2	0.0611271430476141\\
0	0.0440827595296368\\
2	0.0404148474292624\\
4	0.0355142088252107\\
6	0.0322001189241047\\
8	0.0214231966370225\\
10	0.00438398539808158\\
12	0.00181259645342573\\
14	0.00123559999203086\\
16	0.000805343002557788\\
18	0.000314551191769716\\
20	0.0002436718123819\\
22	0.000379683594450413\\
24	3.63975191450951e-05\\
26	1.26433487556646e-05\\
28	1.78156277920729e-05\\
30	3.02674106575001e-05\\
32	1.53252712189874e-06\\
34	1.47505735482754e-05\\
36	1.47505735482754e-05\\
};
\addplot[ybar interval, fill=mycolor1, fill opacity=0.3, dashed, draw=black, area legend] table[row sep=crcr] {%
x	y\\
-14	0.0250188883967774\\
-12	0.0217996236113389\\
-10	0.0152622460410993\\
-8	0.029459385732479\\
-6	0.0848361038869485\\
-4	0.0793655567594408\\
-2	0.0613711979917765\\
0	0.0434467607740488\\
2	0.0400495312765798\\
4	0.0361383304956039\\
6	0.0322629525361025\\
8	0.0215975215971385\\
10	0.00455639469929519\\
12	0.00187255657707002\\
14	0.00119498802330054\\
16	0.000750746723840145\\
18	0.000233710386089558\\
20	0.000208423688578229\\
22	0.000345776431878403\\
24	6.53239685709338e-05\\
26	1.93481549139716e-05\\
28	2.60529610722786e-05\\
30	8.04576738996838e-05\\
32	2.26047750480064e-05\\
34	1.55168371092247e-05\\
36	1.55168371092247e-05\\
};
\end{axis}
\end{tikzpicture}%
%

        \caption{$\alpha_{n}=\ang{65}$ ($\alpha_{s} = \ang{74.41}$)}
    \end{subfigure}\\
    \setlength\abovecaptionskip{0cm}
    \caption{Estimated \gls{pdf} of the $G_{TX2S}$ of the interfering rays for different satellite nadir angles in the Backhaul scenario.}
    \label{fig:gtx_backhaul}
\end{figure*}

\paragraph{Network Load Factor}

In \cref{fig:duty_cycle} we show the total \gls{rfi} for different network load factors.
We define the network load factor through the activation probability $\rho$, which determines whether a \gls{gnb} sector is active, i.e., on average a fraction $\rho$\ of the possible communication links is active.
For instance, the aggregated interference reported in \cref{fig:Idbw} assumes that all the \gls{gnb} sectors are active (either transmitting or receiving), corresponding to $\rho=1$.
\cref{fig:duty_cycle} shows the mean and the 95\textsuperscript{th} percentile of the aggregated \gls{rfi} for $\rho\in\{0.2,0.5,1\}$.
A load factor of 0.2 or 0.5 effectively decreases the average interference $E[I]$ below the ITU threshold for all the considered scenarios.
Conversely, the 95\textsuperscript{th} percentile $I_{95}$ is less affected, remaining above the threshold when considering high density ($\lambda_g=45,100$~\glspl{gnb}/km\textsuperscript{2}) and small nadir angles ($\alpha_{n}$=\ang{10}). Note that \gls{sthz} networks with ultra-wide bandwidths are more likely to operate in scenarios with a small $\rho$, as transmissions can leverage high data rates and thus occupy the channel for reduced periods of time~\cite{akyildiz2022terahertz}.

\begin{table}[b]
    \centering
    \scriptsize
    \[
        \begin{tabular}{c|ccc}
            \multicolumn{1}{c}{} & \multicolumn{3}{c}{\text{$\alpha_{n}$ [deg]}}              \\
                                 & 10                                            & 35    & 65 \\\hline
            10                   & 0.023                                         & 0.028 & 0  \\
            45                   & 0.655                                         & 0.527 & 0  \\
            Real                 & 0.74                                          & 0.527 & 0  \\
            100                  & 0.9138                                        & 0.872 & 0
        \end{tabular}
    \]
    \setlength\belowcaptionskip{-.1cm}
    \setlength\abovecaptionskip{-.3cm}
    \caption{Probability that the \gls{rfi} is greater than the ITU threshold for the Backhaul scenario.}
    \label{tab:backhaul}
\end{table}

\subsection{Backhaul}
\label{ssec:res_backahul}
\paragraph{Interfering Nodes Gain}
For the backhaul scenario, links are established between \glspl{gnb}, i.e., between nodes at similar heights.
Thus, the transmitting beams are generally more aligned to the horizontal axis than those of their cellular counterpart.
This behavior can be clearly observed in \cref{fig:gtx_backhaul}, where the histogram for the transmitter gain is reported, considering  the TEMPEST satellite.
We can observe how, due to the horizontal orientation, the beamforming effectively suppresses the interference, particularly when the satellite is above the network area ($\alpha_{BF}\in\{\ang{10},\ang{35}\}$).
As it lowers toward the horizon, larger gains become more probable.
Furthermore, for the same reason, the gain of the \gls{los} and reflected rays present almost the same distribution.


\paragraph{Aggregated Power}
\cref{fig:I_dbW_backhaul} reports the \gls{ecdf} of the aggregated \gls{rfi} in the considered scenarios, for the TEMPEST satellite at 178~GHz.
We can observe that the aggregated power here is much greater than in the cellular scenario.
This is mainly due to the increase in the transmitting power, which shifts all the distributions by about 20~dB.

The probability of exceeding the threshold in the different scenarios is reported in \cref{tab:backhaul}.
Comparing it with the results reported in \cref{tab:P_th}, we can observe how the backhaul scenario generates much greater interference than the Urban Cellular one.

In conclusion, although the narrow beams suppress the power leaking toward the satellite, a backhaul network with the considered frequency can be potentially harmful to the incumbent satellites.

\section{Conclusions and Future Works}

In this paper, we introduced analytical and simulation methodologies for the evaluation of \gls{rfi} that next-generation terrestrial networks may introduce in passive sensing satellite systems.
We developed a single-link analysis that shows the effect of beam amplification through the combined effect of the problem geometry, of the beam of the terrestrial \gls{tx} and of the satellite.
We then extended this into a large-scale data-driven simulation which relies on topologies for networks and buildings based on real-world data.
The results show that---despite the high propagation and absorption loss at \gls{subthz} frequencies---it is possible to generate \gls{rfi} above ITU thresholds with specific network and satellite configurations.

These insights provide a foundation for our future work, which will focus on developing coexistence methods in a realistic data-driven framework. In addition, we will further extend our analysis by considering models for how \gls{rfi} propagates into the passive sensors measurements.

\bibliographystyle{IEEEtran}
\bibliography{bibl}

\begin{thebibliography}{10}
\providecommand{\url}[1]{#1}
\csname url@samestyle\endcsname
\providecommand{\newblock}{\relax}
\providecommand{\bibinfo}[2]{#2}
\providecommand{\BIBentrySTDinterwordspacing}{\spaceskip=0pt\relax}
\providecommand{\BIBentryALTinterwordstretchfactor}{4}
\providecommand{\BIBentryALTinterwordspacing}{\spaceskip=\fontdimen2\font plus
\BIBentryALTinterwordstretchfactor\fontdimen3\font minus
  \fontdimen4\font\relax}
\providecommand{\BIBforeignlanguage}[2]{{%
\expandafter\ifx\csname l@#1\endcsname\relax
\typeout{** WARNING: IEEEtran.bst: No hyphenation pattern has been}%
\typeout{** loaded for the language `#1'. Using the pattern for}%
\typeout{** the default language instead.}%
\else
\language=\csname l@#1\endcsname
\fi
#2}}
\providecommand{\BIBdecl}{\relax}
\BIBdecl

\bibitem{akyildiz2022terahertz}
I.~F. Akyildiz, C.~Han, Z.~Hu, S.~Nie, and J.~M. Jornet, ``Terahertz band
  communication: {A}n old problem revisited and research directions for the
  next decade,'' \emph{IEEE Trans. on Communications}, vol.~70, no.~6, pp.
  4250--4285, Jun. 2022.

\bibitem{itu-p-676}
ITU, ``Attenuation by atmospheric gases and related effects,'' Rec. ITU-R
  P.676-13, 2022.

\bibitem{petrov2020ieee}
V.~Petrov, T.~Kurner, and I.~Hosako, ``{IEEE 802.15.3d: First Standardization
  Efforts for Sub-Terahertz Band Communications toward 6G},'' \emph{IEEE
  Communications Magazine}, vol.~58, no.~11, pp. 28--33, Nov. 2020.

\bibitem{polese2021coexistence}
M.~Polese, X.~Cantos-Roman, A.~Singh, M.~J. Marcus, T.~J. Maccarone,
  T.~Melodia, and J.~M. Jornet, ``{Coexistence and spectrum sharing above 100
  GHz},'' \emph{Proc. of the IEEE (to appear)}, 2023.

\bibitem{kummerow2022hyperspectral}
{D. C. Kummerow et al.}, ``Hyperspectral microwave sensors—advantages and
  limitations,'' \emph{IEEE Journal of Selected Topics in Applied Earth
  Observations and Remote Sensing}, vol.~15, pp. 764--775, Dec. 2022.

\bibitem{polese2022dynamic}
{M. Polese et al.}, ``{Dynamic spectrum sharing between active and passive
  users above 100 GHz},'' \emph{Communications Engineering}, vol.~1, no.~1,
  p.~6, May 2022.

\bibitem{xing2021terahertz}
Y.~Xing and T.~S. Rappaport, ``{Terahertz Wireless Communications: Co-Sharing
  for Terrestrial and Satellite Systems Above 100 GHz},'' \emph{IEEE
  Communications Letters}, vol.~25, no.~10, pp. 3156--3160, Oct. 2021.

\bibitem{reising2015overview}
{S. Reising et al.}, ``{Overview of temporal experiment for storms and tropical
  systems (TEMPEST) CubeSat constellation mission},'' in \emph{IEEE MTT-S
  International Microwave Symposium}.\hskip 1em plus 0.5em minus 0.4em\relax
  IEEE, 2015.

\bibitem{waters2006earth}
{J.W. Waters et al.}, ``{The Earth observing system microwave limb sounder (EOS
  MLS) on the Aura Satellite},'' \emph{IEEE Trans. on Geoscience and Remote
  Sensing}, vol.~44, no.~5, pp. 1075--1092, May 2006.

\bibitem{park2019modeling}
J.~Park, E.~Lee, S.-H. Park, S.-S. Raymond, S.~Pyo, and H.-S. Jo, ``{Modeling
  and Analysis on Radio Interference of OFDM Waveforms for Coexistence
  Study},'' \emph{IEEE Access}, vol.~7, pp. 35\,132--35\,147, 2019.

\bibitem{marcus2014harmful}
M.~J. Marcus, ``Harmful interference and its role in spectrum policy,''
  \emph{Proc. of the IEEE}, vol. 102, no.~3, pp. 265--269, Feb. 2014.

\bibitem{bosso2021ultrabroadband}
{C. Bosso et al.}, ``Ultrabroadband spread spectrum techniques for secure
  dynamic spectrum sharing above 100 ghz between active and passive users,'' in
  \emph{2021 IEEE International Symposium on Dynamic Spectrum Access Networks
  (DySPAN)}, 2021, pp. 45--52.

\bibitem{guidolin2015study}
F.~Guidolin, M.~Nekovee, L.~Badia, and M.~Zorzi, ``{A study on the coexistence
  of fixed satellite service and cellular networks in a mmWave scenario},'' in
  \emph{IEEE International Conference on Communications (ICC)}, 2015, pp.
  2444--2449.

\bibitem{su2014coexistence}
C.~Su, X.~Han, X.~Yan, Q.~Zhang, and Z.~Feng, ``{Coexistence Analysis between
  IMT-Advanced System and Fixed Satellite Service System},'' in \emph{IEEE
  Military Communications Conference}, 2014, pp. 1692--1697.

\bibitem{gasiewski2002impacts}
A.~J. Gasiewski, C.~S. Ruf, M.~Younis, and W.~Wesbeck, ``{Impacts of mobile
  radar and telecommunications systems on Earth remote sensing in the 22-27 GHz
  range},'' in \emph{IEEE International Geoscience and Remote Sensing
  Symposium}, vol.~3.\hskip 1em plus 0.5em minus 0.4em\relax IEEE, 2002, pp.
  1679--1681.

\bibitem{hassan2017feasibility}
W.~A. Hassan, H.-S. Jo, and A.~R. Tharek, ``{The Feasibility of Coexistence
  Between 5G and Existing Services in the IMT-2020 Candidate Bands in
  Malaysia},'' \emph{IEEE Access}, vol.~5, pp. 14\,867--14\,888, 2017.

\bibitem{hattab2018interference}
{Ghaith Hattab et al.}, ``{Interference Analysis of the Coexistence of 5G
  Cellular Networks with Satellite Earth Stations in 3.7-4.2GHz},'' in
  \emph{IEEE Int. Conference on Communications (ICC) Workshops}, 2018.

\bibitem{alkhateeb2013hybrid}
A.~Alkhateeb, O.~El~Ayach, G.~Leus, and R.~W. Heath, ``Hybrid precoding for
  millimeter wave cellular systems with partial channel knowledge,'' in
  \emph{Information Theory and Applications Workshop (ITA)}.\hskip 1em plus
  0.5em minus 0.4em\relax IEEE, 2013.

\bibitem{zhong2020feasibility}
{Liyuan Zhong et al.}, ``{The Feasibility of Coexistence between IMT-2020 and
  Inter-Satellite Service in 26 GHz band},'' in \emph{Int. Wireless
  Communications and Mobile Computing (IWCMC)}, 2020, pp. 1006--1011.

\bibitem{winter2019statistics}
S.~P. Winter and A.~Knopp, ``{Statistics of Terrestrial Fixed Service
  Interference in the Aeronautical SATCOM Channel},'' in \emph{IEEE
  International Conference on Communications (ICC)}, 2019.

\bibitem{cho2018spectral}
Y.~Cho, H.~Kim, E.~E. Ahiagbe, and H.-S. Jo, ``{Spectral Coexistence of
  IMT-2020 with Fixed-Satellite Service in the 27-27.5 GHz Band},'' in
  \emph{International Conference on Information and Communication Technology
  Convergence (ICTC)}, 2018.

\bibitem{cho2019modeling}
Y.~Cho, H.~Kim, D.~K. Tettey, K.-J. Lee, and H.-S. Jo, ``{Modeling Method for
  Interference Analysis between IMT-2020 and Satellite in the mmWave Band},''
  in \emph{IEEE Globecom Workshops}, 2019.

\bibitem{cho2020coexistence}
Y.~Cho, H.-K. Kim, M.~Nekovee, and H.-S. Jo, ``{Coexistence of 5G With
  Satellite Services in the Millimeter-Wave Band},'' \emph{IEEE Access},
  vol.~8, pp. 163\,618--163\,636, 2020.

\bibitem{ayoubi2023imt}
{R. Aghazadeh et al.}, ``{IMT to Satellite Stochastic Interference Modeling and
  Coexistence Analysis of Upper 6 GHz Band Service},'' \emph{IEEE Open Journal
  of the Communications Society}, May 2023.

\bibitem{jakes1994microwave}
W.~C. Jakes and D.~C. Cox, \emph{Microwave mobile communications}.\hskip 1em
  plus 0.5em minus 0.4em\relax Wiley-IEEE Press, 1994.

\bibitem{han2014multi}
C.~Han, A.~O. Bicen, and I.~F. Akyildiz, ``{Multi-ray channel modeling and
  wideband characterization for wireless communications in the terahertz
  band},'' \emph{IEEE Trans. on Wireless Communications}, vol.~14, no.~5, pp.
  2402--2412, Dec. 2014.

\bibitem{kotz2004experimental}
{D. Kotz et al.}, ``{Experimental Evaluation of Wireless Simulation
  Assumptions},'' in \emph{Proc. of the 7th ACM International Symposium on
  Modeling, Analysis and Simulation of Wireless and Mobile Systems}, Venice,
  Italy, 2004, p. 78–82.

\bibitem{fund2016how}
F.~Fund, R.~Lin, T.~Korakis, and S.~S. Panwar, ``{How bad is the Flat Earth
  assumption? Effect of topography on wireless systems},'' in \emph{14th
  International Symposium on Modeling and Optimization in Mobile, Ad Hoc, and
  Wireless Networks (WiOpt)}, 2016.

\bibitem{loyka2001using}
S.~Loyka and A.~Kouki, ``Using two ray multipath model for microwave link
  budget analysis,'' \emph{IEEE Antennas and Propagation Magazine}, vol.~43,
  no.~5, pp. 31--36, Oct. 2001.

\bibitem{feuerstein1994path}
{M.J. Feuerstein et al.}, ``{Path loss, delay spread, and outage models as
  functions of antenna height for microcellular system design},'' \emph{IEEE
  Trans. on Vehicular Technology}, vol.~43, no.~3, pp. 487--498, Aug. 1994.

\bibitem{dottling2001two}
M.~Dottling, A.~Jahn, D.~Didascalou, and W.~Wiesbeck, ``{Two- and
  three-dimensional ray tracing applied to the land mobile satellite (LMS)
  propagation channel},'' \emph{IEEE Antennas and Propagation Magazine},
  vol.~43, no.~6, pp. 27--37, Dec. 2001.

\bibitem{soler1994determination}
T.~Soler and D.~W. Eisemann, ``{Determination of look angles to geostationary
  communication satellites},'' \emph{Journal of surveying engineering}, vol.
  120, no.~3, pp. 115--127, Aug. 1994.

\bibitem{piesiewicz2007scattering}
R.~Piesiewicz, C.~Jansen, D.~Mittleman, T.~Kleine-Ostmann, M.~Koch, and
  T.~Kurner, ``{Scattering analysis for the modeling of THz communication
  systems},'' \emph{IEEE Trans. on Antennas and Propagation}, vol.~55, no.~11,
  pp. 3002--3009, Nov. 2007.

\bibitem{itu-r-2040}
ITU, ``{Effects of building materials and structures on radiowave propagation
  above about 100 MHz},'' Rec. ITU-R P.2040-2, 2021.

\bibitem{du2021sub}
K.~Du, O.~Ozdemir, F.~Erden, and I.~Guvenc, ``{Sub-Terahertz and mmWave
  Penetration Loss Measurements for Indoor Environments},'' in \emph{IEEE Int.
  Conference on Communications (ICC) Workshops}, 2021.

\bibitem{saggese2022efficient}
F.~Saggese, F.~Chiariotti, K.~Kansanen, and P.~Popovski, ``{Efficient URLLC
  with a Reconfigurable Intelligent Surface and Imperfect Device Tracking},''
  \emph{arXiv preprint arXiv:2211.09171}, 2022.

\bibitem{itu-r-2017}
ITU, ``Performance and interference criteria for satellite passive remote
  sensing,'' Rec. ITU-R RS.2017-0, 2012.

\bibitem{cofield2006design}
R.~Cofield and P.~Stek, ``{Design and field-of-view calibration of 114-660-GHz
  optics of the Earth observing system microwave limb sounder},'' \emph{IEEE
  Trans. on Geoscience and Remote Sensing}, vol.~44, no.~5, pp. 1166--1181, May
  2006.

\bibitem{boeing2017osmnx}
G.~Boeing, ``{OSMnx: New methods for acquiring, constructing, analyzing, and
  visualizing complex street networks},'' \emph{Computers, Environment and
  Urban Systems}, vol.~65, pp. 126--139, Sep. 2017.

\bibitem{3GPP38913}
3GPP, ``{TR 38.913, Study on Scenarios and Requirements for Next Generation
  Access Technologies, V14.1.0},'' 2017.

\end{thebibliography}

\end{document}